\documentclass[twocolumn,preprint]
{aastex631}
\usepackage{amsmath, amstext, natbib, apjfonts, verbatim, graphicx, xcolor, longtable, movie15, animate, float}

\newcommand{\OII}{\hbox{{\rm [O}\kern 0.1em{\sc ii}{\rm ]}}}
\newcommand{\NeIII}{\hbox{{\rm [Ne}\kern 0.1em{\sc iii}{\rm ]}}}
\newcommand{\OIII}{\hbox{{\rm [O}\kern 0.1em{\sc iii}{\rm ]}}}
\newcommand{\Hb}{\hbox{{\rm H}$\beta$}}
\newcommand{\Ha}{\hbox{{\rm H}$\alpha$}}
\newcommand{\NII}{\hbox{{\rm [N}\kern 0.1em{\sc ii}{\rm ]}}}
\newcommand{\SII}{\hbox{{\rm [S}\kern 0.1em{\sc ii}{\rm ]}}}
\newcommand{\Lya}{\hbox{{\rm Ly}$\alpha$}}
\newcommand{\HeI}{\hbox{{\rm He}\kern 0.1em{\sc i}}}
\newcommand{\HeII}{\hbox{{\rm He}\kern 0.1em{\sc ii}}}
\newcommand{\HII}{\hbox{{\rm H}\kern 0.1em{\sc ii}}}

\begin{document}

\title{\large \bf A Census from JWST of Extreme Emission Line Galaxies Spanning the Epoch of Reionization in CEERS}

\shorttitle{CEERS Extreme Emission-Line Galaxies}
\shortauthors{Davis et al.}

\author[0000-0001-8047-8351]{Kelcey Davis}
\altaffiliation{NSF Graduate Research Fellow}
\affiliation{Department of Physics, 196A Auditorium Road, Unit 3046, University of Connecticut, Storrs, CT 06269, USA}

\author[0000-0002-1410-0470]{Jonathan R. Trump}
\affil{Department of Physics, 196A Auditorium Road, Unit 3046, University of Connecticut, Storrs, CT 06269, USA}

\author[0000-0002-6386-7299]{Raymond C. Simons}
\affil{Department of Physics, 196A Auditorium Road, Unit 3046, University of Connecticut, Storrs, CT 06269, USA}

\author[0000-0001-8688-2443]{Elizabeth J.\ McGrath}
\affiliation{Department of Physics and Astronomy, Colby College, Waterville, ME 04901, USA}

\author[0000-0003-3903-6935]{Stephen M.~Wilkins} 
\affiliation{Astronomy Centre, University of Sussex, Falmer, Brighton BN1 9QH, UK}
\affiliation{Institute of Space Sciences and Astronomy, University of Malta, Msida MSD 2080, Malta}

\author[0000-0002-7959-8783]{Pablo Arrabal Haro}
\affiliation{NSF's National Optical-Infrared Astronomy Research Laboratory, 950 N. Cherry Ave., Tucson, AZ 85719, USA}

\author[0000-0002-9921-9218]{Micaela B. Bagley}
\affiliation{Department of Astronomy, The University of Texas at Austin, Austin, TX, USA}

\author[0000-0001-5414-5131]{Mark Dickinson}
\affiliation{NSF's National Optical-Infrared Astronomy Research Laboratory, 950 N. Cherry Ave., Tucson, AZ 85719, USA}

\author[0000-0003-0531-5450]{Vital Fernández}
\affiliation{Michigan Institute for Data Science, University of Michigan, 500 Church Street, Ann Arbor, MI 48109, USA}

\author[0000-0001-5758-1000]{Ricardo O. Amor\'{i}n}
\affiliation{ARAID Foundation. Centro de Estudios de F\'{\i}sica del Cosmos de Arag\'{o}n (CEFCA), Unidad Asociada al CSIC, Plaza San Juan 1, E--44001 Teruel, Spain}
\affiliation{Departamento de Astronom\'{i}a, Universidad de La Serena, Av. Juan Cisternas 1200 Norte, La Serena 1720236, Chile}

\author[0000-0001-8534-7502]{Bren E. Backhaus}
\affil{Department of Physics, 196A Auditorium Road, Unit 3046, University of Connecticut, Storrs, CT 06269, USA}

\author[0000-0001-7151-009X]{Nikko J. Cleri}
\affiliation{Department of Physics and Astronomy, Texas A\&M University, College Station, TX, 77843-4242 USA}
\affiliation{George P.\ and Cynthia Woods Mitchell Institute for Fundamental Physics and Astronomy, Texas A\&M University, College Station, TX, 77843-4242 USA}

\author[0000-0003-1354-4296]{Mario Llerena}
\affiliation{INAF-Osservatorio Astronomico di Roma, via Frascati 33, 00078, Monteporzio Catone, Italy}
\affiliation{Departamento de Astronom\'{i}a, Universidad de La Serena, Av. Juan Cisternas 1200 Norte, La Serena 1720236, Chile}

\author[0000-0001-6776-2550]{Samantha W. Brunker}
\affiliation{Department of Physics, 196A Auditorium Road, Unit 3046, University of Connecticut, Storrs, CT 06269, USA}

\author[0000-0001-6813-875X]{Guillermo Barro}
\affiliation{Department of Physics, University of the Pacific, Stockton, CA 90340 USA}

%Alpha order

\author[0000-0003-0492-4924]{Laura Bisigello}
\affiliation{Dipartimento di Fisica e Astronomia "G.Galilei", Universit\'a di Padova, Via Marzolo 8, I-35131 Padova, Italy}
\affiliation{INAF--Osservatorio Astronomico di Padova, Vicolo dell'Osservatorio 5, I-35122, Padova, Italy}

\author[0000-0001-5384-3616]{Madisyn Brooks}
\affil{Department of Physics, 196A Auditorium Road, Unit 3046, University of Connecticut, Storrs, CT 06269, USA}

\author[0000-0001-6820-0015]{Luca Costantin}
\affiliation{Centro de Astrobiolog\'{\i}a (CAB), CSIC-INTA, Ctra. de Ajalvir km 4, Torrej\'on de Ardoz, E-28850, Madrid, Spain}

\author[0000-0002-6219-5558]{Alexander de la Vega}
\affiliation{Department of Physics and Astronomy, University of California, 900 University Ave, Riverside, CA 92521, USA}

\author[0000-0003-4174-0374]{Avishai Dekel}
\affiliation{Racah Institute of Physics, The Hebrew University of Jerusalem, Jerusalem 91904, Israel}

\author[0000-0001-8519-1130]{Steven L. Finkelstein}
\affiliation{Department of Astronomy, The University of Texas at Austin, Austin, TX, USA}

\author[0000-0001-6145-5090]{Nimish P. Hathi}
\affiliation{Space Telescope Science Institute, 3700 San Martin Drive, Baltimore, MD 21218, USA}

\author[0000-0002-3301-3321]{Michaela Hirschmann}
\affiliation{Institute of Physics, Laboratory of Galaxy Evolution, Ecole Polytechnique Fédérale de Lausanne (EPFL), Observatoire de Sauverny, 1290 Versoix, Switzerland}

\author[0000-0001-9187-3605]{Jeyhan S. Kartaltepe}
\affiliation{Laboratory for Multiwavelength Astrophysics, School of Physics and Astronomy, Rochester Institute of Technology, 84 Lomb Memorial Drive, Rochester, NY 14623, USA}

\author[0000-0002-6610-2048]{Anton M. Koekemoer}
\affiliation{Space Telescope Science Institute, 3700 San Martin Drive, Baltimore, MD 21218, USA}

\author[0000-0003-1581-7825]{Ray A. Lucas}
\affiliation{Space Telescope Science Institute, 3700 San Martin Drive, Baltimore, MD 21218, USA}

\author[0000-0001-7503-8482]{Casey Papovich}
\affiliation{Department of Physics and Astronomy, Texas A\&M University, College Station, TX, 77843-4242 USA}
\affiliation{George P.\ and Cynthia Woods Mitchell Institute for Fundamental Physics and Astronomy, Texas A\&M University, College Station, TX, 77843-4242 USA}

\author[0000-0003-4528-5639]{Pablo G. P\'erez-Gonz\'alez}
\affiliation{Centro de Astrobiolog\'{\i}a (CAB), CSIC-INTA, Ctra. de Ajalvir km 4, Torrej\'on de Ardoz, E-28850, Madrid, Spain}

\author[0000-0003-3382-5941]{Nor Pirzkal}
\affiliation{ESA/AURA Space Telescope Science Institute}

\author[0000-0002-9415-2296]{Giulia Rodighiero}
\affiliation{Department of Physics and Astronomy, Università degli Studi di Padova, Vicolo dell’Osservatorio 3, I-35122, Padova, Italy}
\affiliation{INAF - Osservatorio Astronomico di Padova, Vicolo dell’Osservatorio 5, I-35122, Padova, Italy}

\author[0000-0002-8018-3219]{Caitlin Rose}
\affil{Laboratory for Multiwavelength Astrophysics, School of Physics and Astronomy, Rochester Institute of Technology, 84 Lomb Memorial Drive, Rochester, NY 14623, USA}

\author[0000-0003-3466-035X]{{L. Y. Aaron} {Yung}}
\altaffiliation{NASA Postdoctoral Fellow}
\affiliation{Astrophysics Science Division, NASA Goddard Space Flight Center, 8800 Greenbelt Rd, Greenbelt, MD 20771, USA}

%get author blocks for:  everyone should be acounted for :)

\author{CEERS collaborators}

%Please add author blocks right here if I am missing you. I will organize them.

\begin{abstract}

We present a sample of 1165 extreme emission-line galaxies (EELGs) at $4<z<9$ selected using James Webb Space Telescope (JWST) NIRCam photometry in the Cosmic Evolution Early Release Science (CEERS) program. We use a simple method to photometrically identify EELGs with \Hb\ + \OIII\ (combined) or \Ha\ emission of observed-frame equivalent width ($\mathrm{EW}) >5000$\AA. JWST/NIRSpec spectroscopic observations of a subset (34) of the photometrically selected EELGs validate our selection method: all spectroscopically observed EELGs confirm our photometric identification of extreme emission, including some cases where the SED-derived photometric redshifts are incorrect. We find that the medium-band F410M filter in CEERS is particularly efficient at identifying EELGs, both in terms of including emission lines in the filter and in correctly identifying the continuum between \Hb\ + \OIII\ and \Ha\ in the neighboring broad-band filters.
We present examples of EELGs that could be incorrectly classified at ultra-high redshift (z>12) as a result of extreme \Hb\ + \OIII\ emission blended across the reddest photometric filters.
We compare the EELGs to the broader (sub-extreme) galaxy population in the same redshift range and find that they are consistent with being the bluer, high equivalent width tail of a broader population of emission-line galaxies.
%vary in compactness between emission and continuum observed sizes, and exceed rest frame equivalent width theory predictions in the most extreme cases.
%more compact, and have more luminous continua than the background population.
% The blue rest frame U-B color for the continuum of these EELGs suggest a portion of our sample are young galaxies with a central active galactic nuclei (AGN) causing the extreme emission lines while the remainder of our EELGs with a redder U-B color are early star-bursting galaxies, although there is not a clear distinction between the two populations in color space. We report an evolution of the ratio between sizes in extreme emission and continuum capturing images that evolves with rest-frame equivalent width, a potential tool for de-tangling the AGN and starburst populations.
The highest-EW EELGs tend to have more compact emission-line sizes than continuum sizes, suggesting that active galactic nuclei are responsible for at least some of the most extreme EELGs. Photometrically inferred emission-line ratios are consistent with ISM conditions with high ionization and moderately low metallicity, consistent with previous spectroscopic studies.

\end{abstract}
 
\section{Introduction}

JWST \citep[]{Gardner2023} has opened a new window on galaxies that dwell in the epoch of reionization, a phase change in the early intergalactic medium (IGM). Understanding the cause and progression of reionization is a task at the forefront of modern cosmology. Observations of the cosmic microwave background and the \Lya\ forest in $z>5$ quasars offer conclusive evidence that reionization occurred between $z \approx 11$ and  $z \approx 7$ \citep[e.g.][]{Spergel2007, Qin2021, Fan2002}. This reionization dominated evolution of the IGM at early epochs \citep[e.g.][]{Qin2021, Fan2002}. However, the specific catalysts of reionization remain elusive. We expect both star-forming galaxies \citep[e.g.][]{Robertson2010} and quasars \citep[e.g.][]{Madau2015} to contribute to reionization through background ultraviolet emission, but the properties (e.g., mass, star formation rate, morphology, and environmental density) of these sources and relative contributions of different source types that drove reionization are unknown.

Galaxies at $z > 6$ consistently exhibit higher star formation rates (SFRs) than found in galaxies in the local Universe \citep[e.g.][]{Shibuya2015, Matthee2022}. A sharp increase in SFR results in increased nebular emission as shown in the evolution of equivalent width of \Hb\ + \OIII\ and SFR with redshift \citep[]{Smit2014}.  In the most extreme cases, the nebular emission is so bright that it significantly impacts broadband photometry, and we call these systems ``Extreme Emission-Line Galaxies'' \citep[EELGs;][]{vanderWel2011}.
The strength of a galaxy's emission lines at $5<z<7$ are observed to correlate with \Lya\ escape \citep[e.g.][]{Tang2023, Endsley2023}, indicating that EELGs play an important and outsized role in driving reionization. Understanding EELGs is then critically important for a global picture of galaxy formation and evolution through cosmic history.

EELGs at high redshift were first inferred from excess emission in Spitzer photometry \citep[]{Chary2005, Zackrisson2008, Caputi2017} and later confirmed with Hubble Space Telescope (HST) near-infrared spectroscopy \citep{vanderWel2011, Maseda2014}. Previous studies have used optical spectroscopy to identify EELGs as the low-metallicity, high-SFR tail of compact blue dwarf galaxies associated with spatially concentrated star formation activity \citep[]{vanderWel2011, Amorin2014, Amorin2015, Calabro2017}. In the nearby universe, at $z<0.1$, EELGs have been historically referred to as ``\HII\ galaxies'' due to spectroscopic similarities to \HII\ regions \citep[]{Terlevich1991, Kniazev2004}. EELGs have been reported in the Sloan Digital Sky Survey (SDSS) at $0.35 > z > 0.1$ and classified as ``green peas'' for their compactness and green appearance in SDSS photometric filters due to unusually strong \OIII\ lines falling in the $r$-band \citep[e.g.][]{Cardamone2009, Izotov2011, Amorin2010, Amorin2012}. However, low-redshift EELGs are $\sim$1 order of magnitude less common than at higher redshift.
In this work we find 1165 EELGs in 10 pointings of 9.7 square arcminutes each, or about 12 per square arcminute, as compared to 4 per square arcminute at $z<0.5$ \citep[]{vanderWel2011}). 

EELGs have been identified throughout the $z>3$ universe in other JWST deep-field searches, with consistently demonstrated high SFRs
% , extreme nebular emission, 
and an increased number density with redshift \citep[]{Endsley2023, Matthee2022, Perez-Gonzalez2023}. By increasing the census of known high-redshift EELGs, we work towards a map of extreme nebular emission and eventually an understanding of how these sources impacted an early, evolving universe. %We
A larger census of EELGs across cosmic time will
expand our understanding of how these early galaxies differ from background populations in stellar mass, SFRs, morphology, gas conditions (e.g., ionization, metallicity, gas density), environment, and Ly$\alpha$ escape. Because EELGs are likely the dominant sources of reionizing photon production, studying their demographics in the early Universe can reveal the relative importance of star formation and AGN in driving reionization. 
Early JWST spectroscopy work indicates high-ionization, low-metallicity interstellar medium (ISM) conditions in high redshift galaxies that are consistent with an impact from strong nebular emission \citep[]{Backhaus2023, Sanders2023, Trump2023, Cleri2023, Brinchmann2023}. Spectroscopy of high-redshift galaxies has revealed a surprising population of broad-line and high-ionization AGN, most of which have extreme emission lines affecting their photometry \citep[]{Larson2023, Maiolino2023a, Maiolino2023b, Brinchmann2023}. Early JWST spectroscopy has also identified \Lya\ emission at $z = 10.6$ \citep[]{Bunker2023} that suggests rapid reionization of the IGM in certain regions. Identifying and characterizing EELGs will fill in our picture of strong nebular emission in the early universe.

We present a census of the EELGs in the Cosmic Evolution Early Release Science Survey (CEERS; PI: S.Finkelstein) using imaging and spectroscopy from JWST. Future work will explore alternative selection methods and characterization of physical properties of EELGs (Llerena et al., in prep) as well as spatial correlations and potential improvements in photometric redshift fitting. 
In Section 2, we present the CEERS observational data. In Section 3, we describe our photometric selection and spectroscopic verification along with a classification schematic for our sample. In Section 4, we explore the properties of the EELGs.
We assume a flat $\Lambda$CDM cosmology and $H_0 = 67.4\,\,\mathrm{km} \,\mathrm{s}^{-1}\,\mathrm{Mpc}^{-1}$ and $\Omega_\mathrm{M}$ = 0.315 \citep[]{Planck2020}.

\section{CEERS Observational Data}

We use the CEERS NIRCam imaging to identify EELGs, and the partial coverage of NIRSpec spectroscopy in CEERS for spectroscopic validation. This Early Release Science survey from JWST provides a new look at extreme emission sources in the $z>4$ universe.

\subsection{NIRCam Photometry}

CEERS has coverage from NIRCam, NIRSpec, and MIRI. We define our search in NIRCam photometry data and calibrate it with NIRSpec spectroscopy. EELGs in the CEERS field are selected using an internal photometry catalog (Finkelstein et al., in prep). CEERS photometry covers 10 pointings, each covering 9.7 square arcminutes. The photometry filters used, their associated widths, and depths are described in Table \ref{tbl:filters}.

\begin{table}[]
    \begin{centering}
        
    \begin{tabular}{c|c|c|c}
          \hline
         Filter & Depth & Width [$\mu$m] & Band Center [$\mu$m] \\ [0.5ex] 
         \hline\hline
         F115W & 29.15 & 0.225 & 1.154 \\ 
         \hline
         F150W & 28.9 & 0.318 & 1.501 \\
         \hline
         F200W & 28.97 & 0.461 & 1.990 \\
         \hline
         F277W & 29.15 & 0.672 & 2.786 \\
         \hline
         F356W & 28.95 & 0.787 & 3.553 \\
         \hline
         F410M & 28.4 & 0.436 & 4.092 \\
         \hline
         F444W & 28.6 & 1.024 & 4.421 \\ [1ex] 
         \hline
    \end{tabular}

    \end{centering}
    \caption{NIRCam filter data utilized in CEERS and in this photometric survey. Depths in AB Mag \citep[]{Oke1983} are for a 5$\sigma$ point source.}
    \label{tbl:filters}
\end{table}

Detailed descriptions of the processing we carried out with the JWST pipeline \citep{Bushouse2022} to produce our imaging and photometric data can be found in \citet[]{Bagley2023}. For sources in NIRCam photometry, we denote identifiers as nircamX-Y where ``X'' is the NIRCam photometric pointing and ``Y'' is the photometric catalog identifier. Sources with NIRSpec spectroscopy have an additional identifier denoted as CEERS-Z, where ``Z'' is the spectroscopic ID on the NIRSpec multi-shutter array (MSA) design.

We also utilize HST photometry from the Cosmic Assembly Deep Extragalactic Legacy Survey \citep[CANDELS;][]{Grogin2011, Koekemoer2011} to compute photometric redshifts and establish the number of filters with photometric detections. This includes photometry from the ACS F606W and F814W filters and WFC3/IR F105W, F125W, F140W, and F160W filters.

\subsection{Derived Galaxy Properties}

We report photometric redshifts for our sample from Finkelstein et al. (2023, in prep.). Photometric redshifts were calculated with \texttt{EAZY} \citep[]{Brammer2008, Brammer2010} as described in \citet[]{Finkelstein2023} for the 13 broadband filters including both JWST and HST photometry, utilizing the template library from \citet[]{Conroy2010} expanded with high-$z$ templates from \citet[]{Larson22}. \texttt{EAZY} was run with three selective runs to capture the variety of sources in CEERS, (1) with a maximum redshift of $z = 20$ using fiducial Kron aperture corrected photometry, (2) again with the fiducial Kron aperture but with a maximum redshift of $z = 7$, and (3) replacing the Kron fluxes with the flux measured in $d=0.2$ arcsec diameter apertures as described in further detail in Finkelstein et al. (2023, in prep.).

We attempted to derive stellar masses through modified runs of \texttt{EAZY} but 
% were unable to produce realistic masses for this epoch in cosmic time and so we do not report them here. 
found that the default \texttt{EAZY} templates are insufficient for reliably fitting galaxies with extreme emission lines. Specifically, the \texttt{EAZY} templates struggle to reproduce the extreme emission fluxes of the EELG population and \texttt{EAZY} instead attempts to account for the elevated photometry by increasing the continuum flux. This leads to an overestimation of stellar mass. We will revisit stellar masses for EELGs in future work (Llerena et al., in prep). We use continuum luminosity, defined as the mean luminosity of all JWST photometric filters with 1$\sigma$-clipping of outliers (due to emission lines), instead of stellar masses for exploration of EELG properties that is independent of SED modeling.

We measure galaxy sizes using effective semi-major axes from McGrath et al. (2023, in prep.), calculated with \texttt{GALFIT} \citep{Peng2010} v3.0.5 for the wide-band photometric JWST filters. \texttt{GALFIT} was separately run on each of these broadband filters for sources of ${\rm F356W} > 28.5$.

All line fits are performed with \texttt{lmfit} and Spearman correlation coefficients are calculated with \texttt{scipy.stats.spearmanr}.

\subsection{NIRSpec Spectroscopy}

We verify our sample through spectroscopy taken with NIRSpec \citep[]{Jakobsen2022}.
% and by confirming photometric redshift estimations.
CEERS has both PRISM and M-grating coverage that overlaps with a subset of the photometry. 39 of our photometrically selected candidates have NIRSpec coverage from either M-grating or PRISM data. We select NIRSpec targets that overlap with our sample and visually inspect spectra to inform our EELG definition. Description of the NIRSpec data reduction in CEERS can be found in Arrabal Haro et al. (2023, in prep). Spectra were processed using the STScI calibration pipeline version 1.8.5 across three stages as described in \citet[]{Haro2023b}.
%NIRSpec coverage of CEERS overlapping with our sources primarily utilized PRISM/CLEAR data. NIRSpec data of CEERS also includes G140M/F100LP, G235M/F170LP and G395M/F290LP gratings each spanning 1-5 $\mu m$ but these spectra are de-prioritized in our matching favoring PRISM/CLEAR spectra and include 12 of the 40 spectra explored here.
Most (27/39) of our EELGs with NIRSpec observations have prism spectra, while 12 EELGs have spectra with G140M/F100LP, G235M/F170LP and G395M/F290LP.

Only a subset of the CEERS NIRCam photometry footprint is covered by NIRSpec pointings, and the Micro-Shutter Assembly (MSA; \citep[]{Ferruit2022}) constraints mean that only a subset of sources are allocated spectroscopic apertures. Only 3\% (39/1227) of our photometrically selected EELGs were observed by NIRSpec. We assume that EELGs with spectroscopic coverage are broadly representative of the larger population of EELGs with similar SEDs, and we use the spectroscopically observed EELGs to inform the redshifts of EELGs that have similar SED presentations but lack spectroscopy.

\section{EELG Identification}

\begin{figure}
    \centering
    \includegraphics[scale = 0.2]{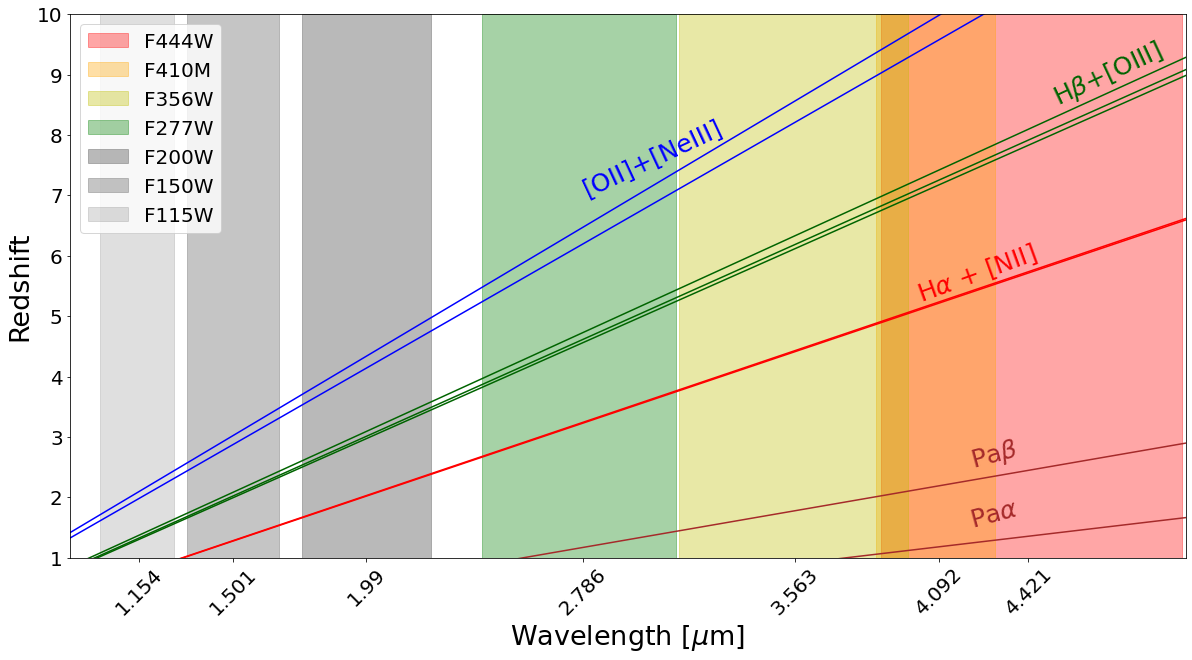}
    \caption{CEERS NIRCam filters and their coverage of extreme emission lines as a function of redshift. Colored filters are utilized in photometric identification while gray filters are not.
    % They fall blueward of where we expect extreme emission at the targeted redshifts.
    We select EELGs using only \Ha\ and \Hb\ + \OIII, but also show Pa$\alpha$, Pa$\beta$, and \NeIII\ + \OII\ to showcase their locations in photometric filters at lower and higher redshifts.}
    \label{fig:lines}
\end{figure}

We identify EELGs using a photometrically inferred equivalent width (EW) selection, requiring an observed-frame equivalent width of $>$5000\AA\ in at least one of the four reddest CEERS NIRCam filters. This EW threshold is motivated by previous work \citep{vanderWel2011} and by the requirement for extreme emission to impact the broadband NIRCam filters. In particular, we identify \Ha\ (potentially blended with the \NII\ and \SII\ doublets) and/or the combination of \Hb\ + \OIII\ lines in the reddest NIRCam filters of CEERS: F277W, F356W, F410M, or F444W. Figure \ref{fig:lines} indicates the redshift ranges for these emission lines to appear in the relevant NIRCam filters: $2.5<z<6.5$ for \Ha\ and $4<z<9$ for \Hb\ + \OIII. In Section 3.2 we show that our photometric selection is confirmed by NIRSpec spectroscopy. Our EELG photometric selection method and the spectroscopic confirmation are described in detail below.

\subsection{Photometric Selection}

We select EELGs by requiring an observed-frame equivalent width of $>$5000\AA\ as inferred from the photometric flux in a given filter relative to the continuum flux, given by:

\begin{equation}
    EW = \left( \frac{F_\nu - C_\nu}{C_\nu} \right) \Delta \lambda 
    \label{eqn:eq1}
\end{equation}
Here $F_\nu$ is the flux in a specific filter, $\Delta\lambda$ is the filter width, and $C_\nu$ is the continuum flux. An observed-frame equivalent width of 5000\AA\ corresponds to rest-frame equivalent widths of 500-1000\AA\
% (where ${\rm EW_{obs}} = (1+z) {\rm EW_{rf}}$)
over the redshift range, $4<z<9$, for our identified EELGs.

The continuum flux is defined as the mean of the NIRCam photometric fluxes, excluding any filters that differ from the mean by $>$1$\sigma$ (i.e., rejecting outlier photometry caused by extreme emission). We exclude the bluer HST filters from the continuum flux estimate because the HST photometry is less reliable due to being shallower than the NIRCam observations. Visual inspection also revealed that the HST photometry was more likely to suffer from artifacts like detector edges or issues with source deblending. Our continuum definition assumes a flat (in $C_\nu$ vs. $\lambda$) continuum: this is generally a good assumption for the young, blue stellar populations generally associated with extreme emission line galaxies (e.g., \citealt{vanderWel2011,Endsley2023}).

JWST has revealed a surprising population of galaxies with both red continua and bright emission lines (e.g., \citealt{Kocevski2023, Barro2023, Haro2023a, Labbe2023, Perez-Gonzalez2023}): these systems are not well-described by our assumption of a flat continuum and so they are incorrectly classified as EELGs under the assumption of a flat, blue continua. We return to the topic of emission-line galaxies with red continua in Section 3.6.
% and do not report them as part of our EELG sample although they pass all described selection criteria.
%, 3455 candidates pass the initial EW cutoff of >5000. 
%***

We use additional selection criteria to remove photometric artifacts and/or poorly measured sources. We required that sources not lie within 100 pixels of detector edges and also excluded sources that are within $1\arcsec$ of any other source that is $>$100~nJy brighter than the target source.
%***
%>10nJy in 5 filts 1849
To remove noisy and potentially spurious detections, we require that sources be brighter than 10~nJy in a minimum of five detected filters. 

%post chi2 cut, 3261 remain 

We apply an additional technique to remove red galaxies from our sample, incorrectly included in the initial EW calculation (Equation \ref{eqn:eq1}) due to the assumption of a flat continuum. We first define a linear red continuum model for each galaxy as a straight line connecting the F200W and F444W filters. For galaxies that have a linear continuum with positive (red) slope, we measure a $\chi^2$ between the NIRCam photometry and the linear continuum model.
Our red-galaxy rejection criterion is defined as:

\begin{equation}
    \chi^2 =  \sum \frac{(F_i - M_i)^2}{\sigma_i^2} < 20
    \label{eqn:chi}
\end{equation}

Here $F_i$ and $\sigma_i$ are the flux density and uncertainty in each filter, and $M_i$ is the sloped red continuum model.  Visual inspection of the spectral energy distributions (SEDs) indicated that requiring $\chi^2 > 20$ is the appropriate threshold to reliably reject low-redshift red galaxies without removing genuine high-redshift EELGs.
We visually inspect all of the EELG candidates and reject an additional 62 red-continuum galaxies that formally pass the $\chi^2$ cut above due to emission lines that affect the photometry but are not extreme We further discuss these red emission line galaxies in section 3.6.

As a final step we visually inspected SEDs and photometric images of the remaining sources to ensure sample reliability. We remove artifacts and sources near detector edges. All sources were categorized by their inferred extreme emission line wavelengths and corresponding inferred redshift. Our EW selection, photometric quality requirements, and visual inspection identified 1165 candidate EELGs with observed frame $\mathrm{EW} > 5000\,$\AA\ in at least one of the F277W, F356W, F410M, or F444W filters. Of these, 702 fall in our two highest confidence tiers.
%associated with our galaxy property reports.
We discuss these confidence tiers in further detail in Section 3.3.

\subsection{Spectroscopic Confirmation}

\begin{figure*}[ht]
    \centering
    \setlength\unitlength{1cm}
    \includegraphics[scale = 0.3]{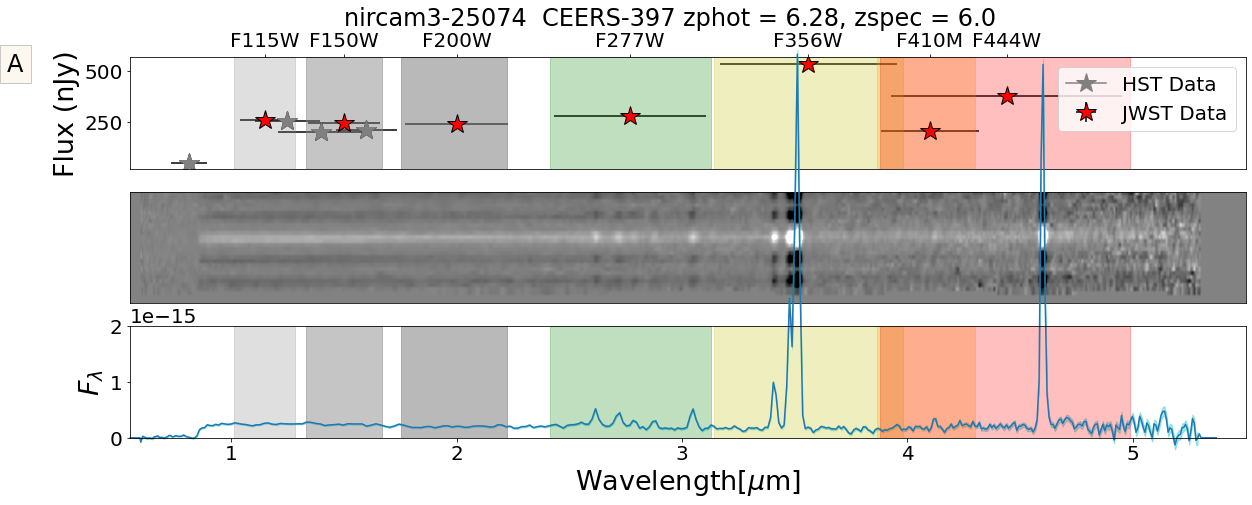}
    \includegraphics[scale = 0.3]{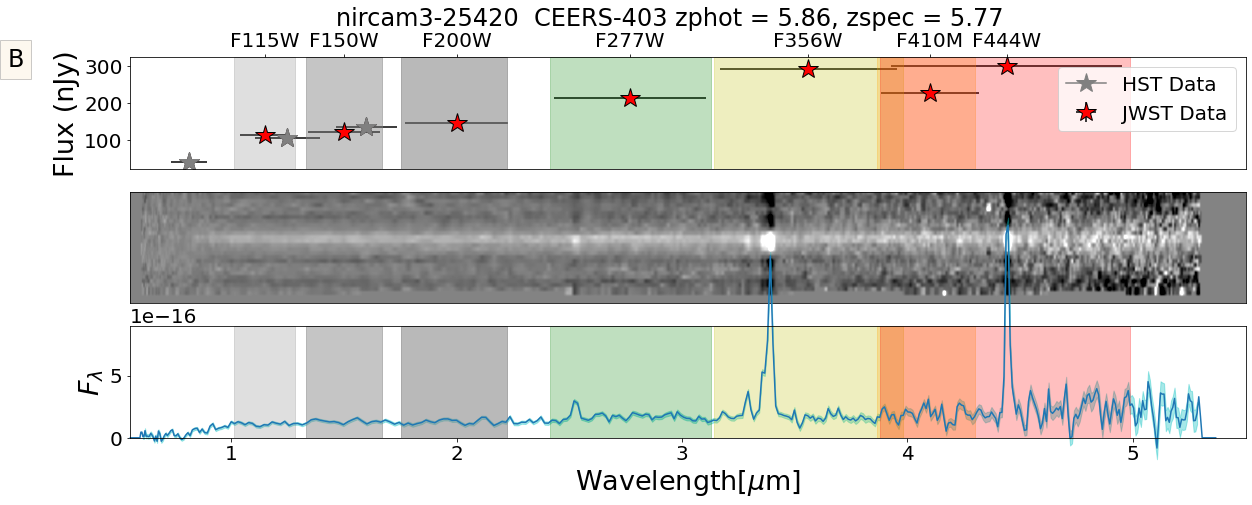}
    \caption{Example of two spectroscopically confirmed EELGs with \Hb + \OIII\ in the F356W filter and \Ha\ in the F444W filter. Shaded regions represent NIRCam filters as defined in Figure \ref{fig:lines}. Filter widths displayed are the bandwidths for the associated filters.}
    
    \label{fig:t1_spec}
\end{figure*}

%1 Undetected, 28 PRISM, 12 M grating, 4 reddened for:
%40 total, 35 EELGs
% 515: solved, matched to nearby source accidentaly
% 545 is Ha in 410, O3 split

We use CEERS NIRSpec spectroscopy to validate and confirm our photometric selection of EELGs.
Of our 1165 photometrically selected EELGs (excluding red emission-line galaxies), 36 were observed with NIRSpec. 

%Two examples of the photometrically selected EELGs that are spectroscopically confirmed are shown in Figure \ref{fig:t1_spec}.
In 34 cases, we spectroscopically confirm the presence of extreme emission lines consistent with the redshift range and the photometric filter identified in our EELG selection. Notably, two
% MPT IDs 3585, 933 
of the 34 spectroscopically confirmed EELGs have catastrophically underestimated photometric redshifts (CEERS-3585 and CEERS-933). In these cases, our EW-based selection is correct while the more sophisticated template-based photometric redshift returns a catastrophic outlier due in part to placement of the $\sim$2$\mu$m stellar bump at the filters that include extreme emission lines. The incorrect photometric redshifts particularly suffer from the lack of a medium-band filter to sample the continuum between extreme emission lines.

Two of the 36 spectra were dominated by noise, with no apparent emission lines or continuum, consistent with the faint photometry in the galaxy ($<$50~nJy continuum, $<$100~nJy emission). An additional 3 galaxies were spectroscopically identified to have a red continuum and emission lines that are bright enough to impact broadband photometry but not meeting our definition of ``extreme'' (i.e., their emission lines have observed-frame EW$<$5000\AA): these systems are discussed in more detail in Section 3.6. None of the 36 EELGs with spectra were revealed to be at a lower redshift than implied by the EELG selection. That is, in zero cases do the spectra disagree with our EELG prediction. 

\begin{figure*}
    \centering
    \includegraphics[scale = 0.27]{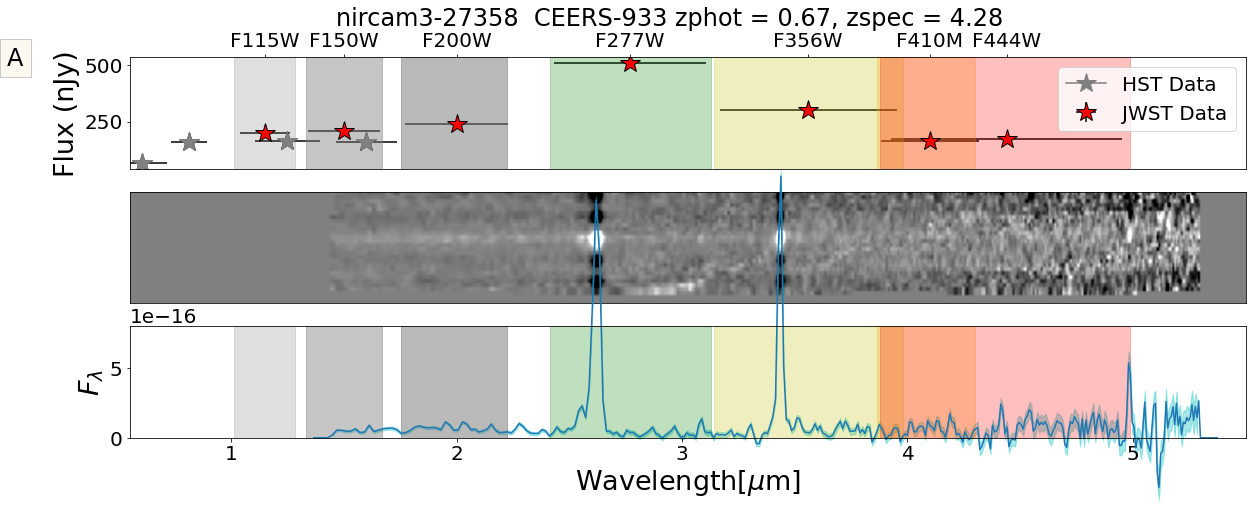}
    \includegraphics[scale = 0.27]{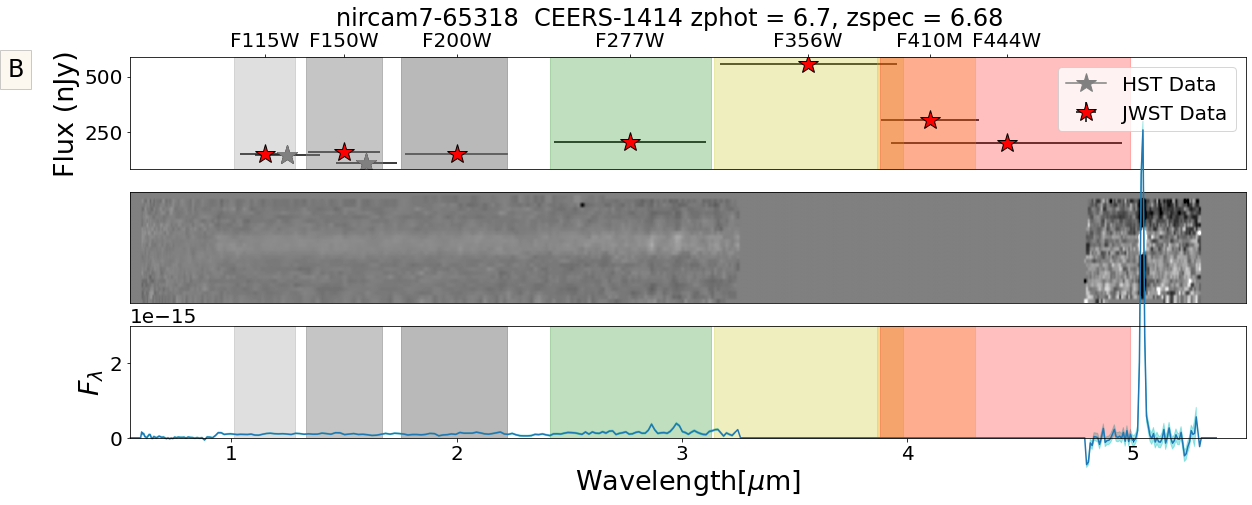}
    \includegraphics[scale = 0.27]{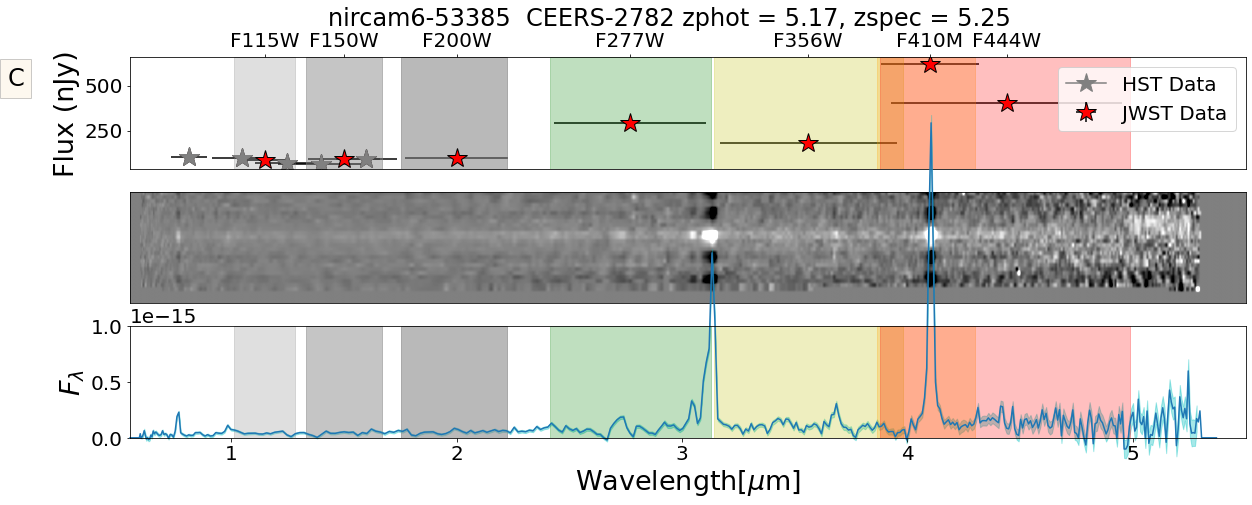}
    \includegraphics[scale = 0.27]{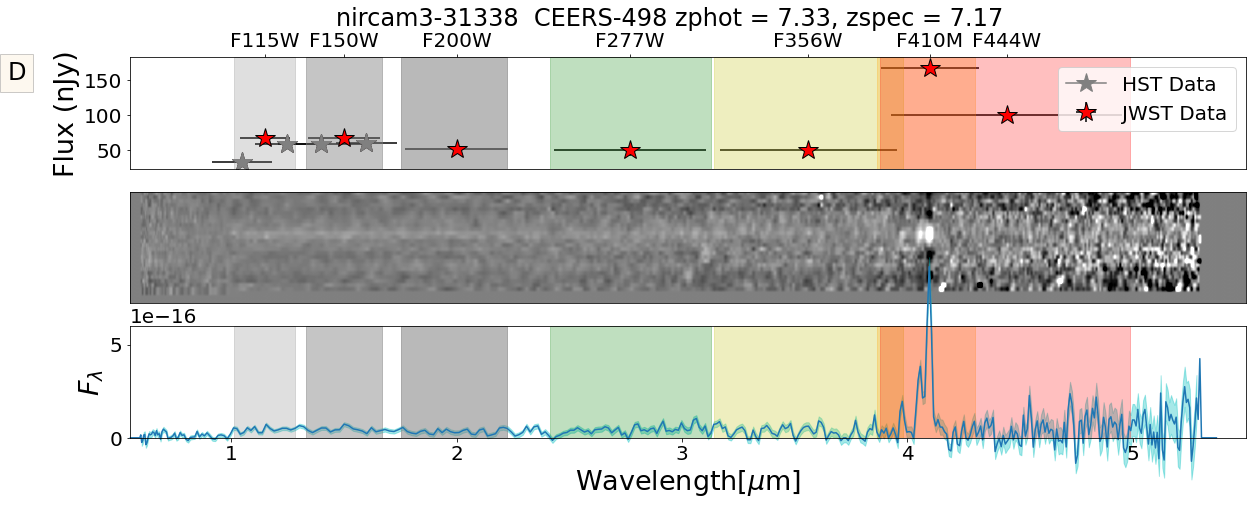}
    \includegraphics[scale = 0.27]{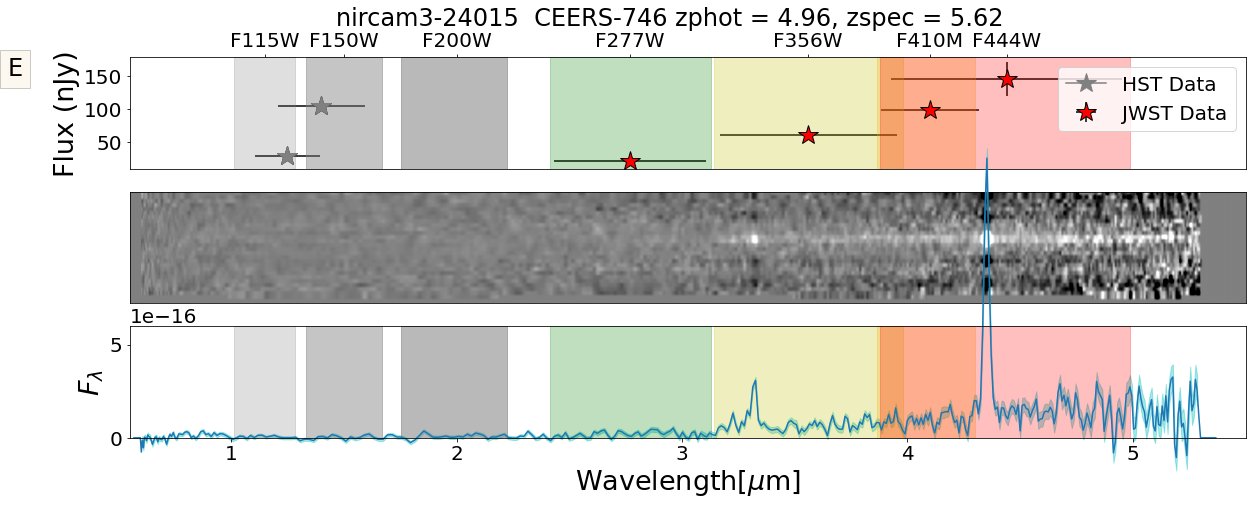}
    \caption{The EELGs in each panel (A-E) are discussed in detail in the text of Section 3.2.
    }
    \label{fig:other_specs}
\end{figure*}

We present a few examples of spectroscopically confirmed EELGs in Figures \ref{fig:t1_spec} and \ref{fig:other_specs}. Figure \ref{fig:t1_spec} shows two sources that have extreme nebular emission from both \Hb\ + \OIII\ in F356W and \Ha\ in F444W. These emission lines are separated by F410M, making them straightforward to isolate from a well-defined continuum. Although both sources present the same emission lines in the same filters, they have a 
noticeable difference in their continuum emission. The source in Figure \ref{fig:t1_spec} panel A has a blue continuum while the source in panel B does not. This demonstrates the diversity among EELGs that otherwise share similar SED shapes.
% , in this case extreme emission-line filter location and spectroscopic redshift.
Figure \ref{fig:other_specs} presents additional spectroscopically confirmed EELGs that highlight the breadth of properties of our spectroscopically confirmed sample. The individual sources shown in each panel of Figure \ref{fig:other_specs} are discussed in detail below.

\begin{itemize}

    \item \textbf{Panel A} showcases an example of catastrophic disagreement ($\Delta z >3$) between photometric and spectroscopic redshift. 
    In this case, \Hb\ + \OIII\ falls in F277W and \Ha\ falls in F356W. However, the lack of medium-band filter between F277W and F356W prevents reliable measurement of the continuum. This causes confusion for SED-based photometric redshift fitting and result in a lower, incorrect redshift. We discuss other sources like this in Section 3.3. 

    \item \textbf{Panel B} shows an example of a galaxy selected to have extreme emission in a single filter, \Hb\ + \OIII\ in F356W. The spectroscopic data has a detector gap spanning most of F356W, F410M, and F444W, but even with this missing wavelength coverage, NIRSpec still includes the \Ha\ line lying just redward of the F444W photometric filter. This is a great illustration of our ability to identify single-line sources from the photometry and indicates how additional MIRI photometry would recover redder emission lines and enable a better understanding of EELGs at $z>7$.
    %If we want to lose a source from the figure to account for room, it is probably this one.

    \item \textbf{Panel C} is another example of a multi-filter line detection. It includes \Hb\ + \OIII, however the line falls between F277W and F356W, distorting the extreme emission between two photometric filters. \Ha\ falls within F410M and passes our observed-frame $\mathrm{EW} > 5000$\AA\ selection both through our photometric identification and by spectroscopic line measurements. Although both lines are extreme, the source only passes our selection because \Ha\ is isolated by the F410M filter. In the absence of a F300M medium-band photometry filter between F277W and F356W, galaxies in the redshift range ($5.0 \lesssim z 5.4$ for \Hb\ + \OIII) have emission lines that blend between the F277W and F356W filters and can be missed by our selection. We also note that this galaxy has \Lya\ emission detected at the bluest end of the spectra.

    \item \textbf{Panel D} showcases a single-line EELG. This is another demonstration of the utility of F410M, with \Hb\ + \OIII\ falling in the medium-band filter such that it is easily isolated from the continuum. This is another source that would benefit from MIRI coverage, as \Ha\ falls redward of both our NIRSpec and NIRCam coverage.
    
    %I know we talked about removing this source but it supports a lot of our other discussion in the text so I would
    %like to keep it.
    \item \textbf{Panel E} shows an EELG with a red continuum. The galaxy is so reddened that it is a dropout in filters blueward of F277W.
    Visual inspection of the HST F125W and F160W images shows the measured photometry displayed in the SED is incorrect and the source is instead undetected in the images.
    Unlike red galaxies with non-extreme emission lines that are rejected as ``Tier R'' (discussed in Section 3.6), the \Ha\ emission of this source passes our observed-frame EW selection criterion and it is classified as an EELG. Similar galaxies at a slightly lower redshift, with \Hb\ +\OIII\ in F277W and \Ha\ spanning all of F356W, F410M, and F444W, can be mistaken for high-$z$ galaxies with a Lyman break between F200W and F277W (similar to the galaxy CEERS-93316 characterized by \citealp{Haro2023a} and discussed in Section 3.4). This EELG and the EELG in panel C are known broad-line AGN and have been discussed in detail by \citet{Kocevski2023}.
    % We recognize this final source as a true EELG, not a reddened interloper like sources discussed in Section 3.5.

\end{itemize}

\begin{figure}[t]
    \centering
    \includegraphics[scale = 0.32]{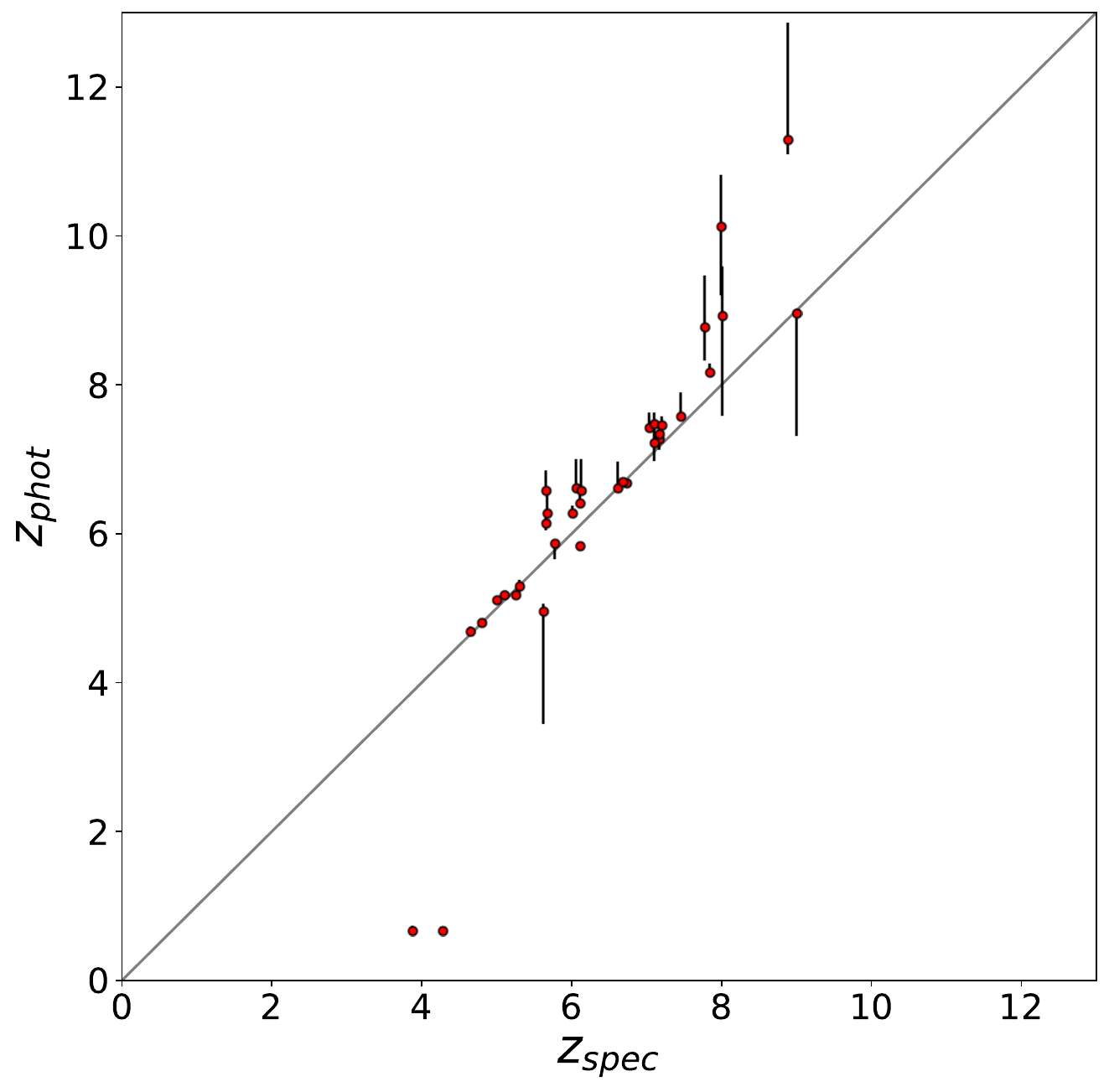}
    \includegraphics[scale = 0.3]{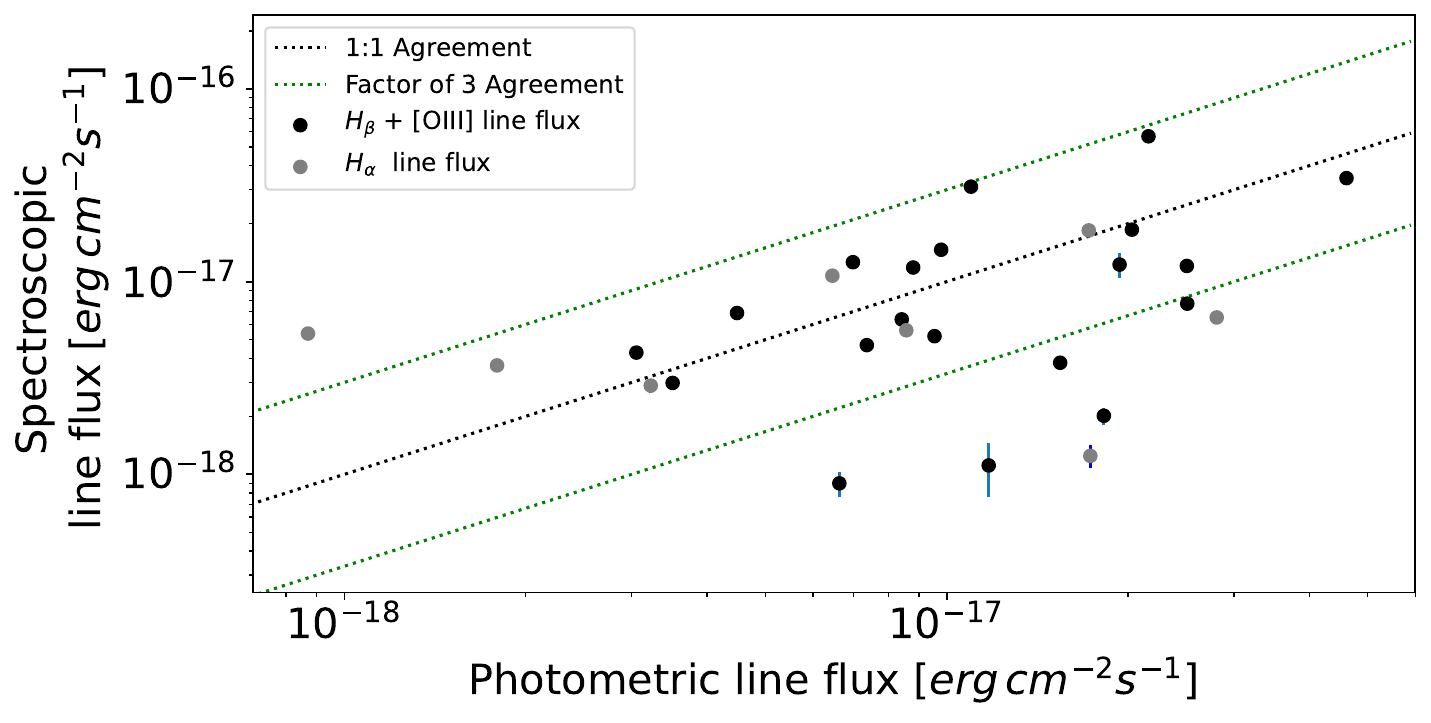}
    \caption{\textbf{Top: }The comparison between spectroscopic and photometric redshifts for spectroscopically confirmed EELGs. In all cases the spectroscopy confirms our EELG selection, although 4/34 sources have incorrect $z_\mathrm{phot}$ that disagrees with the spectroscopic redshift. Error bars represent 68\% confidence ranges.
    \textbf{Bottom:} Comparison of the spectroscopic and photometrically-inferred emission-line flux for \Ha\ and the sum of \Hb\ + \OIII, for the 34 EELGs with NIRSpec spectra. Most line fluxes agree within a factor of 3, indicating the reliability of our photometric EELG selection and EW measurements. A small number (7) of the spectroscopic line measurements are underestimated by a factor of $>$3 compared to the photometry, likely due to aperture losses affecting the spectra.
    }
    \label{fig:specdist}
\end{figure}

%6 in tier 2:2 with high 2, 2 in z gap, 2 outliers
The top panel of Figure \ref{fig:specdist} compares the photometric and spectroscopic redshifts for the 34 EELGs with NIRSpec redshifts (excluding the 2 objects with featureless spectra consistent with noise and the red galaxies). Most (32/34) of the EELGs have an accurate $z_\mathrm{phot}$ that are broadly consistent with their $z_\mathrm{spec}$. But there are two $z \sim 4$ galaxies that have catastrophically under-reported photometric redshifts: in both cases we correctly identified the extreme emission lines in the F277W and F356W filters that are present in the spectra. One of these galaxies with an incorrect $z_\mathrm{phot}$ is shown in Panel A of Figure~\ref{fig:other_specs}.
The bottom panel of Figure \ref{fig:specdist} compares the spectroscopically-measured emission-line fluxes for \Ha\ and the sum of \Hb\ + \OIII\ with the same line fluxes inferred from our photometric selection for the 34 EELGs with NIRSpec spectra. The line fluxes generally agree within a factor of 3, except for 7 galaxies that are extended and likely to suffer from aperture losses in their spectroscopic measurements. The agreement between the spectroscopic and photometrically-inferred emission-line fluxes gives further confidence in our methods for EELG selection and characterization.

Table \ref{tbl:tspec} reports the redshifts and confidence tiers of our sample spectroscopic overlap. We further define these confidence tiers in the next section.
% Of the 39 spectra, 34 represent the EELG population, 4 represent red galaxies we omit from our population, and 2 are undetected spectra.
Table \ref{tbl:tspec} includes the 34 EELGs with good NIRSpec spectra, 2 additional EELGs with spectra that are consistent with noise, and 3 red-continuum emission-line galaxies that are rejected from our EELG sample.
Of the 34 EELG spectra, 6 are classified in our lower (Tier 2) confidence tiers. Of these, 2 low-confidence sources are the catastrophic outliers in Figure \ref{fig:specdist} and 2 are the high-redshift overestimation in Figure \ref{fig:specdist}. The other 2 low-confidence EELGs have spectroscopic redshifts that agree with the photometric redshift but have \Hb\ + \OIII\ in the filter gap between F277W and F356W, causing the lines to be blended across the two filters. We incorrectly classify these EELG as higher redshift sources, mistaking the \Ha\ line for \Hb\ + \OIII\ and missing the actual \Hb\ + \OIII\ emission. These sources are classified as low confidence because the photometric redshift disagrees with the higher-redshift EELG identification.

\subsection{Confidence Tiers}
%These represent final counts

\begin{table}[t]
    \begin{centering}
    \begin{tabular}{|c|c|c|c|c|c|}
          \hline
         & MPT ID & Photometry ID & $z_\mathrm{phot}$  & $z_\mathrm{spec}$  &  Tier\\ [0.5ex] 
         \hline\hline
         1 & 3 &  nircam1-4774 & 8.92 & 8.0 & 1B\\ 
         \hline
         2 & 4 &  nircam1-4777 & 10.12 & 7.99 & 2B\\ 
         \hline
         3 & 20 &  nircam3-23084 & 8.77 & 7.77 & 1B\\ 
         \hline
         4 & 23 &  nircam6-61381 & 11.29 & 8.88 & 2B\\ 
         \hline
         5 & 24 &  nircam6-61419 & 8.95 & 8.99 & 1B\\ 
         \hline
         6 & 38 &  nircam1-1021 & 7.57 & 7.45 & 1B\\ 
         \hline
         7 & 323 &  nircam2-19984 & 6.28 & 5.67 & 1A\\ 
         \hline
         8 & 386 &  nircam3-23965 & 6.61 & 6.61 & 1B\\ 
         \hline
         9 & 397 &  nircam3-25074 & 6.28 & 6.0 & 1A\\ 
         \hline
         10 & 403 &  nircam3-25420 & 5.86 & 5.77 & 1A\\ 
         \hline
         11 & 407 &  nircam3-25552 & 7.42 & 7.03 & 1B\\ 
         \hline
         12 & 439 &  nircam3-27280 & 7.27 & 7.17 & 1B\\ 
         \hline
         13 & 498 &  nircam3-31338 & 7.33 & 7.17 & 1B\\ 
         \hline
         14 & 515 &  nircam6-52374 & 6.58 & 5.66 & 1A\\ 
         \hline
         15 & 545 &  nircam6-54324 & 6.13 & 5.66 & 1A\\ 
         \hline
         16 & 603 &  nircam6-58270 & 6.61 & 6.05 & 1A\\ 
         \hline
         17 & 613 &  nircam6-58955 & 6.67 & 6.73 & 1B\\ 
         \hline
         18 & 746 &  nircam3-24015 & 4.96 & 5.62 & 2B\\ 
         \hline
         19 & 749 &  nircam1-2282 & 7.48 & 7.09 & 1B\\ 
         \hline
         20 & 933 &  nircam3-27358 & 0.67 & 4.28 & 2A\\ 
         \hline
         21 & 1027 &  nircam6-59920 & 8.17 & 7.83 & 1B\\ 
         \hline
         22 & 1038 &  nircam8-79680 & 7.45 & 7.19 & 1B\\ 
         \hline
         23 & 1374 &  nircam9-87370 & 5.11 & 5.00 & 1A\\ 
         \hline
         24 & 1414 &  nircam7-65318 & 6.70 & 6.68 & 1B\\ 
         \hline
         25 & 2355 &  nircam1-8674 & 6.40 & 6.11 & 1A\\ 
         \hline
         26 & 2362 &  nircam3-31319 & 5.29 & 5.30 & 2B\\ 
         \hline
         27 & 2782 &  nircam6-53385 & 5.17 & 5.25 & 1A\\ 
         \hline
         28 & 355 &  nircam3-21394 & 6.58 & 6.13 & 1A\\ 
         \hline
         29 & 428 &  nircam3-26436 & 5.83 & 6.10 & 1A\\ 
         \hline
         30 & 44 &  nircam1-1253 & 7.21 & 7.10 & 1B\\ 
         \hline
         31 & 1912 &  nircam1-2166 & 5.17 & 5.10 & 1A\\ 
         \hline
         32 & 2000 &  nircam2-16056 & 4.81 & 4.80 & 1A\\ 
         \hline
         33 & 3584 &  nircam1-2149 & 4.69 & 4.64 & 1A\\ 
         \hline
         34 & 3585 &  nircam1-5040 & 0.67 & 3.87 & 2A\\ 
         \hline
         35 & 2411 &  nircam3-27576 & 3.25 & 3.23 & R\\ 
         \hline
         36 & 3129 &  nircam7-66543 & 1.24 & 1.01 & R\\ 
         \hline
         37 & 34103 &  nircam1-4118 & 1.24 & 1.23 & R\\ 
         \hline
         38 & 94 & nircam1-5545 & 1.63 & ? & 3A\\ 
         \hline
         39 & 670 &  nircam6-62174 & 6.28 & ? & 1A\\ %50e, 30cont
         \hline

    \end{tabular}

    \end{centering}
    \caption{NIRSpec overlap with our photometrically selected EELG population.}
    \label{tbl:tspec}
\end{table}

\begin{table}[t]
    \begin{centering}
    \begin{tabular}{c|c|c|c}
          \hline
         Tier & $z_\mathrm{phot}$ agreement & Num of EELs & Num Sources  \\ [0.5ex] 
         \hline\hline
         1A & yes & 2 & 480\\ 
         \hline
         1B & yes & 1  & 222\\ 
         \hline
         2A & no & 2 & 251\\ 
         \hline
         2B & no & 1 & 107\\ 
         \hline
         3 & ? & ?  &  105\\ 
         \hline
         R & - & 0  & 62
         \\ 
         \hline

    \end{tabular}

    \end{centering}
    \caption{ Number counts for confidence tier descriptors for EELGs. We report 1165 EELGs in tiers 1-3 and 1227 in all tiers. }
    \label{tbl:t2}
\end{table}

We categorize our EELG candidates into different confidence tiers according to the reliability of both the photometric identification and the equivalent width estimate of the extreme emission, with 1 being our highest confidence and 3 being our lowest. Confidence is primarily determined from comparing the SED-derived photometric redshift with the redshift inferred by our EELG selection.
% a factor of photometric redshift derived from SEDs as discussed in Section 2.2.
We additionally assign letters to each tier to denote the number of extreme lines we observe, "A" for two distinct lines and "B" for one. The number of EELGs in each confidence tier is given by Table \ref{tbl:t2}. In most of our analysis of the general EELG population and its properties in Section 4, we use only the ``Tier 1'' sources that have the highest-confidence emission-line identification and EW measurements.

\begin{figure}[t]
    \centering
    \includegraphics[scale = 0.28]{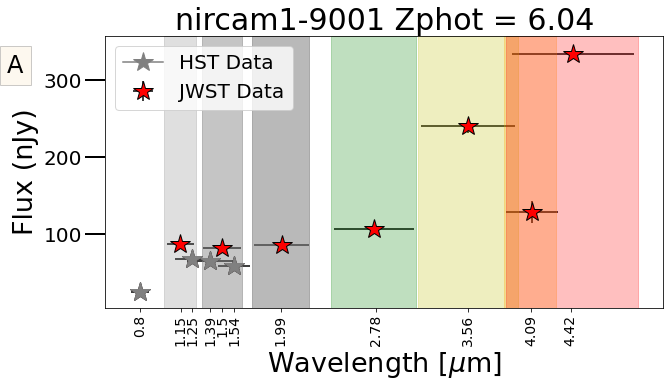}
    \includegraphics[scale = 0.185]{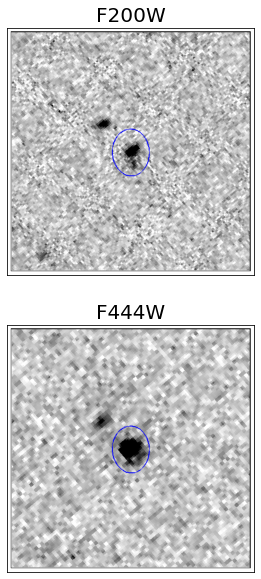}
    \includegraphics[scale = 0.28]{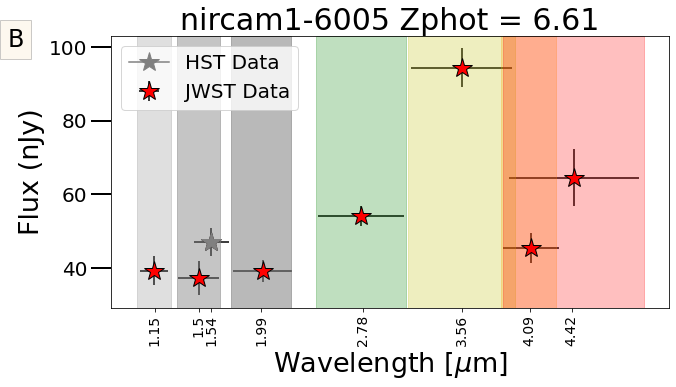}
    \includegraphics[scale = 0.185]{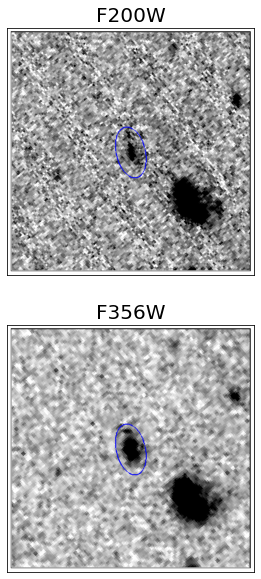}
    \includegraphics[scale = 0.28]{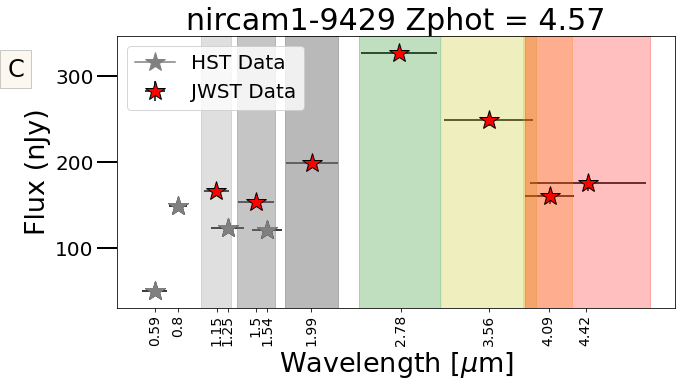}
    \includegraphics[scale = 0.185]{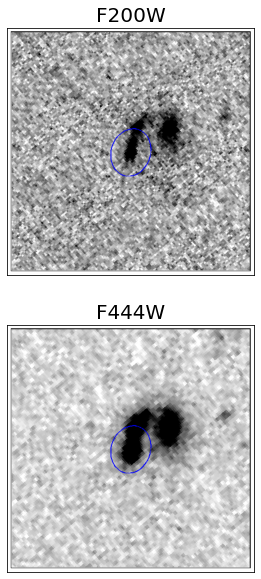}
    \caption{Examples of sources classified in confidence Tier 1A, with photometric redshifts that are consistent with our identification of multiple extreme emission features. Photometry postage stamps show a continuum filter (top) and an extreme emission filter (bottom), with North up and East left. All postage stamps are 4\arcsec\ in width and height. {\bf{Panel A}}: EELG with \Hb\ + \OIII\ in F356W and \Ha\ in F444W. {\bf{Panel B}}: EELG with \Hb\ + \OIII\ in F356W and \Ha\ in F444W, with likely \OII\ + \NeIII\ contribution in F277W. {\bf{Panel C}}: EELG with \Hb\ + \OIII\ in F277W and \Ha\ in F356W.
    % While many EELGs occupy clustered environments, we note that the source in this panel C
    This source has two close neighbors that also exhibit bright emission in F277W, potentially indicating a cluster of three emission-line galaxies at the same redshift. 
    }
    \label{fig:tier1A}
\end{figure}

\begin{figure}[t]
    \centering
    \includegraphics[scale = 0.28]{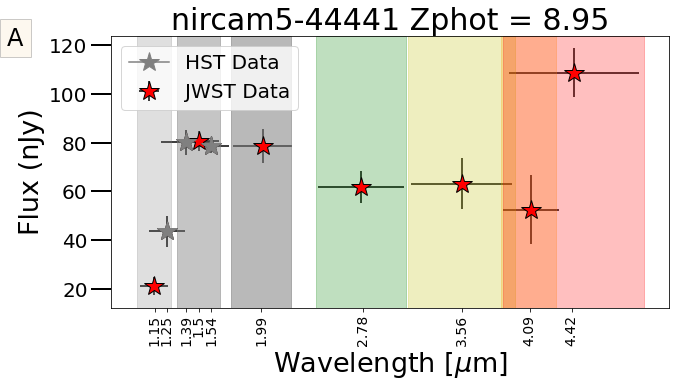}
    \includegraphics[scale = 0.185
    ]{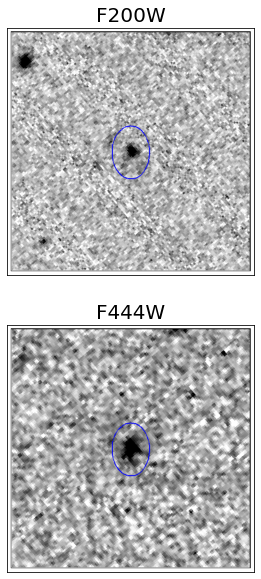}
    \includegraphics[scale = 0.28]{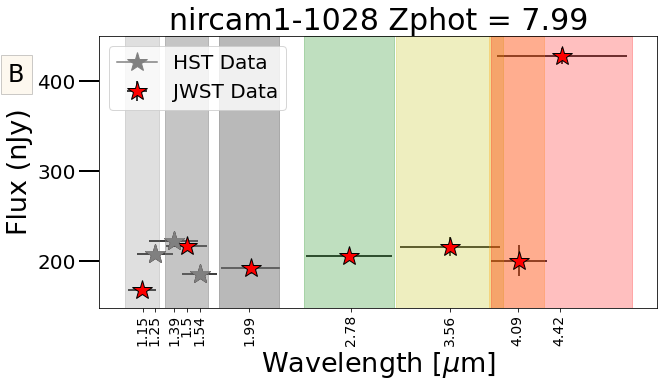}
    \includegraphics[scale = 0.185
    ]{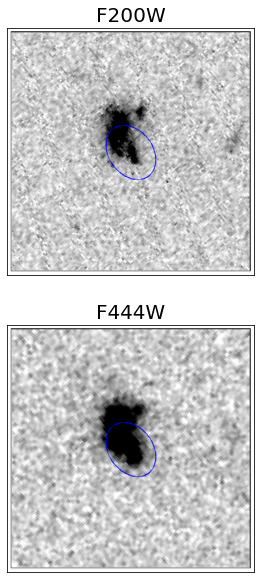}
    \includegraphics[scale = 0.28]{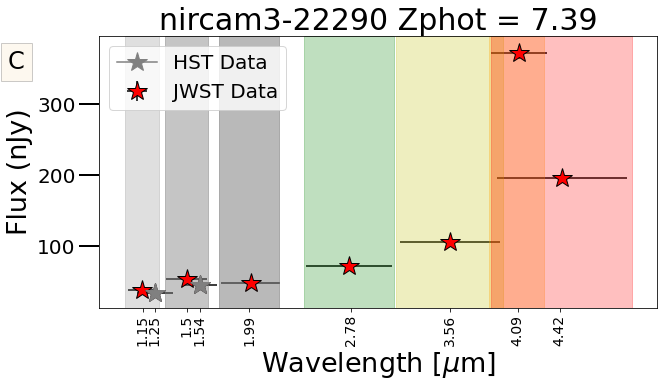}
    \includegraphics[scale = 0.185]{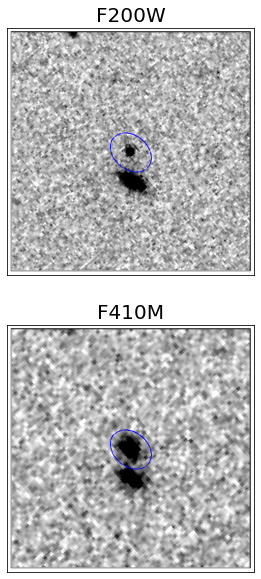} 

    \caption{Examples of sources with a single extreme emission feature that is consistent with their $z_\mathrm{phot}$, classified in confidence Tier 1B. Photometry postage stamps show a continuum filter (top) and an extreme emission filter (bottom), with North up and East left. All postage stamps are 4\arcsec\ in width and height. {\bf{Panel A}}: Extreme emission in F444W consistent with \Hb\ + \OIII\ and a Lyman break between the F150W and F115W filters. {\bf{Panel B}}: Extreme emission in F444W consistent with \Hb\ + \OIII. The images indicate at least one nearby companion. {\bf{Panel C}}: Extreme emission in F410M consistent with \Hb\ + \OIII. While this EELG has a close neighbor in its image, this companion galaxy does not exhibit the same dramatic differences between continuum and emission-line filter morphologies as the EELG.
    %it does not appear more compact in the continuum filter as the EELG does, indicating that the nearby neighbor is not an EELG with similar properties.
    }
    \label{fig:tier1B}
\end{figure}

Tier 1 includes sources with clear extreme emission in the broadband photometry and with photometric redshifts that are consistent with the photometrically inferred emission identified by our EELG selection. Example SEDs of sources in Tiers 1A and 1B are given in Figures \ref{fig:tier1A} and \ref{fig:tier1B}, respectively. Figure \ref{fig:tier1A} presents several source with extreme emission from \Hb\ + \OIII\ and \Ha\ in separate broadband filters. Figure \ref{fig:tier1B} represents sources with extreme emission in one filter, \Hb\ + \OIII\ in all cases, with \Ha\ presumed to fall redward of the NIRCam photometry. Panels A and B of Figure \ref{fig:tier1B} capture the extreme \Hb\ + \OIII\ emission only in F444W while Panel C has \Hb\ + \OIII\ emission in both F410M and F444W. EELGs classified in Tier 1B with a single (extreme) emission-line feature always have $6.7<z<9.0$ such that \Hb\ + \OIII\ is in at least one of the NIRCam filters but \Ha\ falls redward of the F444W filter.

%Add a few more example SEDs to this figure

Tier 2 contains sources with distinct extreme emission identified in the photometric filters but with photometric redshifts that disagree with the redshift inferred from the photometric filter that includes the extreme emission feature. The mismatched photometric redshifts are often low-redshift $z_\mathrm{phot}$ solutions, due in part to placement of the $\sim$2$\mu$m stellar bump at the filters that include extreme emission lines and the absence of a medium band filter to separate lines falling in F277W and F356W. Other Tier 2 EELGs have broad probability distribution functions for a high-redshift $z_\mathrm{phot}$ solution.
We further subdivide incorrect-$z_\mathrm{phot}$ galaxies with 2 filters containing extreme emission lines (\Hb\ + \OIII\ and \Ha\ at $4<z<6.5$) as ``Tier 2A'' and galaxies with only 1 extreme emission line (\Hb\ + \OIII\ at $6.5<z<9.5$) as ``Tier 2B.''

\begin{figure}[t]
    \centering
    \includegraphics[scale = 0.28]{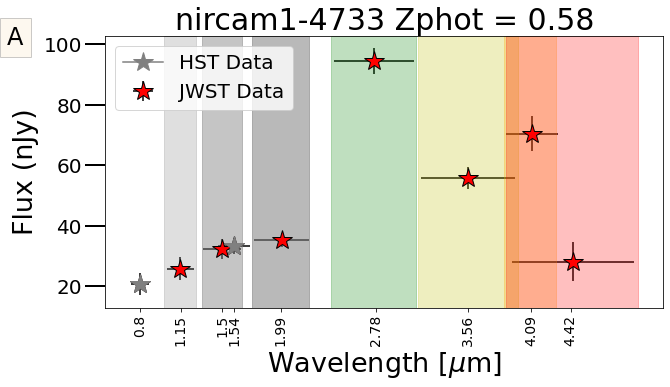}
    \includegraphics[scale = 0.185]{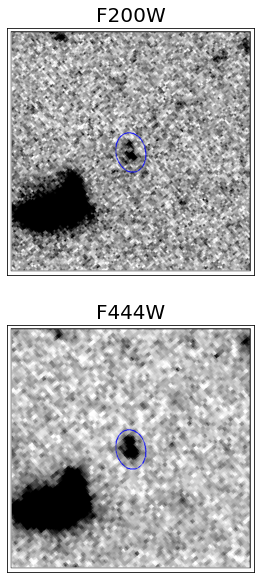}
    \includegraphics[scale = 0.28]{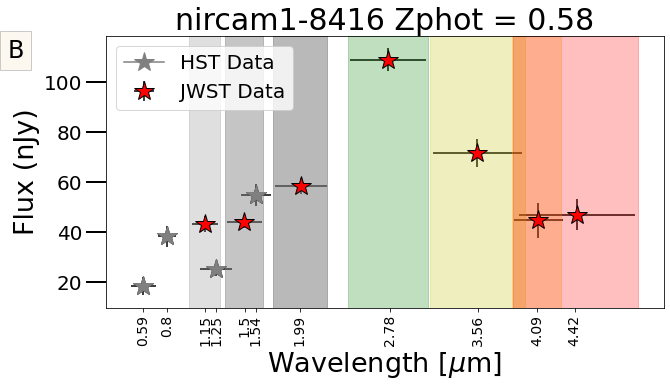}
    \includegraphics[scale = 0.185]{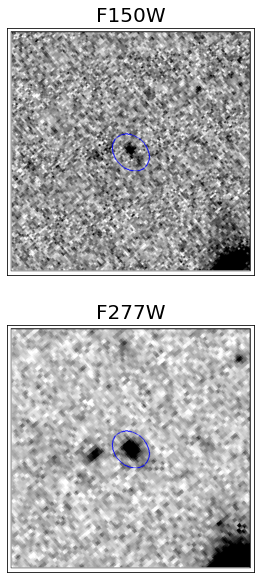}
    \includegraphics[scale = 0.28]{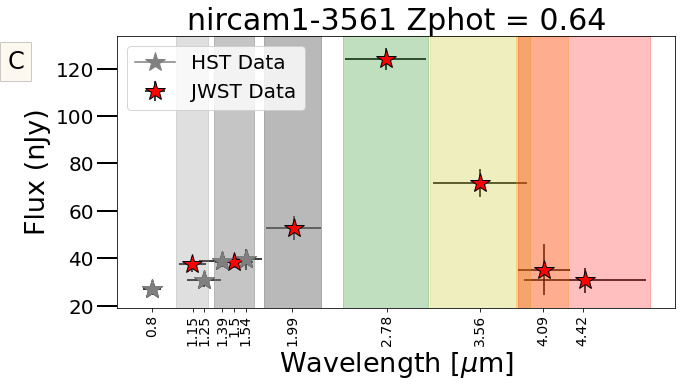}
    \includegraphics[scale = 0.185]{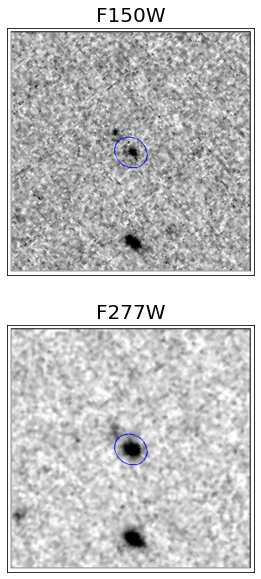}
    \caption{
    Examples of EELGs in confidence Tier 2A, identified to have extreme emission in two filters, but with low-redshift $z_\mathrm{phot}$ solutions. Photometry postage stamps show a continuum filter (top) and an extreme emission filter (bottom) with North up and East left. All postage stamps are \arcsec\ in width and height. {\bf{Panel A}}: Extreme emission associated with \Ha\ in F277W and \Hb\ + \OIII\ in F410M, consistent with $z \approx 5$. In the F200W continuum postage stamp, this source appears to be two close-companion EELGs. {\bf{Panels B,C}}: Extreme emission associated with \Ha\ in F277W and \Hb\ + \OIII\ in F356W, consisted with a $z \approx 4-5$. The observed SEDs of these sources are very similar to the spectroscopically confirmed EELG shown in Panel A of Figure \ref{fig:other_specs}, suggesting that they are bona-fide EELGs but with incorrect photometric redshifts. 
    }
    \label{fig:tier2A}
\end{figure}

Figure \ref{fig:tier2A} shows examples of ``Tier 2A'' EELGs. Most sources in this tier are associated with extreme emission from \Hb\ + \OIII\ in F277W and \Ha\ emission (which may or may not be similarly extreme) in F356W or F410M. The lack of a medium-band filter (in this case, F300M between F277W and F356W) to separate the emission-line contribution causes the photometric redshift code to mistakenly assign the rest-frame 2$\mu$m stellar bump to the increased photometric flux that is caused by the emission lines.

Across the entire sample of galaxies in which we identify extreme emission from \Hb\ + \OIII\ in F277W and \Ha\ in F356W, only 45\% of galaxies have correct photometric redshifts, with the remainder typically having low-redshift solutions that instead put the $\sim$2$\mu$m stellar bump in the F277W and F356W filters at $z_\mathrm{phot} \approx 0.6$.
The difficulty of the SED fitting at this redshift is likely due to the lack of medium-band filter observations (i.e., F300M) between the broad-band filters with extreme emission. We include the galaxies with $z_\mathrm{phot}>4$ in our Tier 1 sample, while the galaxies with low-redshift solutions $z_\mathrm{phot}<4$ are in Tier 2. The completeness of our EELG selection at different redshifts is discussed in more detail in Section 4.1.

An example of a galaxy in Tier 2A is shown in Panel A of Figure \ref{fig:other_specs}. In this case, the NIRSpec spectroscopy confirms our $z=4.28$ EELG selection while the SED fit incorrectly assigns a low-redshift solution at $z=0.67$. One additional spectroscopically confirmed EELG also has $z_\mathrm{spec}=3.87$, matching our EELG selection, and incorrect low-redshift $z_\mathrm{phot}=0.67$. 
%This catastrophic redshift underestimation is due in part to the absence of a medium band filter between the F277W and F356W filters. 
In the higher redshift case where the same extreme emission lines are separated by F410M, the SED fitting code more frequently provides an accurate estimate of the photometric redshift.

Three other NIRSpec-observed EELGs with similar photometric excess in F277W and F356W have a correct $z_\mathrm{phot} \simeq 4$ that matches the spectroscopic confirmation of \Hb\ + \OIII\ in F277W and \Ha\ in F356W. These sources have very similar observed SEDs to the spectroscopically confirmed EELG shown in Panel A of Figure \ref{fig:other_specs}. The 5 spectroscopic confirmations of our EELG candidates at this redshift (2 of which refute a low-redshift $z_\mathrm{phot}$) indicate that our EELG identification is generally correct. However the EELGs with low-redshift (and likely incorrect) $z_\mathrm{phot}$ must be assigned lower confidence and excluded from our analysis in Section 4 due to uncertain redshifts and potentially unreliable rest-frame EWs.

Tier 2B generally includes EELGs identified to have \Hb\ + \OIII\ in F410M, F356W, or F444W that have broad photometric redshift probability distributions. For sources with extreme emission in the F444W filter, only $\approx$40\% have a photometric redshift that matches our EELG selection. EELGs with extreme emission identified in F444W and photometric redshifts of $7.5<z<9.5$ are categorized as Tier 1B since this $z_\mathrm{phot}$ agrees with \Hb\ + \OIII\ extreme emission in F444W. The remaining sources are assigned Tier 2B.
 
\begin{figure}[t]
    \centering
    \includegraphics[scale = 0.28]{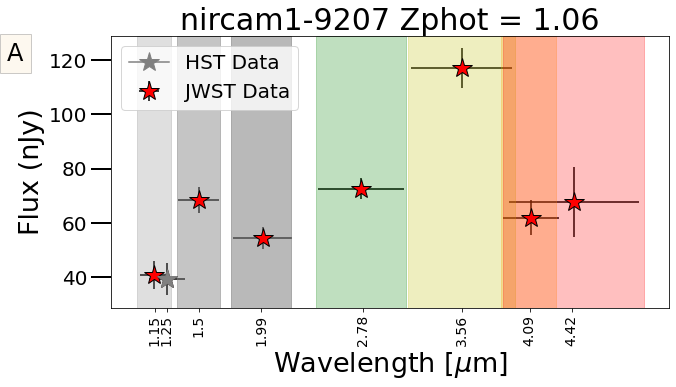}
    \includegraphics[scale = 0.185]{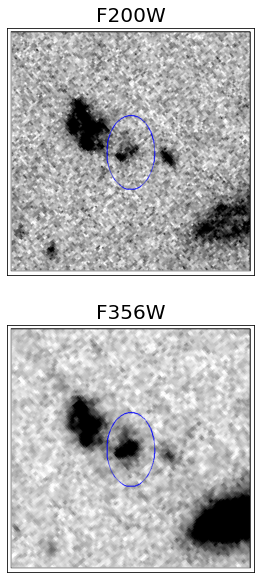}
    \includegraphics[scale = 0.28]{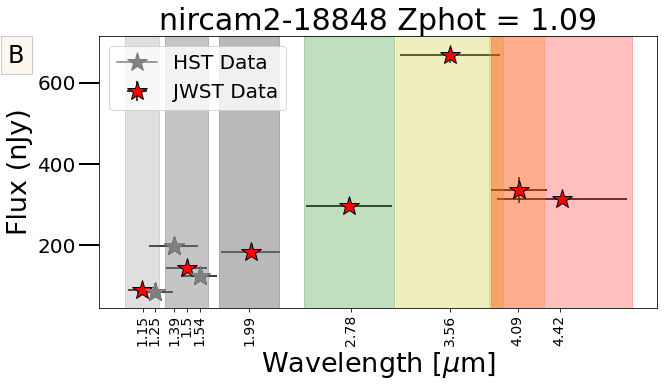}
    \includegraphics[scale = 0.185]{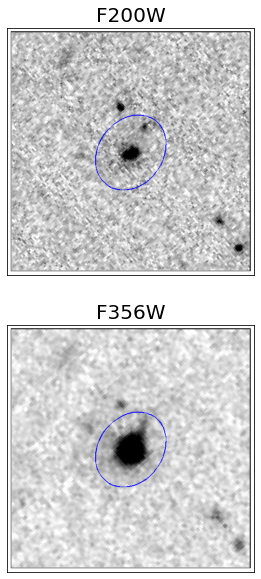}
    \includegraphics[scale = 0.28]{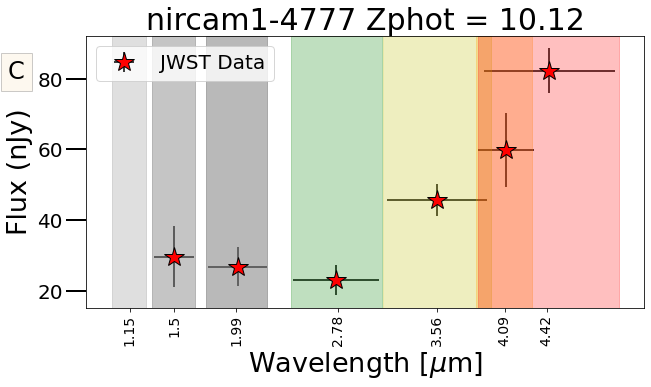}
    \includegraphics[scale = 0.185]{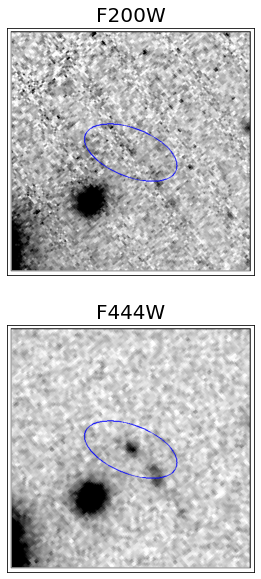}
    \caption{
    Examples of EELGs in confidence Tier 2B, with only one extreme emission feature (\Hb\ + \OIII) identified in a filter that is inconsistent with the SED-derived photometric redshift. Photometry postage stamps show a continuum filter (top) and an extreme emission filter (bottom), with North up and East left. All postage stamps are 4\arcsec\ in width and height. {\bf{Panel A}}: Extreme emission in F356W consistent with \Hb\ + \OIII\ at  $z \approx 6.5$. {\bf{Panel B}}: Extreme emission in F356W consistent with \Hb\ + \OIII\ at  $z \approx 6.5$. The morphology exhibits a significant size difference between the continuum and emission filters, suggesting the presence of an extended ionization region. {\bf{Panel C}}: Emission in F444W consistent with \Hb\ + \OIII. The SED shape is similar to extremely red objects (EROs) reported by \citet{Barro2023}, with a combination of a flat continuum in bluer filters and red emission in redder filters. The inferred \Hb\ + \OIII\ emission implies $z < 9.5$, inconsistent with the higher photometric redshift.}
    
    \label{fig:tier2B}
\end{figure}

Examples of EELGs in confidence Tier 2B are shown in Figure \ref{fig:tier2B}. Panels A and B are cases of \Hb\ + \OIII\ falling in F356W, suggesting a true redshift of $z \approx 6.5$ while SED fitting assigns the sources a much lower redshift. Panel C displays an example SED with \Hb\ + \OIII\ inferred to be in F444W, inconsistent with the photometric redshift of $z_\mathrm{phot} = 10.12$ that would place \Hb\ + \OIII\ redward of the NIRCam filters. 

The photometric agreement tends to be better when extreme \Hb\ + \OIII\ emission falls within both the F410M and F444W filters, with 74\% of EELGs having $z_\mathrm{phot}$ that matches the photometric identification of extreme emission. Panel D of Figure \ref{fig:other_specs} is an example of a spectroscopically confirmed EELG with \Hb\ + \OIII\ within both F410M and F444W and correct $z_\mathrm{phot}$. Figure \ref{fig:tier2B} presents additional examples of EELGs identified to have \Hb\ + \OIII\ in F410M and/or F444W but with disagreeing $z_\mathrm{phot}$, despite the similarity in observed SED to the spectroscopically confirmed EELG shown in Figure \ref{fig:other_specs} Panel D.

We define a low-confidence ``Tier 3'' that includes EELG candidates with EW estimates that are likely to be unreliable. This category includes galaxies with unusual SEDs that imply emission-line contributions blended across multiple filters, lower-redshift galaxies with extreme emission from \HeI$\lambda$1.08$\mu$m and/or Paschen emission lines, or galaxies with bright emission lines superimposed on a red continuum. We discuss the more unusual galaxy SEDs in more detail in the next subsection.

\subsection{Puzzling Sources}

\begin{figure}
    \centering
    \includegraphics[scale = 0.28]{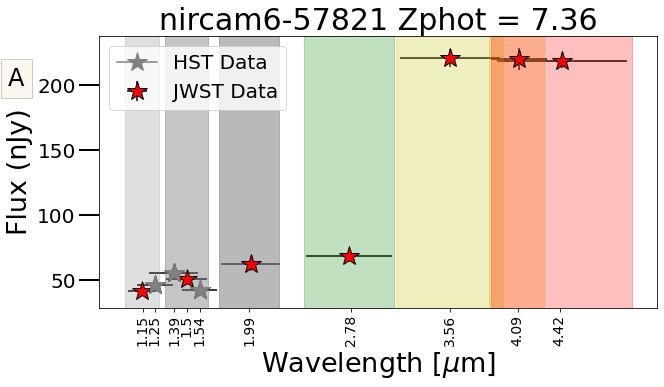}
    \includegraphics[scale = 0.185]{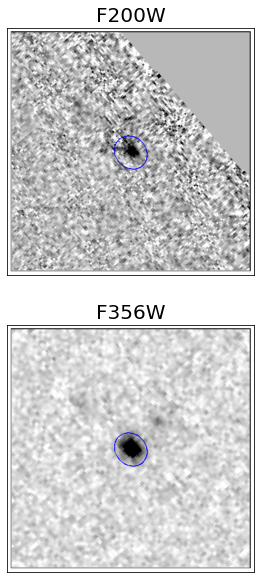}
    \includegraphics[scale = 0.28]{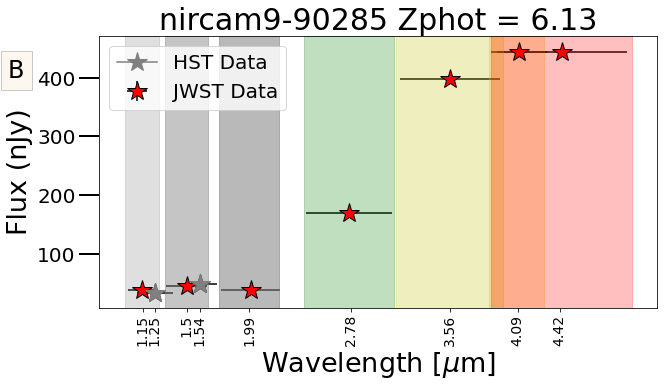}
    \includegraphics[scale = 0.185]{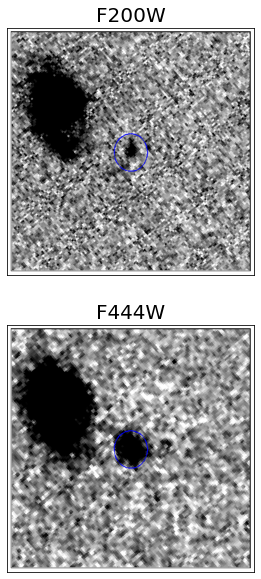}
    \includegraphics[scale = 0.28]{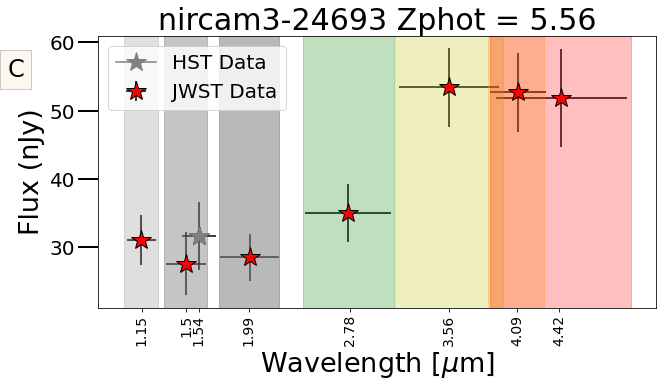}
    \includegraphics[scale = 0.185]{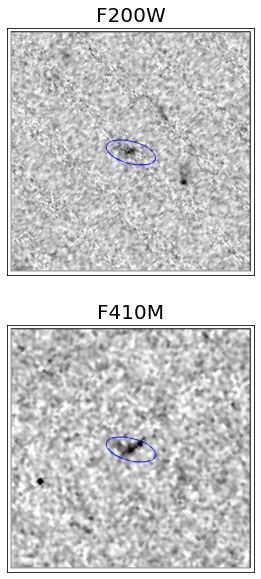}
    \includegraphics[scale = 0.28]{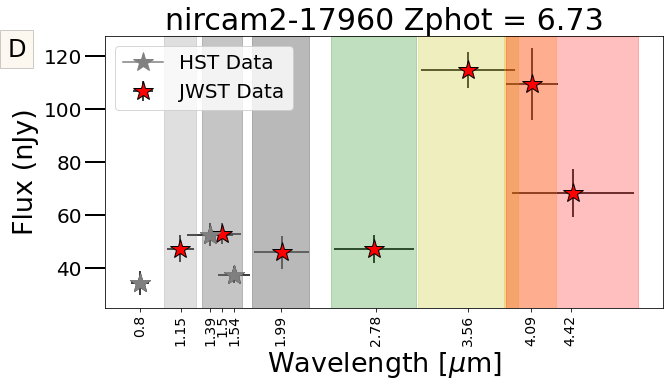}
    \includegraphics[scale = 0.185]{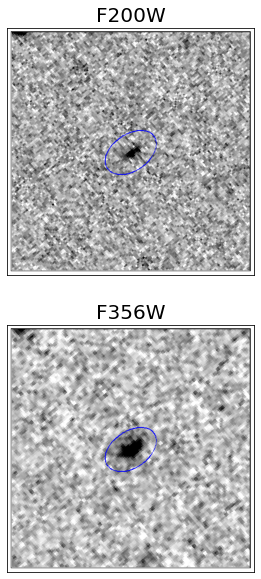}
    \caption{Example SEDs in which single extreme emission features impact more than one photometric filter. Photometry postage stamps show a continuum filter (top) and an extreme emission filter (bottom) with North up and East left. All postage stamps are 4\arcsec\ in width and height. 
    {\bf{Panels A-C}}: Example SEDs from the category of sources with extreme emission in F356W, F410M, and F444W that are consistent with \Hb\ + \OIII\ and/or \Ha\ emission blended across 3 or more (partially overlapping) filters. Deblending the emission line contribution to each filter is nontrivial and our photometric measurement of EW is likely to be unreliable, and so these sources are categorized in the lowest confidence Tier 3. Similar to Panel B of Figure \ref{fig:tier2B}, Panel B of this Figure has an image that is significantly larger in the emission-line filter than the continuum, suggesting the presence of an extended ionization region. 
    {\bf{Panel D}}: \Hb\ + \OIII\ blended across F356W and F410M. This extreme emission only impacts the broadband photometry in this way for a narrow redshift range ($z \approx 6.7$), correctly identified by the photometric redshift code. We assign this source to Tier 1B for extreme emission in a single filter with a correct photometric redshift.} 
    
    \label{fig:falsecont3}
\end{figure}

Our Tier 3 category includes sources with unusual SED shapes. Some of these galaxies have SEDs that are consistent with emission-line features blended across 2 or more filters, with examples shown in Figure \ref{fig:falsecont3}. It is nontrivial to disentangle the potential contribution of \Hb\ + \OIII\ and \Ha\ between each filter and so, in most cases, we cannot reliably measure an EW from the photometry. These galaxies are confidently identified as EELGs, as the very red F277W-F356W colors would require steep Balmer breaks that are implausible given the young age of the Universe at such a high redshift. These sources are assigned to our lowest confidence tier due to their unreliable EW measurements rather than any uncertainty in their designation as an EELGs.

Extreme emission lines that are blended across multiple NIRCam filters
create confusing SED shapes. A dramatic example of this is the galaxy CEERS-93316, which has a relatively constant flux density in the F277W, F356W, F410M, and F444W filters and is undetected blueward of F200W. Such an SED can be mistaken for a high-redshift \Lya\ break and this galaxy was initially identified as a $z=16.5$ candidate by \citet{Donnan2023}. NIRSpec observations in \citet{Haro2023a} instead demonstrated that this object has a combination of a red continuum, making the blue NIRCam filters undetected, and extreme emission-line contribution to several NIRCam filters: \Hb\ + \OIII\ in F277W, and \Ha\ blended across all three of F356W, F410M, and F444W.
% While \citet[]{Haro2023a} suggest that sources like CEERS-93316 are rare due to the specific set of conditions required to generate their SED shape, we find that it is representative of a broader population of sources
We identify a population of sources with similar SED shapes to CEERS-93316, as shown in panels A-C of Figure \ref{fig:falsecont3}. The particular source CEERS-93316 is not included in our EELG sample because it was not detected in the minimum of 5 photometric filters required in this selection, but it spectroscopically confirms the characteristic shape of a photometrically blended extreme \Hb\ + \OIII\ lines across the reddest filters mimicking an ultra-high redshift galaxy.

The spectroscopic confirmation of CEERS-93316 as an EELG with emission features blended across multiple filters suggests that the galaxies with similar unusual SEDs shown in Figure \ref{fig:falsecont3} are also EELGs with similar emission-line blending in their photometry. It is notable that EELGs like CEERS-93316 and the sources shown in Figure \ref{fig:falsecont3} would be missed by simple color-color selection due to the emission features blending nearly equally across different filters. While we report correct photometric redshifts for all sources presented in Figure \ref{fig:falsecont3}, this is a result of detected continua in all cases. If the continuum was slightly dimmer (only by 30nJy in two cases), these SEDs would be difficult to distinguish from CEERS-93316 and could easily be classified as ultra-high redshift candidates. We present this population of photometrically blended emission-line SEDs to showcase how the
% F410M medium-band filter can introduce SEDs that
red NIRCam filters can
mimic bright continuum in the presence of extreme emission lines. The consequences of red-continuum emission-line galaxies for ultra high-$z$ galaxy searches is discussed in more detail in Section 3.5.

As a counterexample to the puzzling SEDs shown in Panels A-C of Figure \ref{fig:falsecont3}, the bottom panel presents a very similar source at a slightly different redshift. Again, an extreme emission feature (at this redshift, \Hb\ + \OIII) affects all three of the F356W, F410M, and F444W filters. But in this case the line is fully (or mostly) encompassed by the F356W filter and we are able to measure the EW from the broad-band photometry, placing this source in confidence Tier 1B as a reliably measured EELG. When only two filters are affected by extreme emission lines, as in this example and in EELGs with similar SEDs, 85\% have photometric redshifts consistent with redshifts inferred from emission line location.

\begin{figure}
    \centering
    \includegraphics[scale = 0.28]{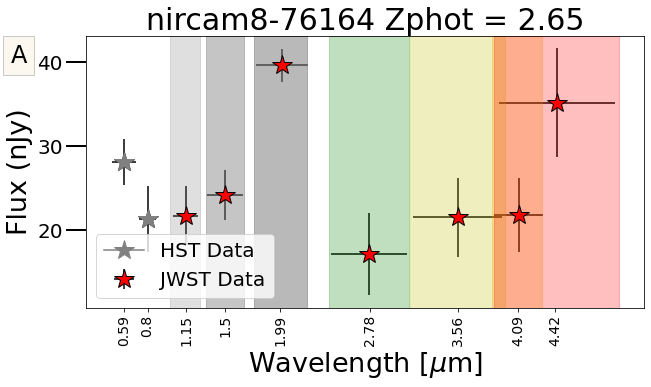}
    \includegraphics[scale = 0.185]{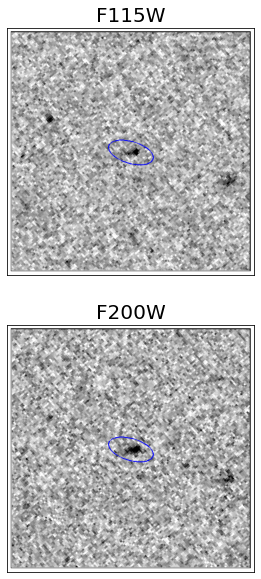}
    \includegraphics[scale = 0.28]{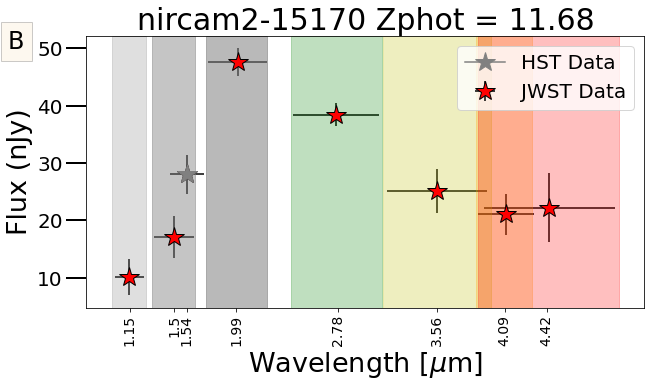}
    \includegraphics[scale = 0.185]{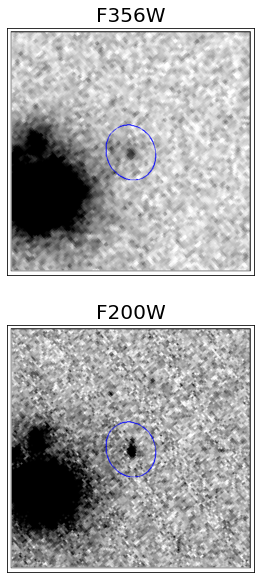}
    \includegraphics[scale = 0.28]{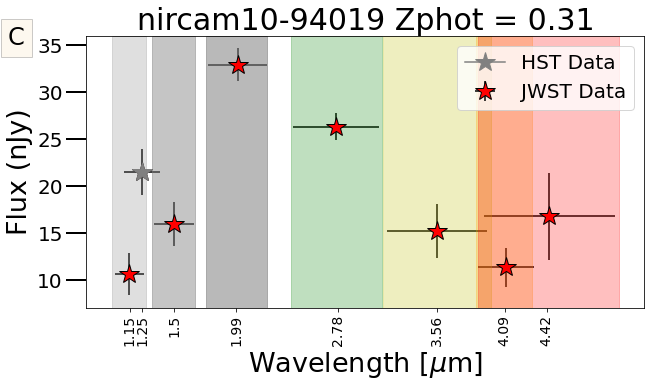}
    \includegraphics[scale = 0.185]{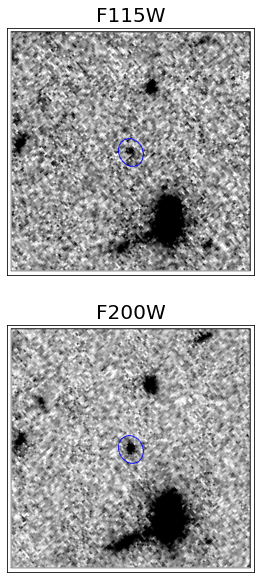}
    
    \caption{Example SEDs for source we classify Tier 3 consistent with emission in filters associated with lower redshift galaxies. Photometry postage stamps show a continuum filter (top) and an extreme emission filter (bottom) with North up and East left. All postage stamps are 4" in width and height. {\bf{Panel A}}: Source with emission in F200W and F444W, likely a combination of \Hb\ + \OIII\ in F200W and \HeI\ or Paschen lines in  F444W.
    {\bf{Panels B and C}}: Galaxies with an emission feature in F200W, likely
    from \Hb\ + \OIII at $z \sim 3$
    and not consistent with the photometric redshift prediction that mistakes the line for a Lyman break in the case of the high redshift solution or a stellar bump in the case of the very low redshift solution. Both of these sources pass our EW selection from the contribution to the F277W filter.
    }
    \label{fig:tier3}
\end{figure}

Tier 3 also includes sources with apparent extreme emission contributing to bluer filters outside of our target selection, with examples shown in Figure \ref{fig:tier3}. These galaxies likely have extreme emission but at lower redshift, with some combination of \Hb\ + \OIII\ and/or \Ha\ in the F200W filter and \HeI$\lambda$1.08$\mu$m and/or Paschen lines in the F444W filter. This work is primarily focused on extreme \Hb\ + \OIII\ and \Ha\ emission in $z>4$ galaxies and so we categorize these potential low-redshift EELGs in '' Tier 3 ''. 

\subsection{Consequences for ultra high-z galaxy searches}

One of the major scientific objectives of JWST is the search for galaxies in the ultra-high redshift universe ($z>12$). Over the past year, Lyman break dropout galaxy searches using NIRCam photometry have exploded \citep{Finkelstein2022, Donnan2023, Yan2023, Finkelstein2023, Perez-Gonzalez2023}. But spectroscopic confirmation of ultra high-$z$ candidates remains somewhat limited, with only a handful of galaxies at $10<z<13.5$ identified to have Lyman breaks in low-resolution NIRSpec prism observations \citep{CurtisLake2023, Haro2023a, Haro2023b, Bunker2023, Harikane2023, Hsiao2023, Fujimoto2023, Wang2023} plus a few more $10<z<13.5$ candidates with ambiguous single-line ALMA detections \citep{Harikane2022, Bakx2023}.

Notably, none of the photometrically selected $z>13.5$ candidates have been spectroscopically confirmed.
% Every spectroscopic observation of a $z>13.5$ galaxy has instead indicated a lower redshift, with most of the ultra high-$z$ candidates turning out to be $z<7$ galaxies with red continua and bright emission lines \citep[]{Fujimoto2023, Haro2023a, Zavala2023}.
The only spectroscopic observation of a $z>13.5$ galaxy candidate, CEERS-93316, instead indicated a lower redshift of $z_{\rm spec} \sim 4.9$ \citep{Haro2023a}. A few other $z>13.5$ candidates have far-infrared detections that indicate $z>7$ dusty galaxy SEDs \citep{Fujimoto2023, Zavala2023}.

In the case of CEERS-93316, \Hb\ + \OIII\ and \Ha\ emission increase the flux in some combination of the F277W, F356W, and F444W filters (see, e.g., Figure \ref{fig:lines}) while dust attenuation of the continuum causes the bluer NIRCam filters to be undetected. The steep red F200W-F277W color and undetected bluer NIRCam filters are then mistaken for a $z>13.5$ Lyman break. The large number of red-continuum emission-line galaxies in our sample, identified as outliers in our EELG selection, presents a cautionary tale for photometric selection of ultra high-$z$ galaxy candidates.

\subsection{Rejected Sources: Red Emission Line Galaxies}

\begin{figure}
    \centering
    \includegraphics[scale = 0.2]{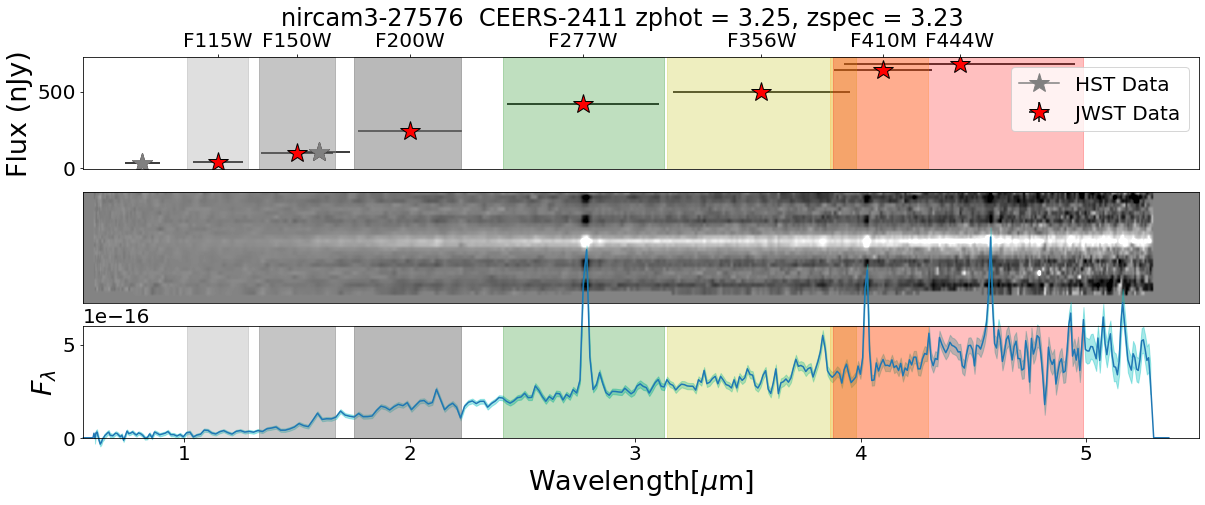}
    \caption{Spectroscopic examples of a Tier R galaxy. We identify \Ha\ emission in F277W, a SIII triplet blended between F356W and F410M, and a He line in F444W. None of these identified lines have sufficient EWs to pass our observed frame selection, but the red continuum incorrectly boosts the EW measurement. 
    %{\bf{Panel B}}: \Ha\ in F150W and and Pa$\alpha$ blended between F356W and F410M.
    }
    \label{fig:dusty}
\end{figure}

\begin{figure}
    \centering
    \includegraphics[scale = 0.28]{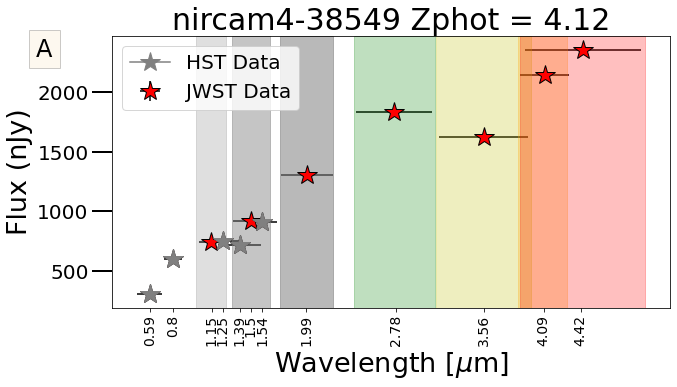}
    \includegraphics[scale = 0.185]{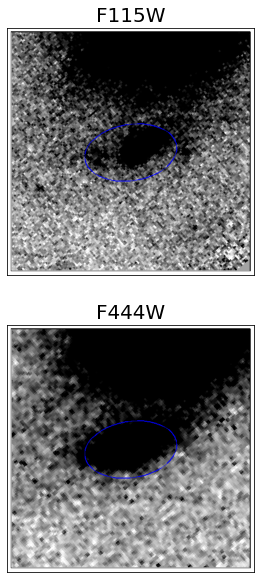}
    \includegraphics[scale = 0.28]{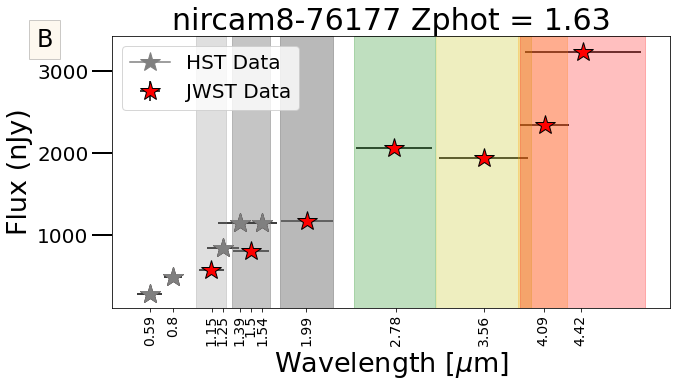}
    \includegraphics[scale = 0.185]{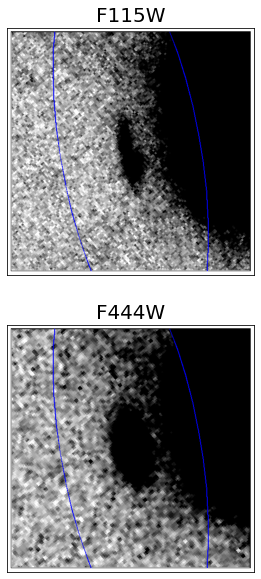}
    \includegraphics[scale = 0.28]{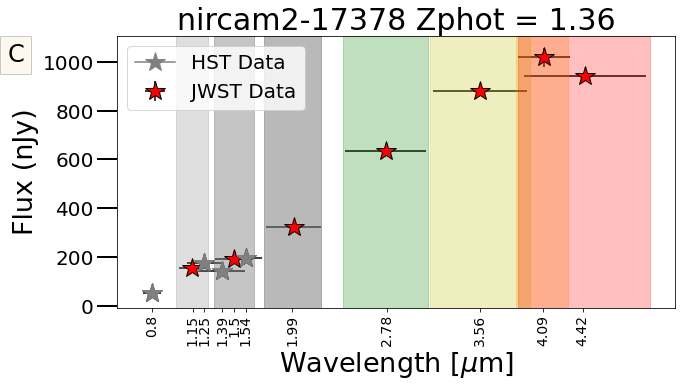}
    \includegraphics[scale = 0.185]{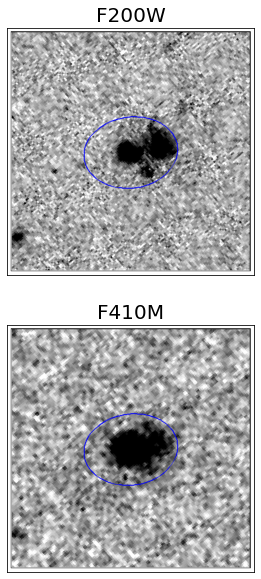}
    \caption{
    Example SEDs of galaxies with red continua and emission lines rejected from our EELG sample and assigned confidence Tier R. Our selection criteria (described in Section 3.1) incorrectly identifies 62 of these galaxies as EELGs because we do not account for their unusually red continua. After visual inspection we remove them from the sample because the photometry implies that their emission lines are bright but not extreme (i.e., $\mathrm{EW}_{OF}<5000$\AA). Photometry postage stamps show a continuum filter (top) and an extreme emission filter (bottom) with North up and East left. All postage stamps are 4\arcsec\ in width and height. {\bf{Panels A, B}}: Both red emission line galaxies have similar SEDs with steep red continua and bumps in the broadband photometry characteristic of nebular emission. Both galaxies are near neighbors to larger, brighter galaxies. The $z_\mathrm{phot} = 1.6$ in Panel B implies Pa$\alpha$ in F444W and Pa$\beta$ in F277W. The $z_\mathrm{phot} = 4.1$ in Panel A suggests \Hb\ + \OIII\ in F277W and \Ha\ falling redward of the F444W filter. {\bf{Panel C}}: A red galaxy with an emission line inferred in F410M. The $z_\mathrm{phot}=1.36$ solution places Pa$\alpha$ in F410M. The image shows three highly clustered sources.
    }
    \label{fig:tierR}
\end{figure}

A subset of our sample includes sources with highly red continua and emission lines that are not extreme. These pass our selection because they do exhibit nebular emission, creating an SED that is not well-fit by a straight line as used in our red-galaxy rejection criterion Equation \ref{eqn:chi}.
However, their nebular emission is not extreme (observed frame $\mathrm{EW} > 5000$\AA), and it is picked up in our selection due to the red continuum (rather than our assumed flat, blue continuum). These are unlike the source in Panel E of Figure \ref{fig:other_specs} that has a red SED but demonstrates sufficiently strong nebular emission to pass our observed-frame EW $>$ 5000\AA\ requirement. We retain sources similar to Panel E of Figure \ref{fig:other_specs} in our EELG sample, while those with highly red continua and emission lines that are not extreme are rejected and categorized as "Tier R". We detect 3 cases of spectroscopic overlap with these sources, two of which are shown in Figure \ref{fig:dusty}. We present a few example SEDs from this confidence tier in Figure \ref{fig:tierR}.

A few of these sources have spectroscopically confirmed emission lines, but with EWs that are too small to be classified as EELGs.  Each galaxy has a very red continuum with emission lines that are bright but do not meet our definition of ``extreme'' (observed-frame $\mathrm{EW}>5000$\AA).
We assume that the 3 spectroscopically observed red emission-line galaxies are representative of the larger population of 62 galaxies that meet our initial EELG criteria but have similarly red photometric SEDs. We reject these red emission-line galaxies from our EELG sample and categorize them as ``Tier R'' in Table \ref{tbl:t2}.

The population of red-continuum emission-line galaxies is surprising because bright emission lines are usually consistent with a young stellar population and a blue continuum. The red continuum may imply significant dust reddening, which could mean that these sources have emission lines that are \text{intrinsically} high-EW but are attenuated such that they are observed to be merely bright rather than extreme. These sources may instead have non-uniform attenuation with patchy dust such that the continuum is reddened while the emission lines have little to no attenuation. An additional explanation is that their SEDs may include a mix of red and blue components, perhaps with reddened AGN emission lines superimposed on young galaxy starlight \citep{Kocevski2023, Barro2023, Labbe2023}.

\section{Properties of CEERS EELGs}

We discuss the properties of the EELG sample with our highest-confidence EW measurements (i.e. the ``Tier 1'' sample). 

\subsection{General Sample Properties}

\begin{figure}[t]
    \centering
    \includegraphics[scale = 0.3]{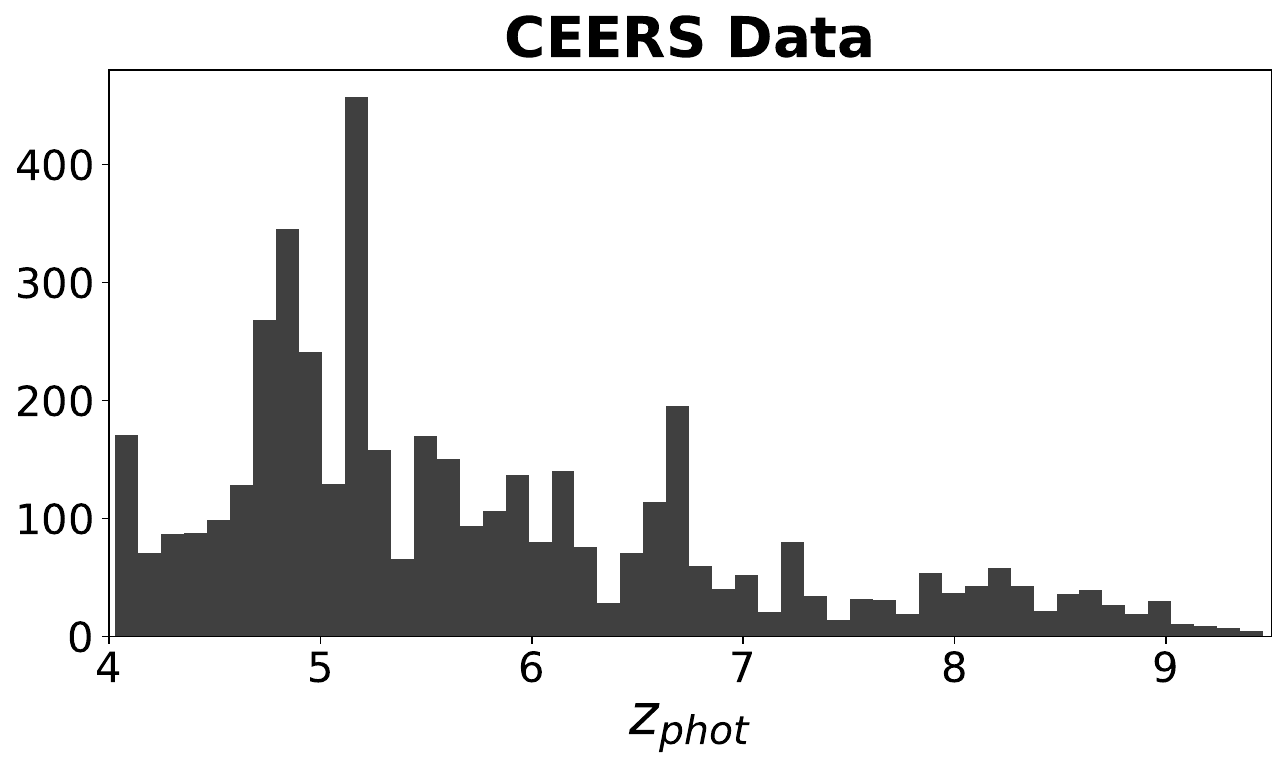}
    \includegraphics[scale = 0.3]{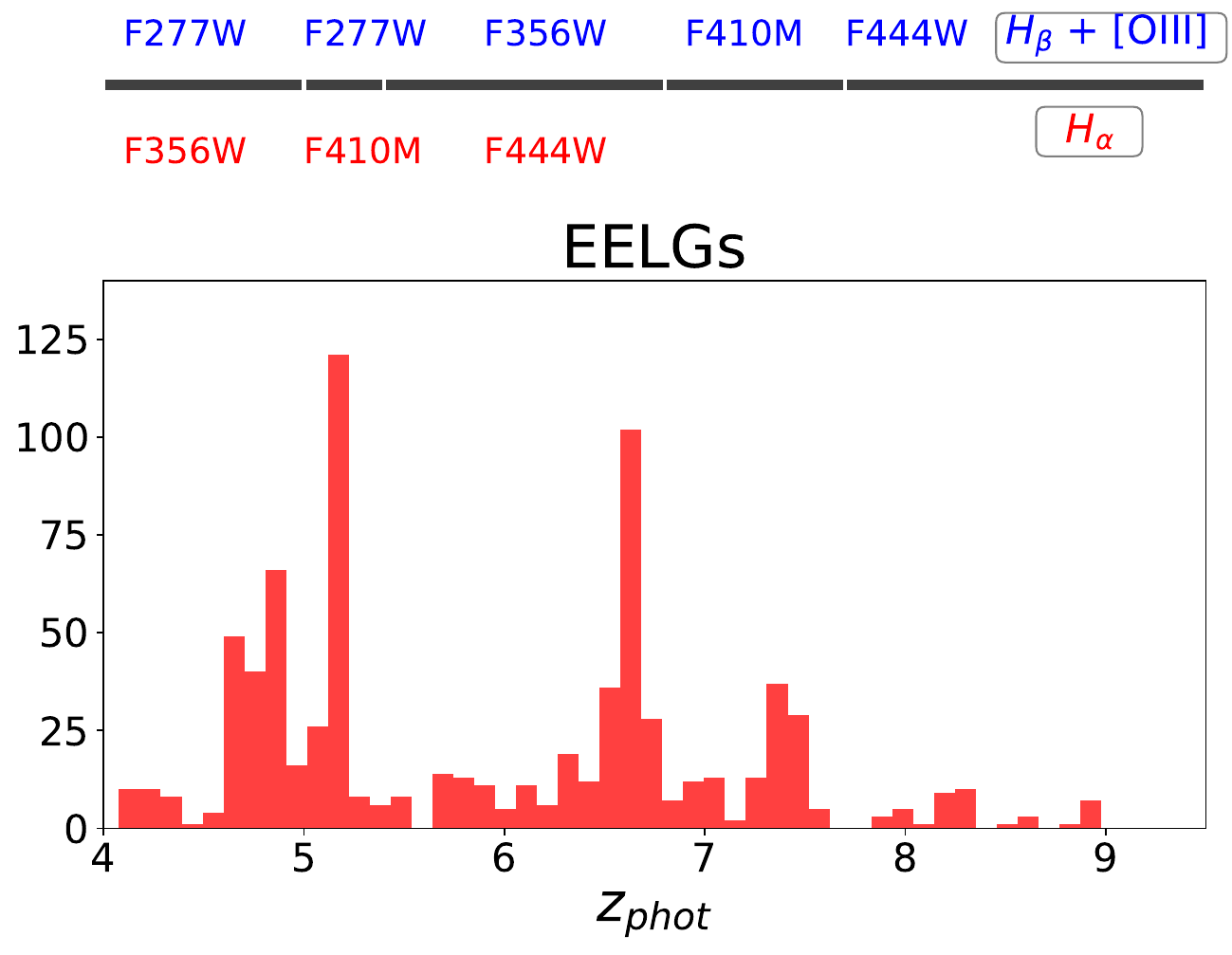}
    \caption{
    The distribution of photometric redshifts for non-EELGs (top) and EELGs (bottom) in CEERS. Both distributions have a peak at $z \sim 5$ that corresponds to \Ha\ in the F410M filter, and another peak at $z \sim 6.7$ that corresponds to \Ha\ at the red edge of F444W and \Hb\ + \OIII\ in F356W, where F410M samples the continuum between emission lines. The non-uniform distribution is likely caused by the photometric redshift estimation accurately identifying emission lines (both extreme and not) in each filter, but with best-fit SED solutions that ``pile up'' at specific redshifts while the true redshift distribution is much smoother. The similar redshift distribution seen in both samples supports the idea that EELGs are the tail of a broader distribution of emission-line galaxies. We indicate which filters include each emission line with horizontal bars that span the appropriate redshift ranges. The completeness for EELG identification in each of these redshift ranges is presented in Figure \ref{fig:percentages}.}
    \label{fig:photoz_dist}
\end{figure}

\begin{figure}[t]
    \centering
    \includegraphics[scale = 0.35]{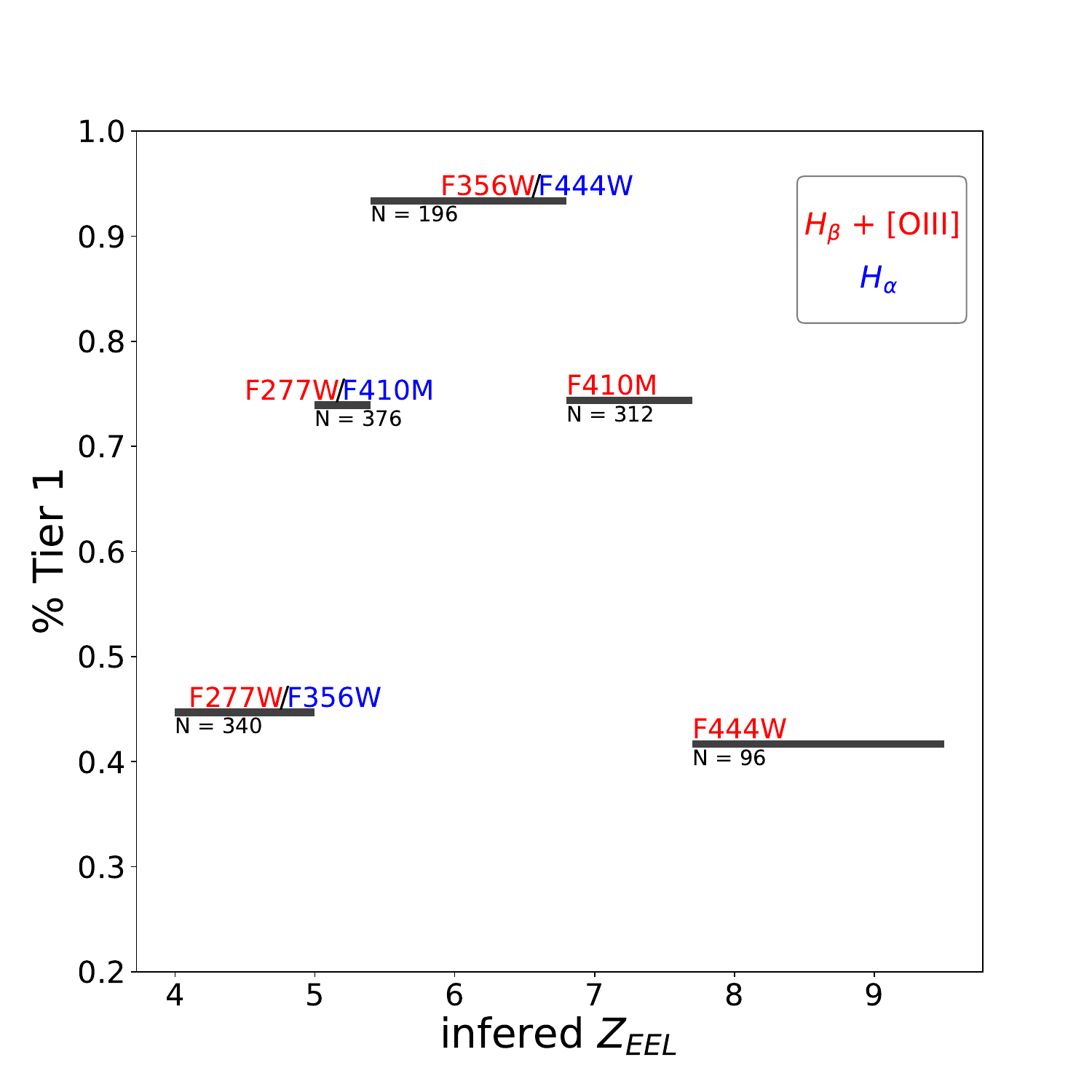}
    \caption{Fraction of EELGs that fall in Tier 1 as a function of the redshift $z_\mathrm{EEL}$ implied by the location of the extreme emission line in the photometric filters. $N$ is the total number of EELGs that fall in Tiers 1 and 2 for a given redsfhift range.
    }
    \label{fig:percentages}
\end{figure}

Figure \ref{fig:photoz_dist} presents the distribution of photometric redshifts for our EELG sample and for the sample of non-EELGs in CEERS  with $F_\nu > 10$ nJy in at least 5 photometric filters (the same as our detection threshold for the EELG sample). The distribution is distinctly non-uniform. The $z_\mathrm{phot}$ distributions of both EELGs and non-EELGs have peaks at $z \sim 5.2$ that correspond to the \Ha\ line falling within the F410M filter and \Hb\ + \OIII\ falling in F277W. The EELGs also have a peak at $z \sim 6.7$ associated with the \Hb\ + \OIII\ lines in the F356W filter and \Ha\ at the red edge of the F444W filter, with F410M sampling the continuum between extreme emission lines.
This implies that the non-uniform distribution is caused by the reliability of the SED fits, with higher photometric redshift accuracy when emission-lines, both extreme and not, are observed in the F410M filter or on either side of it and decreased accuracy at other redshifts.

The galaxy distribution may have an intrinsically clustered distribution in redshift, but it is improbable that this clustering coincides with the two specific redshift ranges associated with emission lines falling within and around F410M. The peaks in the $z_\mathrm{phot}$ distribution are likely associated with ``herding'' of the photometric redshifts into specific SED-fitting solutions at $z=5.2$ and $z=6.6$ while the true distribution is actually much broader and smoother.

Figure \ref{fig:percentages} displays the fraction of EELGs classified in the Tier 1 sample, with photometric redshifts that agree with the inferred redshift from the extreme emission line(s). The fraction of sources in Tier 1 is an effective estimate of our completeness for the EELG selection. The utility of the F410M filter is primarily responsible for the redshift ranges that correspond to high completeness fractions. Our highest completeness sources are associated with ranges where F410M includes either \Ha\ ($z \sim 5.2$) or \Hb\ + \OIII\ ($z \sim 7.2$), or F410M samples the continuum between \Hb\ + \OIII\ in F356W and \Ha\ in F444W ($6 \lesssim z \lesssim 7$). Our lowest completeness fractions are associated with single line detections of \Hb\ + \OIII\ in F444W or \Hb\ + \OIII\ in F277W and \Ha\ in F356W without a medium band filter to reliably sample the continuum between these lines.

\begin{figure*}[h]
    \centering
    \includegraphics[scale = 0.4]{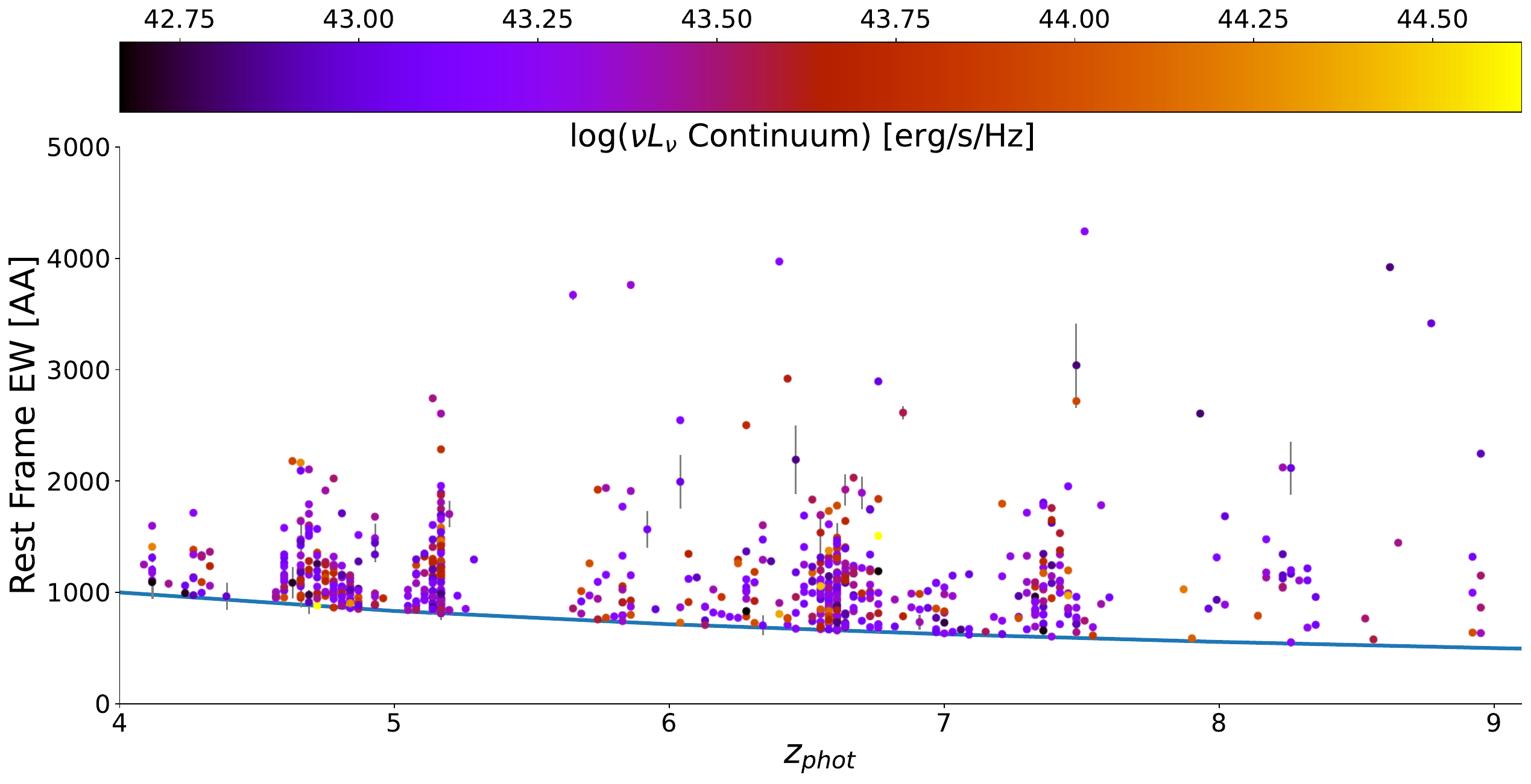}
    \caption{The distribution of rest-frame EWs for our sample of high-confidence (Tier 1) EELGs. The blue line denotes our selection limit of observed-frame $\mathrm{EW}>5000$\AA. Color represents continuum luminosity, measured the photometric filter two blueward of the filter capturing \Hb\ + \OIII. The density of sources near the selection criteria, indicated by the y-axis histogram, suggests that our EELG selection samples the tail of a broader distribution of emission-line galaxies. }
    \label{fig:ewcut}
\end{figure*}

Figure \ref{fig:ewcut} presents the distribution of rest-frame equivalent widths measured for the EELGs in our highest-confidence (Tier 1) sample. Our EELG selection required observed-frame $\mathrm{EW}>5000$\AA\ and this detection limit is shown in blue in the Figure. Most of the EELGs have rest-frame EWs near the detection limit, implying that our EELG sample is the tail of a continuous distribution of high emission-line EWs in galaxies rather than a distinct population. Our rest-frame EW distribution is broadly consistent with previous work indicating a median EW of 700-800\AA\ in $z \sim 7$ galaxies \citep{deBarros2019, Labbe2023, Endsley2023}, since this median roughly corresponds to our selection limit. Our EW distribution is consistent with the log-normal \Hb\ + \OIII\ EW distribution from CEERS reported in \citet[]{Endsley2023}{}.

\begin{figure*}[h]
    \centering
    \includegraphics[scale = 0.18]{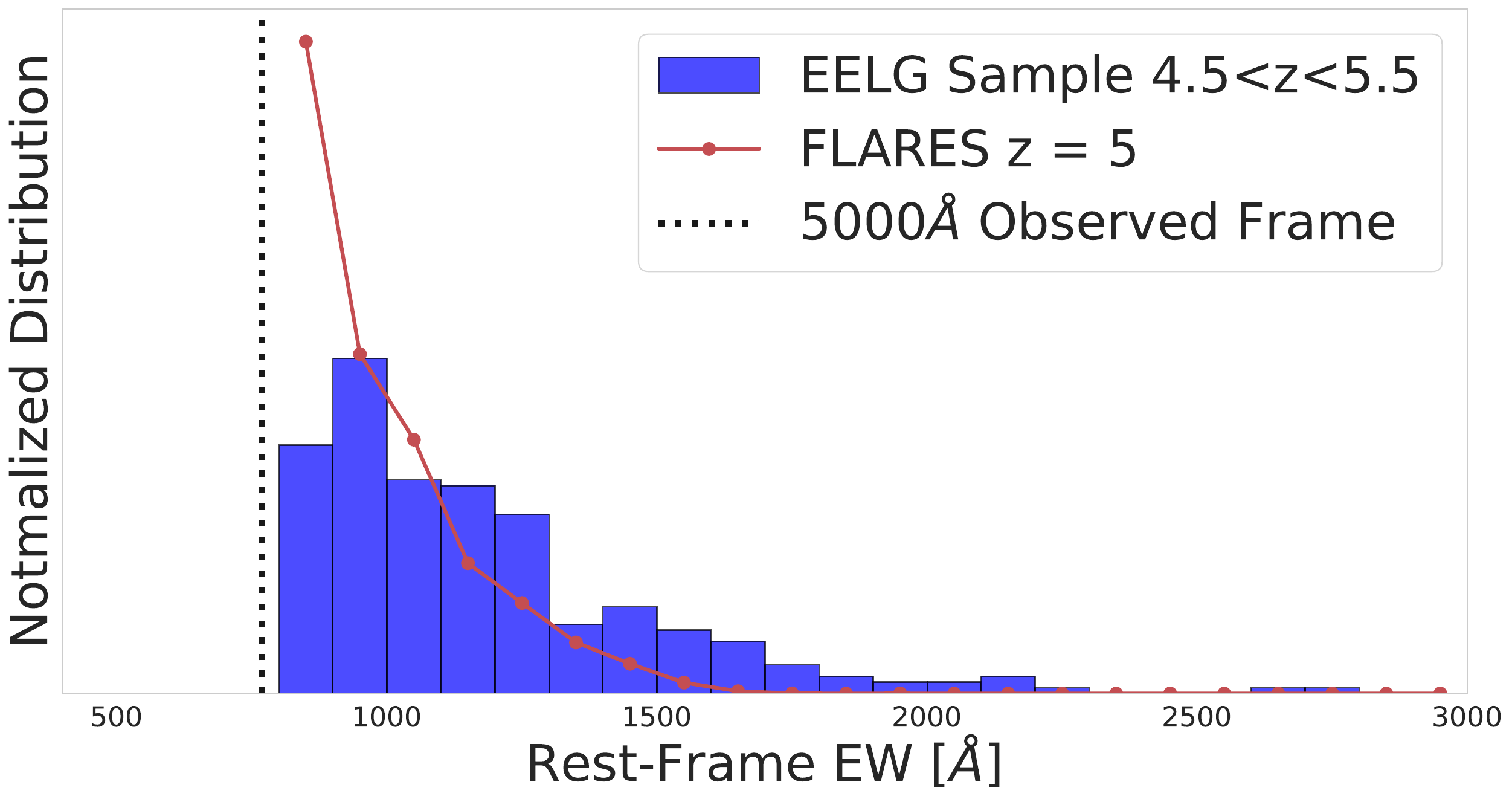}
    \includegraphics[scale = 0.18]{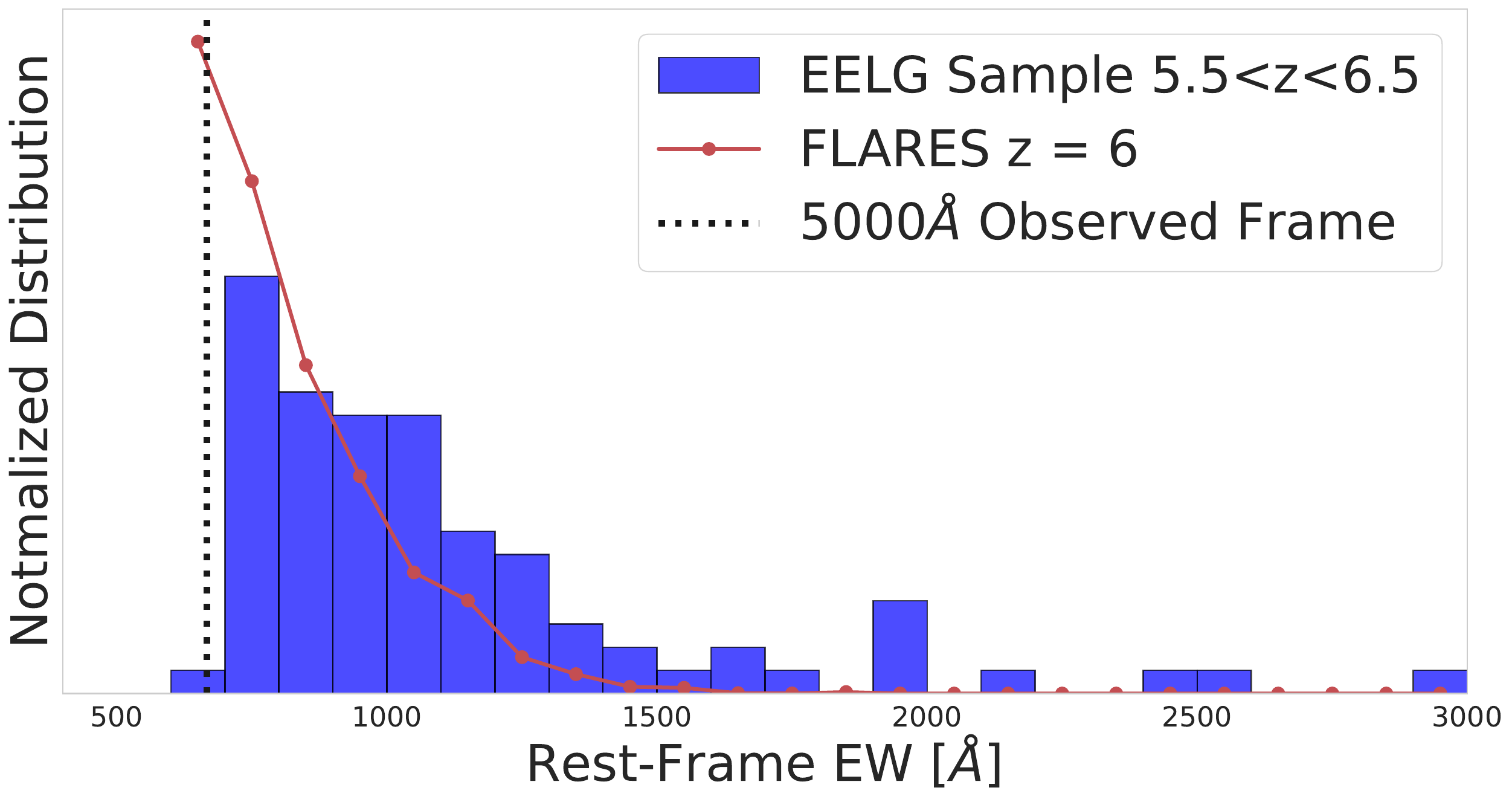}
    \includegraphics[scale = 0.18]{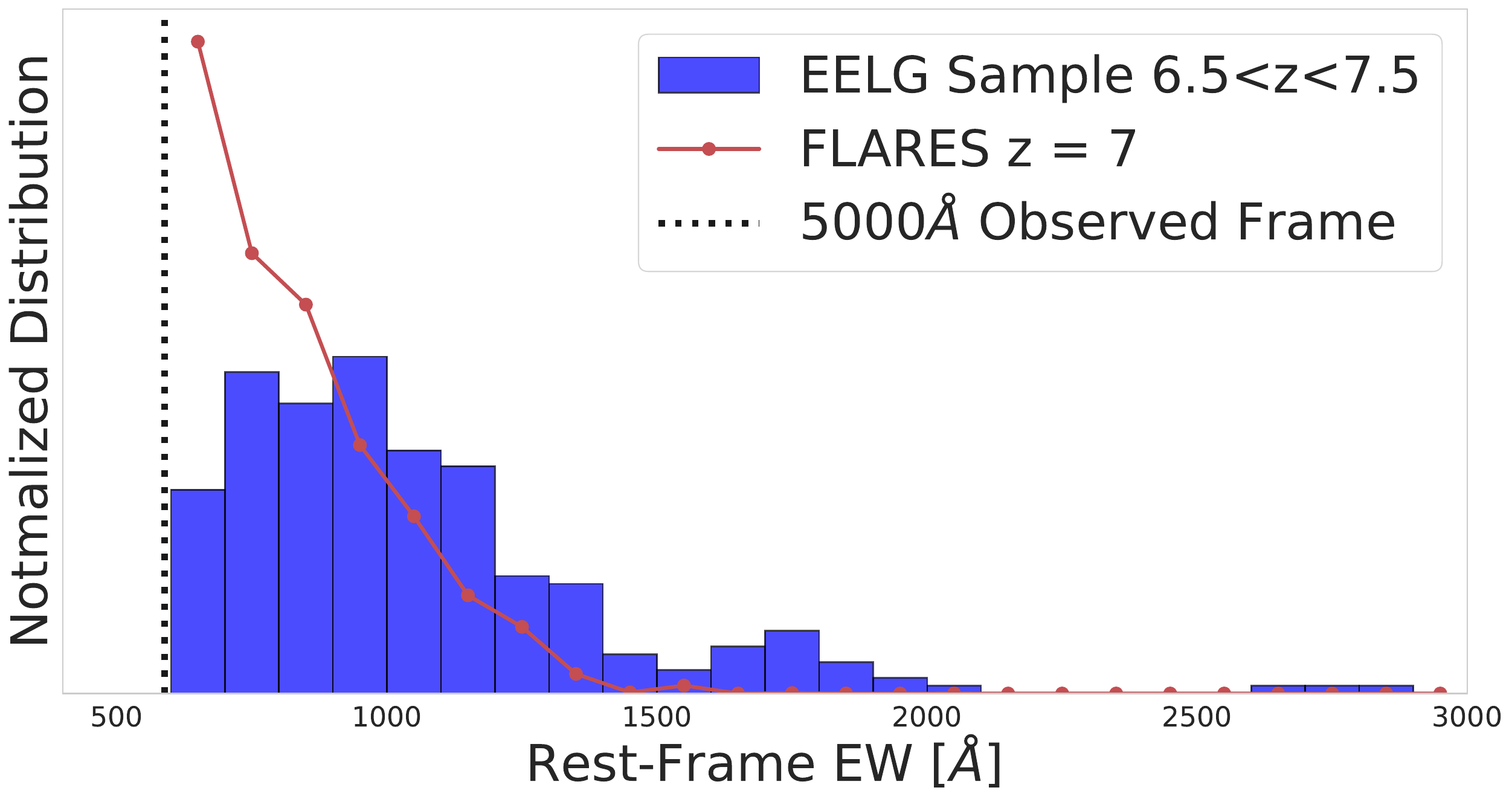}
    \includegraphics[scale = 0.18]{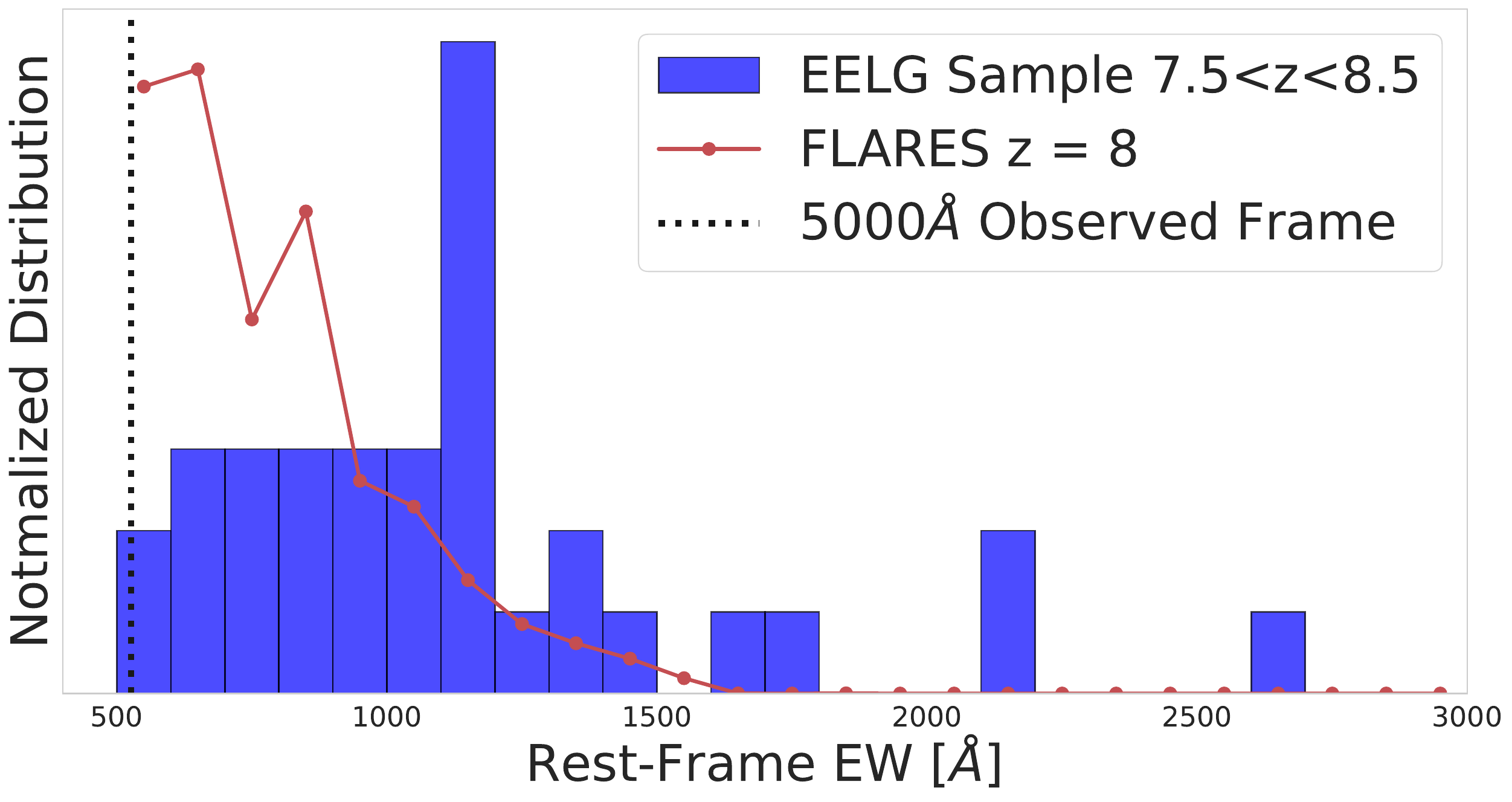}
    \caption{The distribution of rest-frame equivalent widths for our sample of high confidence (Tier 1) EELGs compared to theoretical predictions from FLARES \citep[]{Wilkins2023}. We make this comparison by binning our EELG sample to redshift and equivalent width ranges that match the FLARES simulations, and normalize both the observed and predicted histograms to the same integrated area. Observational biases affect the low-EW part of the distribution, but above this observations and FLARES predictions are broadly consistent at rest-frame $800 \lesssim EW \lesssim 1500$, FLARES underestimates the observed reported number counts for higher EWs, especially at (EW$>2500$\AA). This suggests that AGN (which are excluded from FLARES) may contribute to the most extreme EELGs.}
    \label{fig:ew_theory}
\end{figure*}

\begin{figure}[t]
    \centering
    \includegraphics[scale = 0.28]{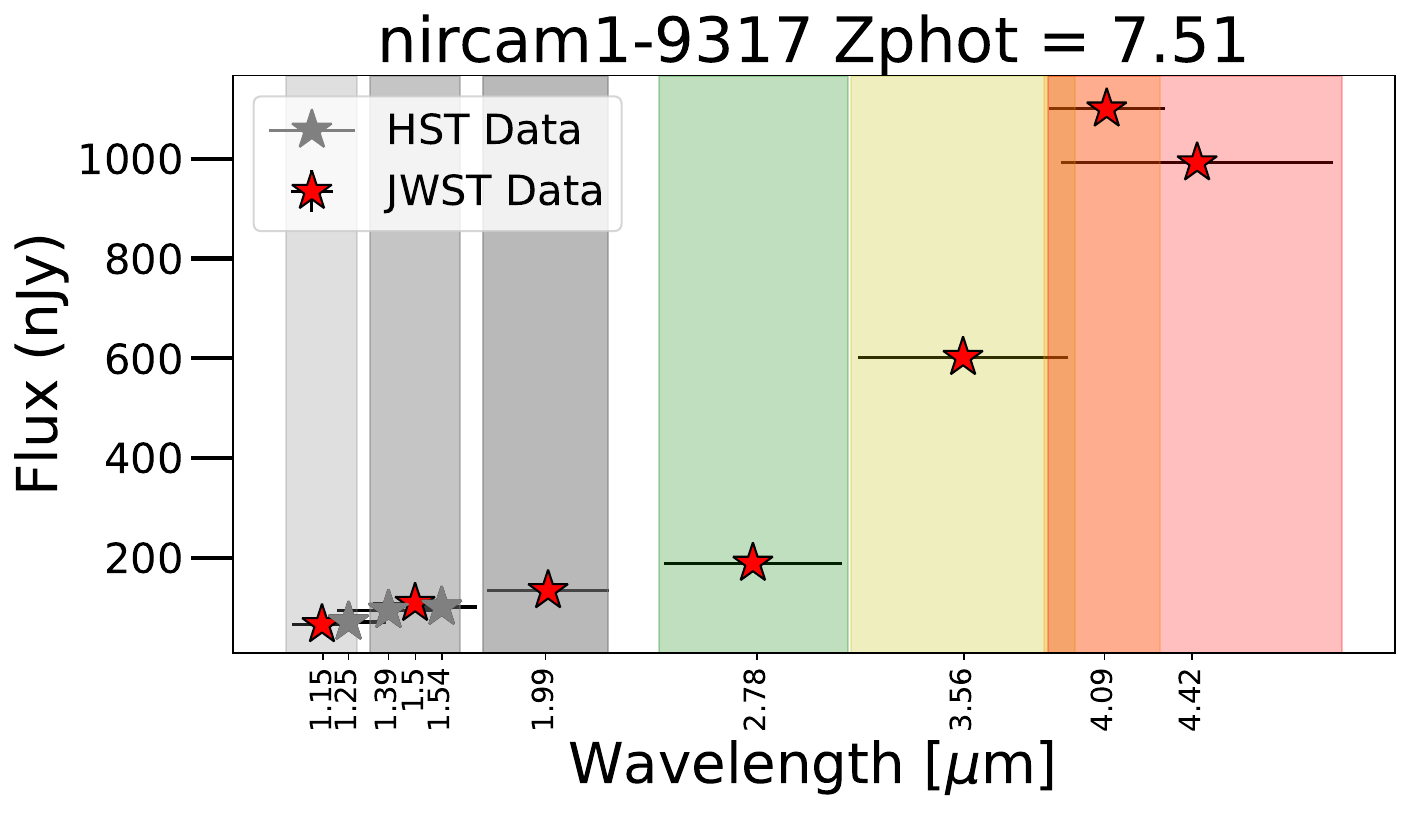}
    \includegraphics[scale = 0.185]{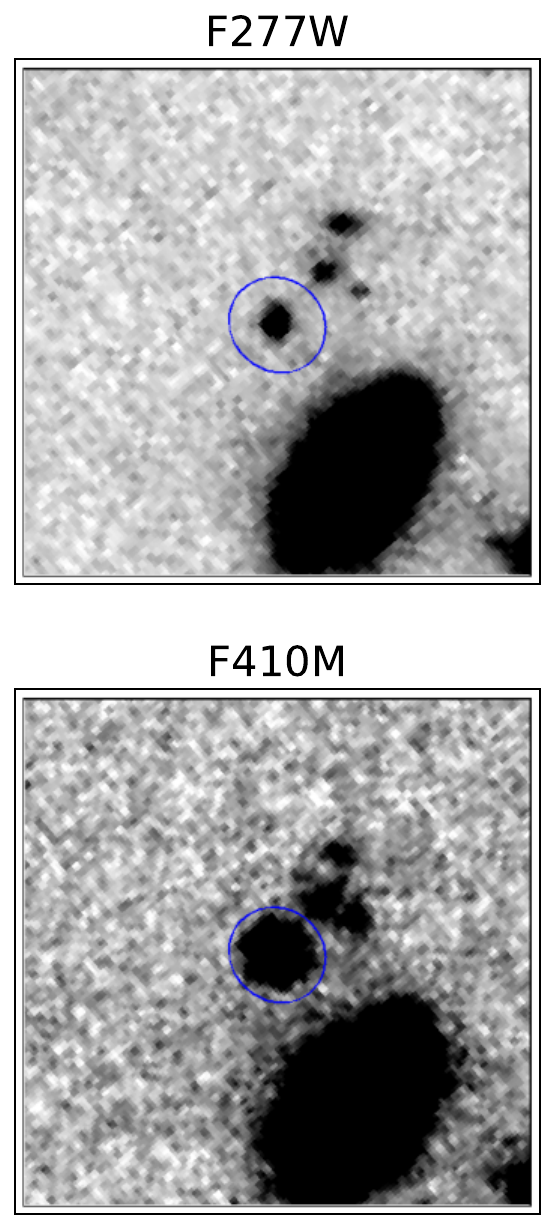}
    \caption{The galaxy with the most extreme rest-frame equivalent width, an EELG with $EW_\mathrm{rf}$ = 4200 \AA\ (and $EW_\mathrm{obs}$ = 36000 \AA) inferred to have \Hb\ + \OIII\ in the F410M and F444W filters (and partially in F356W). The right image shows the galaxy postage stamps for an emission and continuum filter, each 4\arcsec\ on a side
    % , indicating that the galaxy is more extended in the emission-line filter. 
    % While this most-extreme case falls outside of predictions from FLARES, it is consistent with the log-normal \Hb\ + \OIII\ EW distribution from CEERS reported in \citet[]{Endsley2023}{}.
    Our population of highest-EW sources are consistent with the inferred intrinsic distribution of \citet{Endsley2023} but are inconsistent with predictions from FLARES, suggesting that AGN may contribute to the most extreme EELGs.
    }
    \label{fig:big}
\end{figure}

Figure \ref{fig:ew_theory} showcases how this sample compares to theoretical predictions from First Light And Reionisation Epoch Simulations \citep[FLARES;][]{FLARES-I, FLARES-II, Wilkins2023} based on nebular reprocessing of stellar emission (without AGN). The observed distribution of EWs provides a good match to the FLARES predictions for rest-frame EW$<2500$\AA. However, a key difference is that FLARES does not predict the existence of objects with extremely high rest-frame (EW$>2500$\AA) equivalent widths, like our most extreme EELG source shown in Figure \ref{fig:big}, those identified in CEERS by \citet[]{Endsley2023}, and those occupying the tail of our distribution in Figure \ref{fig:ewcut}. The EELG in Figure \ref{fig:big} has a rest-frame $\mathrm{EW} = 4200$\AA\ from \Hb\ + \OIII\ emission in the F410M and F444W filters. This galaxy is unresolved in the NIRCam imaging, implying a very compact morphology for both the continuum (F277W) and emission lines (F356W and F444W) and appears to be in a densely populated environment. This may suggest an additional source of ionizing photons not modeled in FLARES, for example contribution from accreting supermassive black holes. 
%
% An alternative explanation for the high measured rest-frame equivalent width would involve a collapse of our blue continuum assumption. However, the continuum is relatively flat in $F_{\nu}$ between F115W and F277W which would suggest low dust attenuation. At this redshift, the dust required to generate this degree of attenuation is unlikely to have formed, and the addition of dust into the explanation would require a lower redshift solution. This source also drops out from HST photometry blueward of the F115W filter, consistent with a Lyman break at the photometrically inferred redshift. Additionally, these high rest-frame equivalent width sources are not a distinct population and are consistent with the tail end of the distribution, indicating that they represents realistic EW measurements. This source and many of the extremely high rest frame equivalent width sources are excellent candidates for spectroscopic followup. 
In Section 4.2 we discuss additional evidence for AGN contribution to EELGs from an analysis of their morphologies.

\begin{figure}[t]
    \centering
    \includegraphics[scale = 0.35]{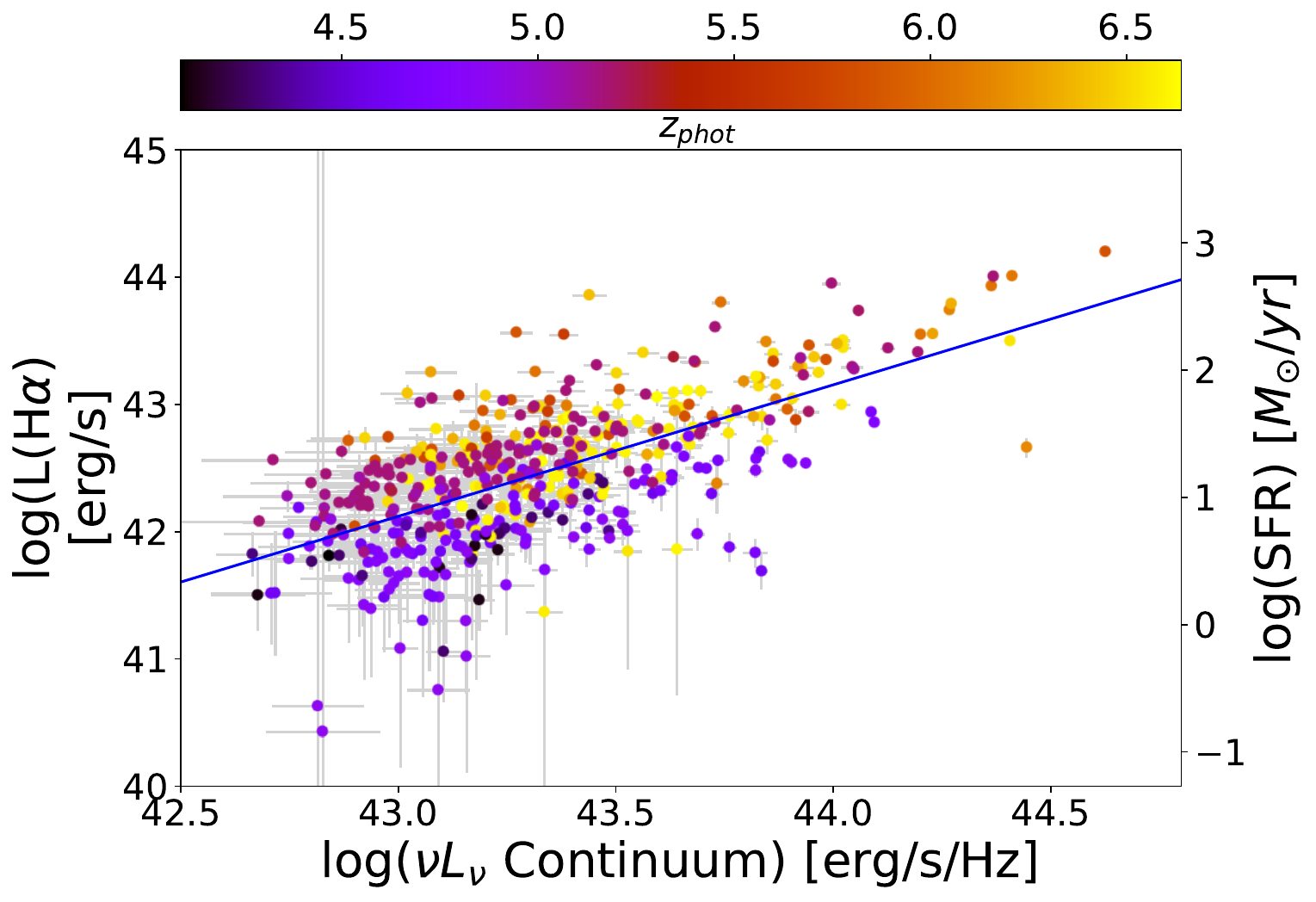}
    \includegraphics[scale = 0.35]{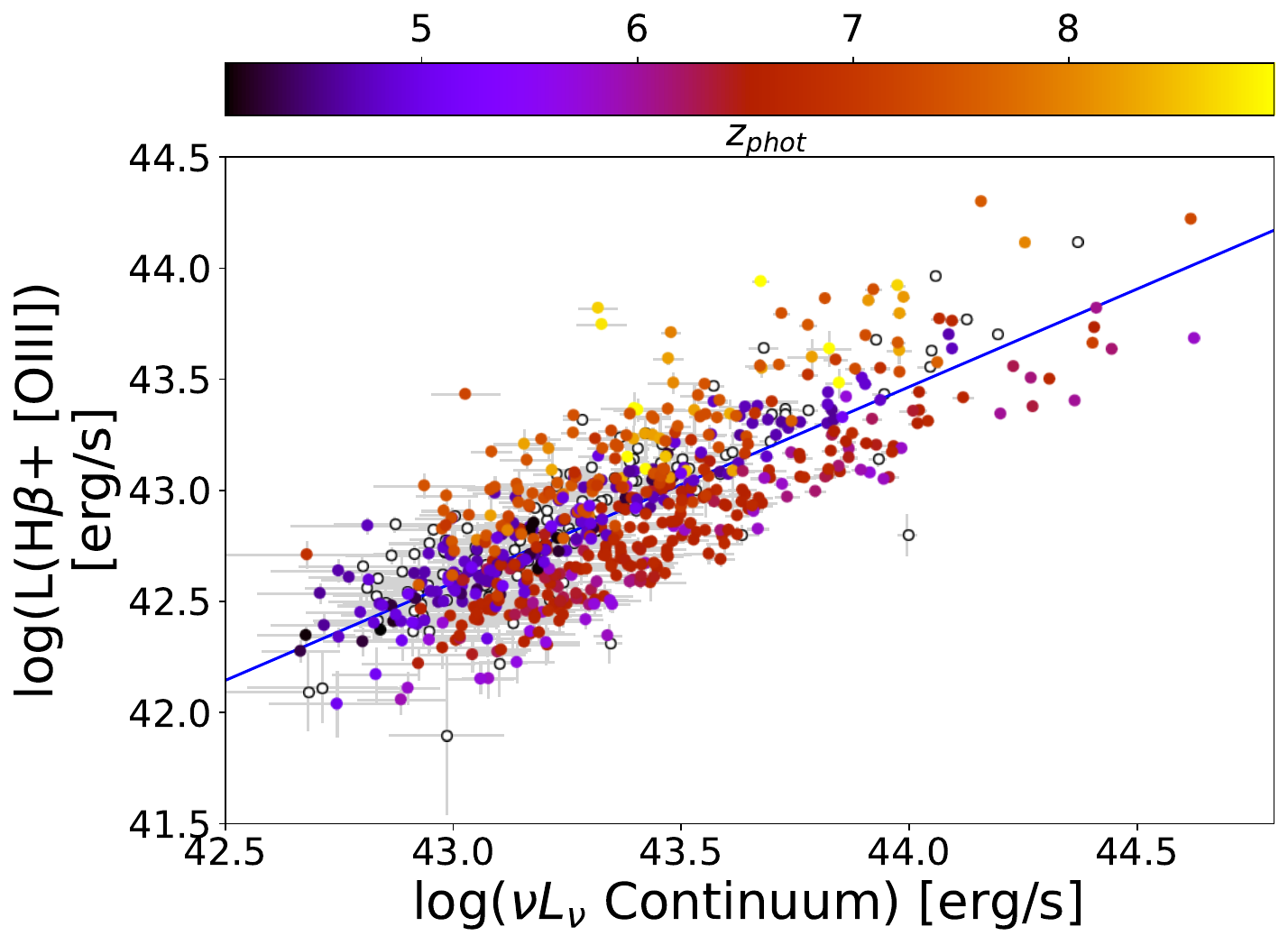}
     \caption{\Ha\ and \Hb\ + \OIII\ Luminosity vs continuum luminosity for the EELGs. Continuum luminosity is measured from the photometric filter lying two filters blueward of the \Hb\ + \OIII\ emission. Color gradient represents photometric redshift. The top panel includes an axis for SFR measured from \Ha\ luminosity using the \citet{Kennicutt2012} relation. The best-fit relation between \Ha\ and continuum luminosity has a slope of $m=1.03 \pm 0.05$ and a Spearman correlation coefficient of 0.62, and the best-fit relation between \Hb\ + \OIII\ and continuum luminosity has a slope of $m=0.88 \pm 0.03$ and a Spearman correlation coefficient of 0.78.
%     \textbf{Top:} best fit line of $y = 1.03 \pm 0.05 x - 2.32 \pm 2.35$ with a Spearman correlation coefficient of 0.62. \textbf{Bottom:} best fit line of $y = 0.88 \pm 0.03 x - 4.71 \pm 1.11$ with a Spearman correlation coefficient of 0.78. In both plots, large error is associated with EELGs that are the least luminous for each line, representing EELGs that were not selected for extreme emission in these associated lines.
     Both lines have slope near 1, in agreement with theoretical predictions for the relationship between emission-line and continuum luminosity from FLARES \citep{Wilkins2023}.} 
    \label{fig:lum}
\end{figure}

Figure \ref{fig:lum} presents the emission-line luminosities inferred for our high-confidence (Tier 1) EELG sample. Luminosity is calculated as:
\begin{equation}
    L = (F_\nu-C_\nu) \frac{c}{\lambda^2} 4 \pi d_L^2  \Delta  \lambda
\end{equation}
where $F_\nu$ is the flux density in the filter with extreme emission, $C_\nu$ is the continuum flux density, $d_L$ is the luminosity distance implied by the photometric redshift,
% as calculated from \texttt{astopy.cosmology.luminosity\_distance},
and $\Delta\lambda$ is the filter width. The \Ha\ line luminosity contains some contribution from the nearby \NII\ and \SII\ doublets, and the \Hb\ + \OIII\ luminosity represents the sum of the three lines (plus potential additional contribution from weaker features like \HeII$\lambda$4686 and higher-order Balmer lines). The \Hb\ + \OIII\ luminosities are generally greater than the \Ha\ luminosities, implying high ionization and/or moderately low metallicity gas conditions: we discuss this in more detail in Section 4.3.
The top panel of \ref{fig:lum} includes an axis indicating the SFR inferred from the \Ha\ luminosities, using the \citet{Kennicutt2012} relation of $\log(\mathrm{SFR}) = \log[L(\Ha)] - 41.27$. The range of $\mathrm{SFR} \sim 1-100$~M$_\odot$~yr$^{-1}$ for the EELGs is consistent with moderate to high starburst activity observed in the most extreme emission-line galaxies at these epochs (e.g., \citealt{Heintz2023}).

Figure \ref{fig:lum} shows how the EELG emission-line luminosities correlate with continuum luminosity. We define the continuum as the flux in the photometric filter lying two filters blueward of the filter with \Hb\ + \OIII\ emission.
% These panels explore how line luminosities correlate with continuum luminosity, both of which are measurements from the photometry that, unlike other derived properties, do not rely on photometric modeling so we chose to report it instead of a stellar mass due to limitations discussed in Section 2.2. 
The best-fit lines in each panel have slopes near unity, in agreement with FLARES predictions from \citet[]{Wilkins2023} that have \Ha\ and \Hb\ + \OIII\ luminosities increasing linearly with continuum luminosity, i.e. a constant rest-frame EW distribution with continuum luminosity.

\subsection{Comparison of EELGs with the broader galaxy distribution}

We compare the properties of our EELGs with the broader sample of CEERS galaxies. Our CEERS comparison sample is constructed from all non-EELGs with photometric redshifts between $4<z<9.5$ and $F_\nu > 10~$nJy in at least five filters, using the same redshift range and continuum brightness limit as our EELG selection. Our implicit assumption for this comparison is that both the EELG sample and the larger CEERS comparison sample are representative of the broader galaxy population. Although our EELG sample is not complete, as shown in Figure \ref{fig:percentages}, our spectroscopic confirmation indicates that it is likely to be a relatively pure sample of EELGs. The similarity between the photometric redshift distribution of EELGs and non-EELGs seen in Figure \ref{fig:photoz_dist} also indicates that the biases of the photometric redshift accuracy are likely to affect both populations in similar ways.

\begin{figure}[t]
    \centering
    \includegraphics[scale = 0.23]{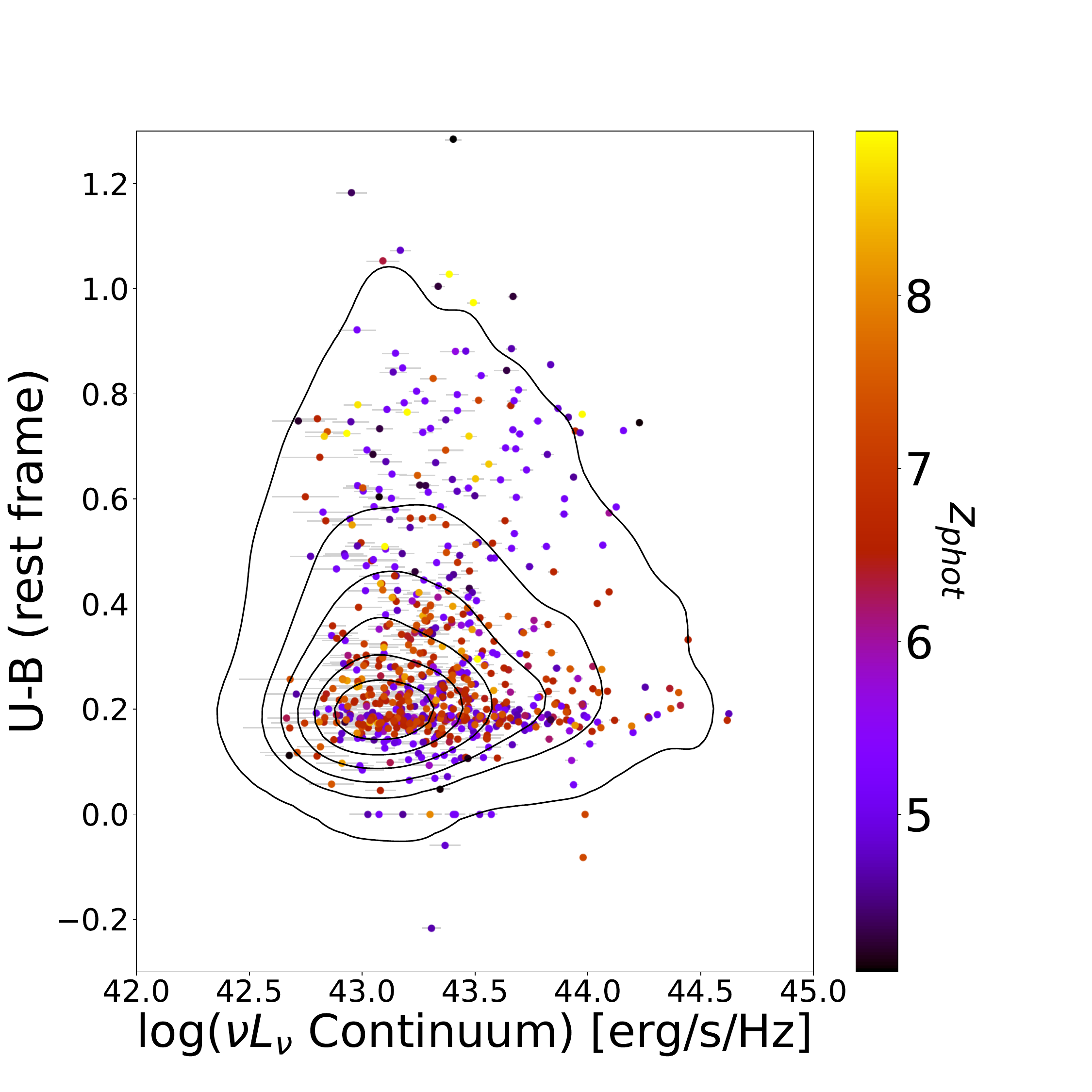}
    \caption{Rest-frame $U-B$ color versus continuum luminosity. Points represent our EELG sample while contours represent the CEERS comparison population. Color indicates photometric redshift. The continuum luminosity is derived from the photometric filter falling two filters blueward of \Hb\ +\OIII\ for both the background population and the EELGs. Most EELGs have blue continua, consistent with young stellar populations, but there is a substantial population of red EELGs, especially at lower redshift ($z_\mathrm{phot}<6$), that may indicate dusty starburst starburst galaxies.
    }
    \label{fig:rfcol}
    
\end{figure}

Figure~\ref{fig:rfcol} presents the rest-frame $U-B$ color versus continuum luminosity for both the EELGs and the CEERS non-EELG comparison sample. As in previous figures, the continuum luminosity is measured from the photometric filter that is two filters blueward of the filter with \Hb\ +\OIII\ emission. Rest-frame $U-B$ colors are measured from \texttt{EAZY}. As discussed in Section 2.2, \texttt{EAZY} struggles to fit SED templates to our extreme emission-line sources. To force \texttt{EAZY} into a solution for rest-frame colors, we replace photometric fluxes that include extreme emission lines with the mean continuum flux, calculated as the mean flux of all filters with fluxes $<1\sigma$ from the mean flux in all JWST photometric filters.

Our EELG population is generally blue in color, especially for our higher redshift population. This is consistent with young stellar populations and extreme fluorescence resulting from OB stars in \HII\ regions. That said, the overall galaxy population in the same redshift range in CEERS has similarly blue colors. This implies that our EELG sample is the high-EW tail of a broader population of star-forming galaxies that dominates the galaxy population at cosmic dawn \citep[e.g.,][]{Madau2014}. However, not all EELGs follow this trend, with many exhibiting redder colors, especially at lower photometric redshifts.
% This could suggest star-forming activity in early starbursting galaxies.
The EELGs with redder $U-B$ colors may be dusty starbursts and, for nebular to stellar attenuation ratios that are greater than unity \citep{Calzetti2000}, would include intrinsic emission lines that are even more luminous than the extreme observed EWs.

\begin{figure*}[t]
    \centering
    \includegraphics[width=\textwidth]{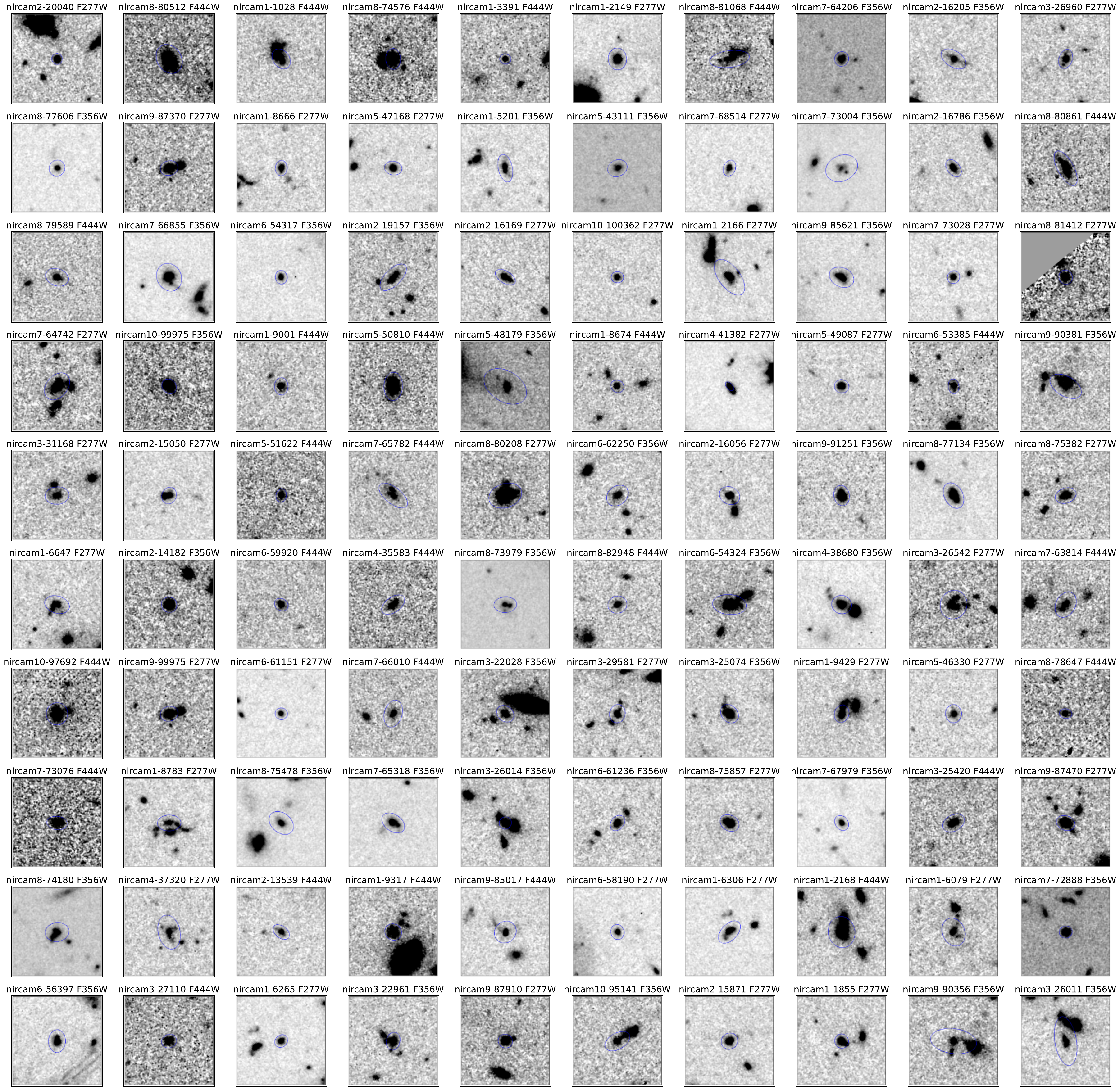}
    \caption{Postage stamps of the 100 brightest of our tier 1 sources in an emission filter. All images are 4" in width and height. North is up and east is left. Figures \ref{fig:morph_emission} and \ref{fig:morph_cont} are a combined animation in the online version which is temporarily available at \url{https://drive.google.com/file/d/1BEb1CRgE5R544Xu3IKMp_442VjfF3HfD/view?usp=sharing} }
    %Add rainbow citation here}
    \label{fig:morph_emission}
\end{figure*}

% \newpage
\begin{figure*}[t]
    \centering
    \includegraphics[width=\textwidth]{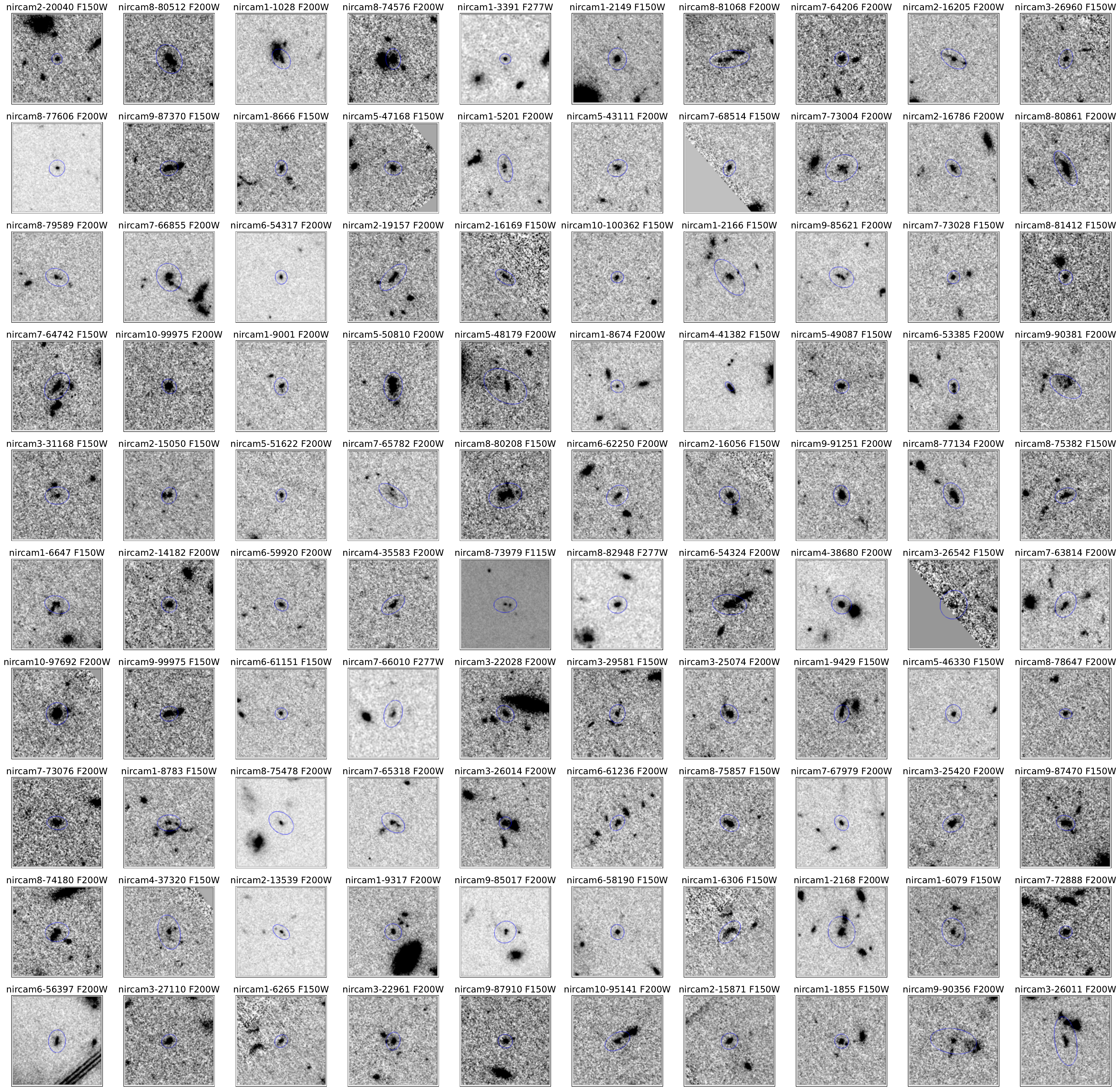}
    \caption{Postage stamps of the 100 brightest of our tier 1 sources in a continuum filter. All images are 4" in width and height. North is up and east is left. Figures \ref{fig:morph_emission} and \ref{fig:morph_cont} are a combined animation in the online version which is temporarily available at \url{https://drive.google.com/file/d/1BEb1CRgE5R544Xu3IKMp_442VjfF3HfD/view?usp=sharing} }
    %Add rainbow citation here}
    \label{fig:morph_cont}
\end{figure*}

% \newpage

Figure \ref{fig:morph_emission} and \ref{fig:morph_cont} showcase the images of our 100 brightest EELGs vary in a photometric filter that captures an extreme emission line and a photometric filter capturing the continuum.
% Overall, most of these EELGs appear to balloon outward in the extreme emission-line filter.
Most EELGs appear to have nearby neighbors. Some, such as nircam9-99975, nircam3-26011, and nircam7-64742 have nearby companions that are similarly brighter in the emission-line filter, supporting the idea that EELGs may be products of their environments fueled by galaxy mergers \citep[]{vanderWel2011}. However, several EELGs such as nircam1-8674, nircam4-38680 and nircam1-855 have many nearby neighbors that do not appear to be brighter in the emission filter. Further, many EELGs do not have any nearby neighbors, namely nircam6-54317 and nircam8-77605. 
%A particularly interesting source,  nircam2-16205, extends in the emission filter in a central region and not in the galaxy it occupies, implying that the extreme emission results from a central AGN and not the surrounding galaxy.
This diversity in the 100 brightest sources implies variety in environmental and spatial distributions for EELGs. 

\begin{figure}[t]
    \centering
    %Pick one of these after adjusting discussion
    \includegraphics[scale = 0.21]{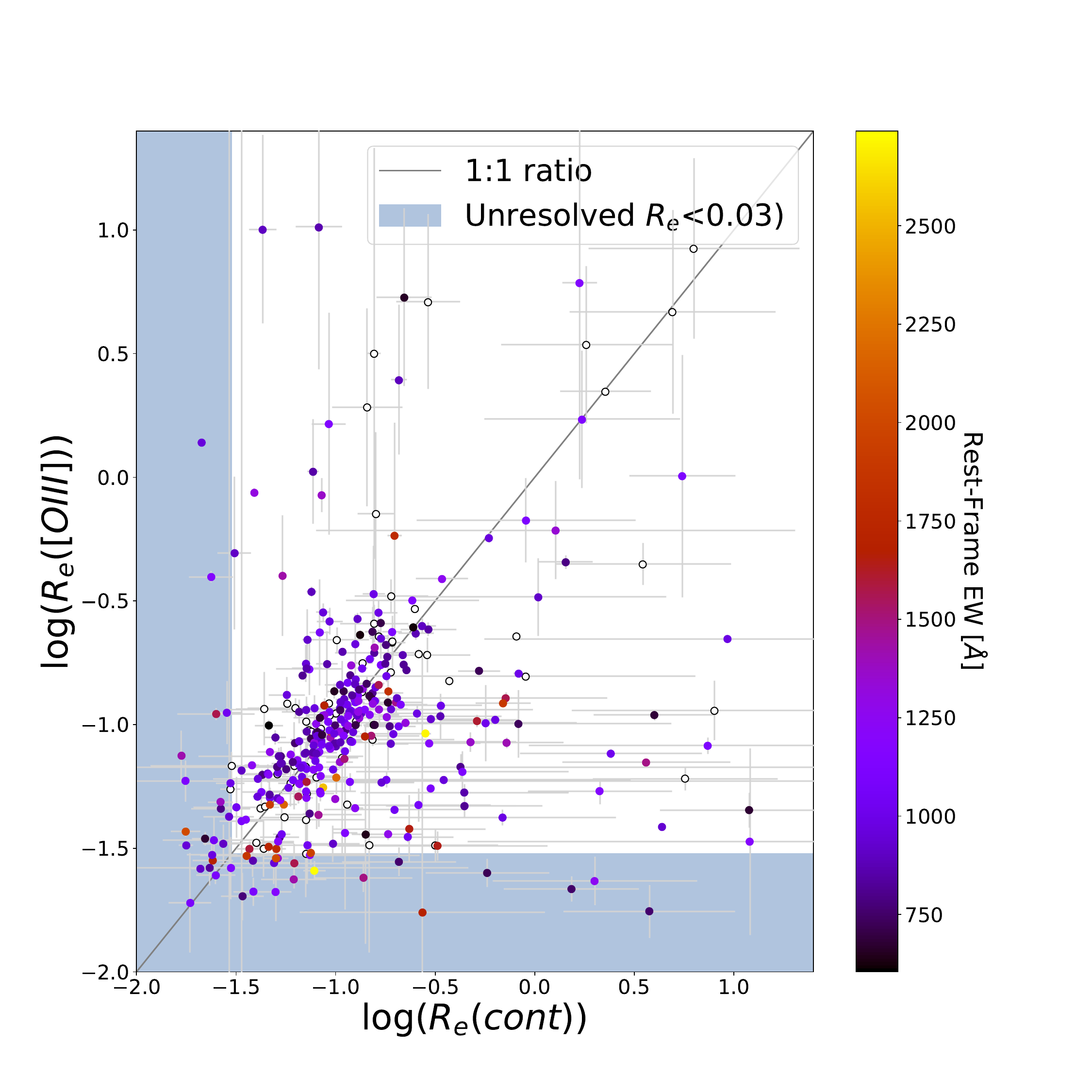}
    %\includegraphics[scale = .38]{example.png} %use other example, report as sed, report ew as well, new fig 24
    %\includegraphics[scale = 0.35]{ref_o3_ew.pdf}
    %includegraphics[scale = 0.35]{ref_o3_lo3.pdf}
   % \includegraphics[scale = 0.2]{ref_rf.pdf}
    \caption{Semi-major axis size in the filter with extreme \Hb\ + \OIII\ vs semi-major axis size in the continuum. Blue shaded regions indicate sizes that are unresolved. Color represents rest-frame EW. Most EELGs have similar sizes in their emission-line and continuum images, but there is a significant population of EELGs with more compact \Hb\ + \OIII\ sizes that is consistent with emission-line contribution from point-source AGN.
    }
    \label{fig:re}
\end{figure}

We next examine the intrinsic sizes of the EELGs, measured as semi-major axes from \texttt{GALFIT} by McGrath et al. (2023, in prep.). Figure \ref{fig:re} compares the semi-major axis sizes of EELGs measured in the filter with extreme \Hb\ + \OIII\ emission with the size measured in the continuum image. The continuum size is measured in a filter that is 2 broad-band filters blueward of \Hb\ + \OIII\ emission and typically corresponds to rest-frame $\sim$4200\AA. Most EELGs have similar sizes between their emission-line and continuum images, but with large scatter and a significant population of galaxies with smaller emission-line sizes. EELGs with compact \Hb\ + \OIII\ sizes may result from emission-line contribution from point-source AGN embedded within larger continuum sizes that are produced by galaxy starlight.
% EELGs that have more compact sizes in continuum filters could indicate star bursting is responsible for driving the extreme emission.
% From visual image inspection, almost all EELGs balloon outward in emission filters, illustrated in Figures \ref{fig:morph_cont} and \ref{fig:morph_emission}, and this semi-major axis size change is a measure of both the extreme emission and the surrounding structure, and how these change in size across images.
% and explore these semi-major axes in the filter containing \Hb\ + \OIII\ as inferred from the photometric redshift and a continuum filter we take as a photometric filter 2 filters blueward of \Hb\ + \OIII\. Figure \ref{fig:re} explores how these semi-major axes relate in proportion to each other. Many of the EELGs cluster around the broader CEERS distribution, but many still occupy both extremes of the plot, showcasing the variability in the EELG population.
The color-coding by rest-frame EW in Figure \ref{fig:re} suggests that the higher equivalent width sources tend to have more compact emission-line sizes than continuum sizes.
% We explore this in more detail in Figure \ref{fig:extended_example}.

\begin{figure}[h]
    \centering
    \includegraphics[scale = 0.25]{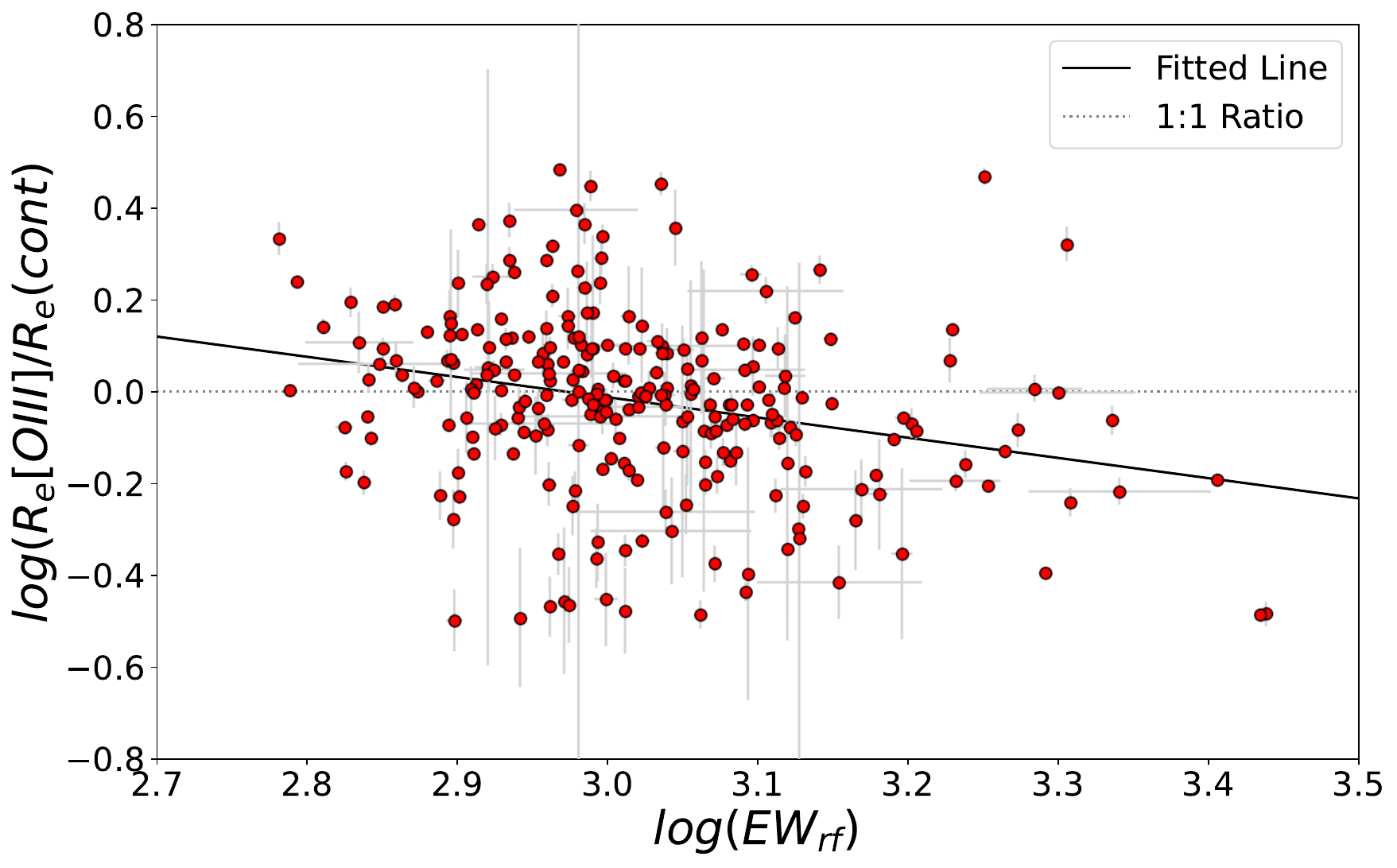}
    \caption{The ratio of the galaxy semi-major axis measured from the emission-line and continuum images compared to the rest-frame EW.
    The best-fit line has a slope $m =  -0.44 \pm  0.10$ and Spearman correlation coefficient of $-0.31$ that indicates a weak but significant anti-correlation. More compact emission-line sizes as a function of EW is consistent with contribution from a central AGN to the most extreme emission lines.
%     The data has a negative correlation associating compactness in the \Hb\ + \OIII\ filter and high rest frame EWs. This presents a potential tool for de-tangling starburst (more compact in the continuum filter) from AGN contribution (more compact in the \Hb\ + \OIII\ filter). The best fit line has slope $m =  -0.44 \pm  0.10$ and y-intercept $b = 1.31 +/- 0.30$ with a Spearmen correlation coefficient of $-0.31$ representing a weak to moderate negative correlation. 
    }
    \label{fig:size_ew}
\end{figure}
%CUT: +/- .5 ratio and >0nJy cont

%[[Model]]
%    Model(linear_model)
%[[Fit Statistics]]
%    # fitting method   = leastsq
%    # function evals   = 7
%    # data points      = 255
%    # variables        = 2
%    chi-square         = 8.77685514
%    reduced chi-square = 0.03469113
%    Akaike info crit   = -855.132074
%    Bayesian info crit = -848.049546
%    R-squared          = 0.06977359
%[[Variables]]
%    slope:     -0.44057906 +/- 0.10113754 (22.96%) (init = 1)
%    intercept:  1.30973865 +/- 0.30570984 (23.34%) (init = 0)
%[[Correlations]] (unreported correlations are < 0.100)
%    C(slope, intercept) = -0.9993
%Correlation matrix:
%[[ 0.0102288  -0.03089623]
% [-0.03089623  0.09345851]]
%Correlation between slope and intercept: -0.0309
%Correlation between intercept and slope: -0.0309
%slope = -0.44+/-0.1 intercept =  1.31+/-0.31

%SpearmanrResult(correlation=-0.3070617531067232, pvalue=4.653514925853434e-07)

%Figure \ref{fig:re} presents the size-luminosity relationship for EELGs and for the larger population of (non-extreme) CEERS galaxies. Both %EELGs and the larger galaxy population generally have compact sizes...

\begin{figure}
    \centering
    \includegraphics[scale = .28]{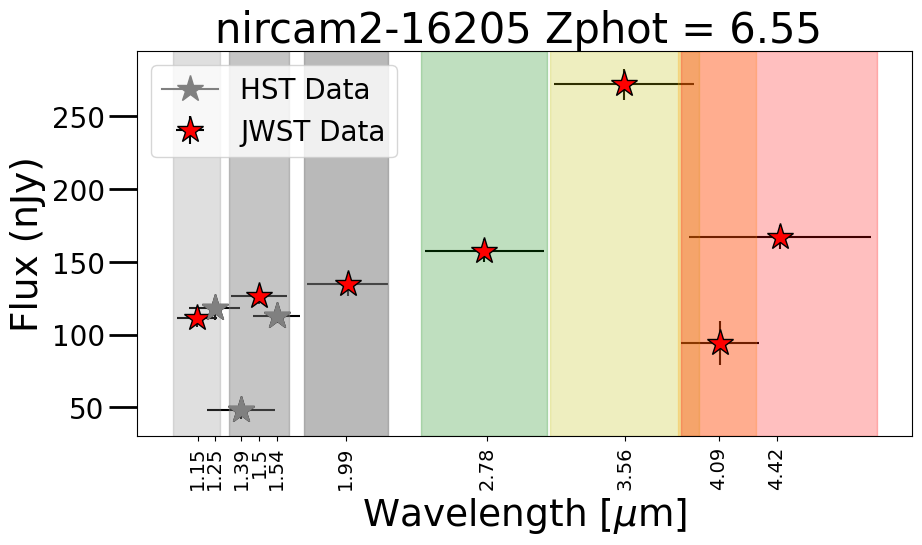}
    \includegraphics[scale = .185]{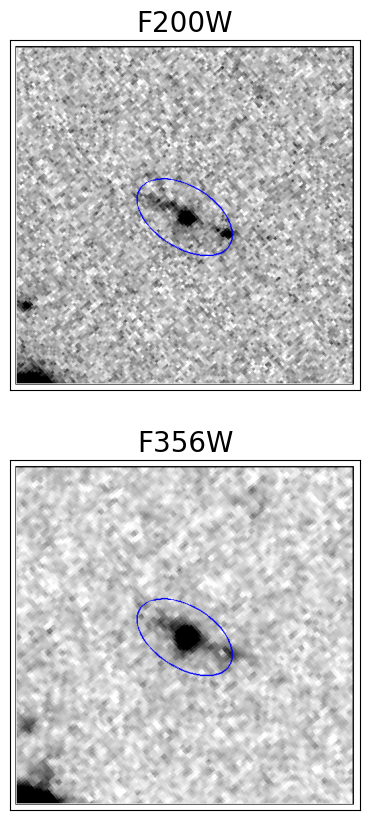}
    \caption{An example of a source that is more compact in the extreme emission filter ($\log(R_e\OIII/R_e\mathrm{(cont)} = -0.15$). The SED on the left indicates extreme \Hb\ + \OIII\ emission in F356W and continuum measured from F200W. Postage stamps show the continuum filter (top) and the extreme emission filter (bottom), with North up and East left. All postage stamps are 4  \arcsec\ in width and height. The continuum morphology includes a bright extended structure in the top image, while the emission-line morphology is more concentrated to a central feature. EELGs like this object are consistent with emission-line contribution from a central AGN.}
    \label{fig:extended_example}
\end{figure}

Figure \ref{fig:size_ew} further explores the relationship between emission-line and continuum size with rest-frame EW. We limit our comparison to resolved   ($R_e>0.03$) \texttt{GALFIT} morphologies and size ratios between $10^{0.5}$ and $10^{-0.5}$, as ratios outside this range typically represent sources that are poorly resolved or have deblending issues.
% These ratios represent our most believable \texttt{GALFIT} size measurements.
Additionally, we exclude sources in the redshift range where \Hb\ + \OIII\ is blended across the F277W and F356W such that we cannot define a reliable emission-dominated filter.
% We also exclude sources not resolved in both emission and continuum images. 
% Compactness in the continuum filter is associated with higher rest-frame EWs, as traced by the best fit line with slope $m =  -0.44 \pm  0.10$ and y-intercept $b = 1.31 +/- 0.31$ with a Spearman correlation coefficient of $-0.31$, representing a negative trend. A Spearman coefficient of -1 would represent a perfectly negative correlation and a coefficient of $-0.31$ represents a weak to moderate correlation. Considering that \texttt{GALFIT} struggles to converge solutions for EELGs in emission filters, flagging about half our sample, a larger population of EELGs may further constrain this trend. While we do not present this as a definitive method for distinguishing between AGN and starbursts, we do report a data trend and note that the EELGs occupying the upper left corner of Figure \ref{fig:size_ew} are the EELGs most likely to represent starbursts while the EELGs in the lower right corner of the figure are the most likely to represent AGN.
A best-fit line between the size ratio and EW indicates a weak but significant anticorrelation, with a slope $m = -0.44 \pm  0.10$ that is $>$4$\sigma$ inconsistent from zero and a Spearman correlation coefficient of $-0.31$ that indicates considerable scatter. This relationship is consistent with emission lines from a central AGN, that is more compact than the extended continuum associated with galaxy starlight, contributing to the most extreme (highest-EW) EELGs.
We show an example of a high-EW galaxy with an emission-line size that is more compact than its continuum size in Figure \ref{fig:extended_example}.

\subsection{ISM Conditions of EELGs}

We use the emission-line fluxes and ratios inferred from the photometry to infer the interstellar medium (ISM) conditions of the EELGs. Emission-line luminosities are measured from the continuum-subtracted photometry in the filter with extreme emission, as described in Equation 3. We further separate the \OIII$\lambda$5008 emission from the blended \Hb\ + \OIII\ by using the ratio of $\OIII\lambda5008/\OIII\lambda4960 = 3$ set by atomic physics \citep{Storey2000} and assuming $\Ha/\Hb \simeq 3$ for the typical temperature and density conditions of ISM gas \citep{Osterbrock1989} and assuming little dust attenuation. 
That is,
\begin{equation}
    \OIII\lambda5008 = \frac{3}{4} [(\Hb + \OIII) -\frac{1}{3} \Ha ]
\end{equation}
% This holds for sources with a flat continuum under the assumption that $\frac{H \alpha}{H \beta} = 3$ (Osterbrock citation). We report these in \ref{fig:o35008}

\begin{figure}[t]
    \centering
    \includegraphics[scale = 0.3]{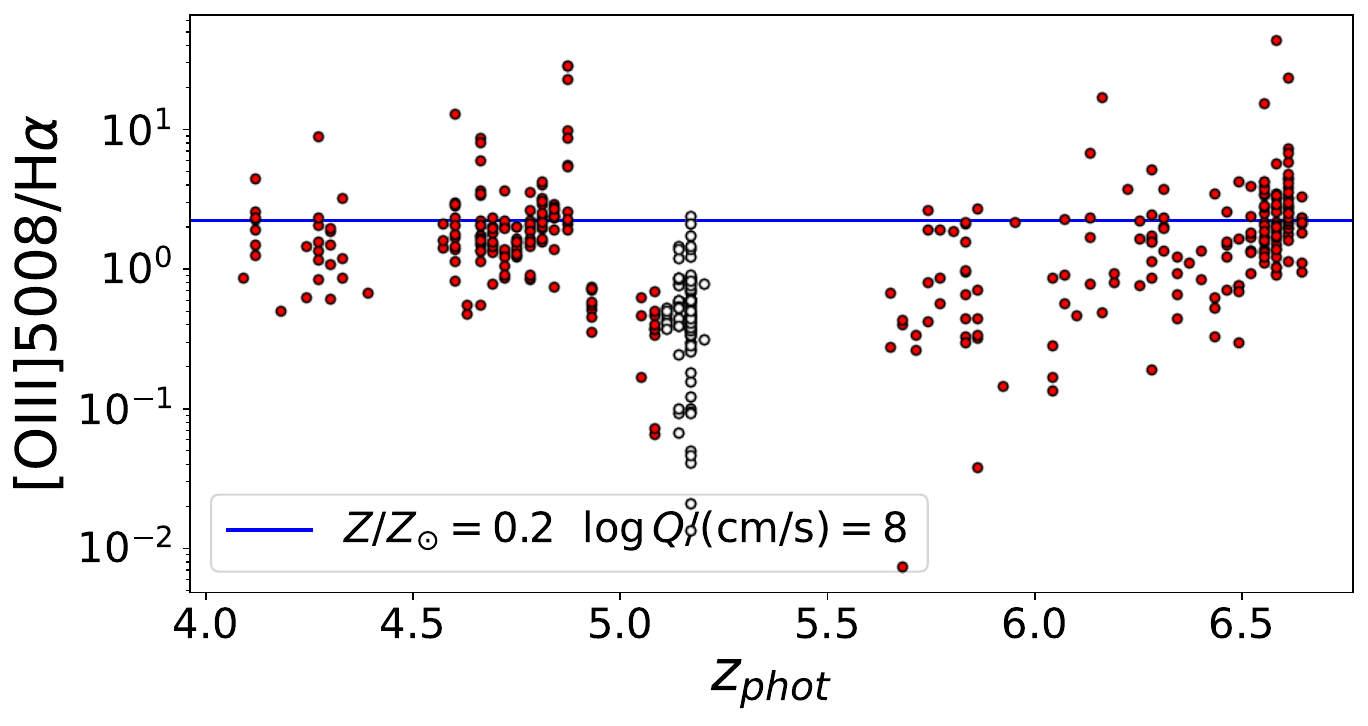}
    \includegraphics[scale = 0.3]{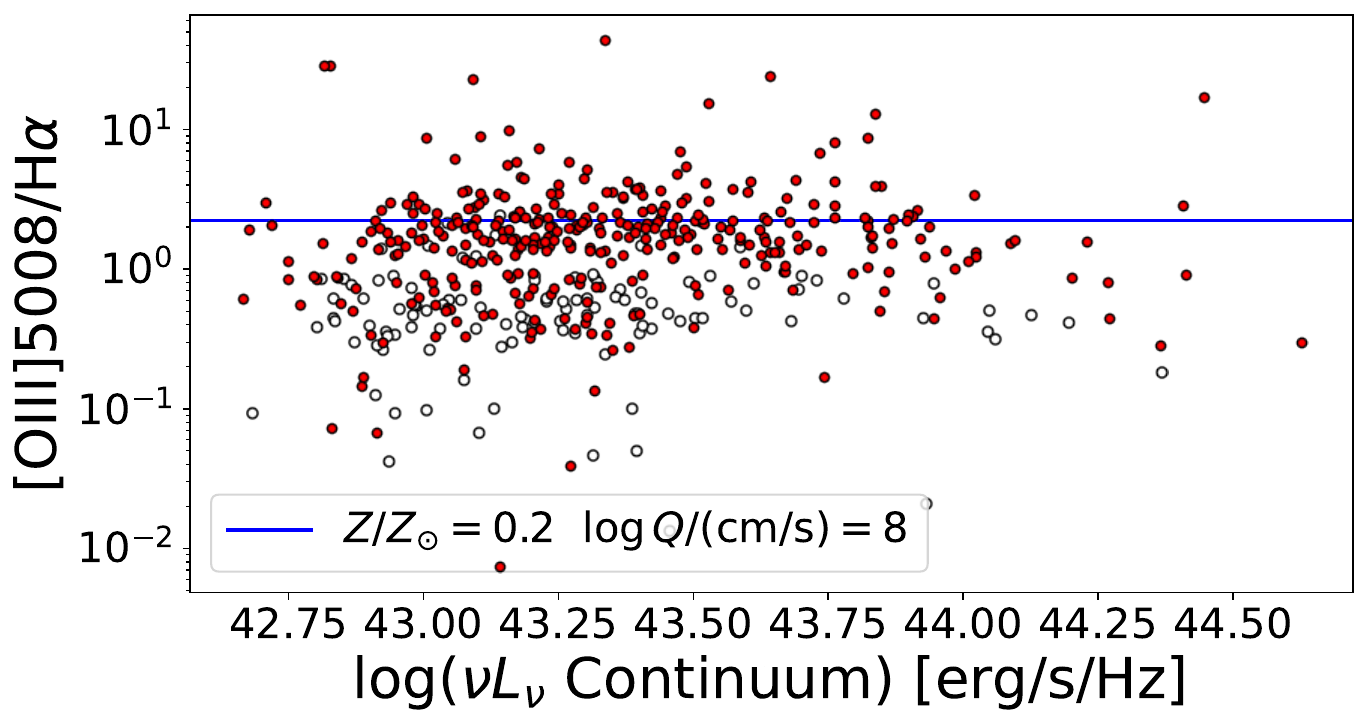}
    \caption{ \textbf{Top:} Photometrically inferred \OIII/\Ha\ ratio with photometric redshift (top) and with continuum luminosity (bottom).
    White circles indicate redshift ranges where \Hb\ + \OIII\ is blended between the F277W and F356W filters and is not reliably measured from the photometry. The inferred \OIII/\Ha\ ratios are consistent with an ISM with moderately low metallicity and high ionization, and indicated by the predicted MAPPINGS~V line ratio for $Z/Z_{\odot} = 0.2$ and $\log Q/\mathrm{(cm/s)} = 8$ (blue line).
     }
    \label{fig:o3ha_line}
\end{figure}

Figure \ref{fig:o3ha_line} presents the photometry-inferred ratio of \OIII$\lambda$5008/\Ha\ for the EELGs, limited to the redshift range where we observe both lines ($z<6.5$). Open symbols indicate the redshift range where the \Hb\ + \OIII\ complex is blended across the F277W and F356W filters and the \OIII$\lambda$5008 luminosity is frequently underestimated. The EELGs generally have $\OIII/\Ha \gtrsim 1$, indicating high ionization and moderately low metallicity \citep[e.g.][]{Kewley2019}. There are no significant relationships between the \OIII/\Ha\ ratio with redshift or continuum luminosity, consistent with predictions of relatively uniform ISM conditions as a function of these quantities in high-redshift galaxies from FLARES \citep{Wilkins2023}.

Figure \ref{fig:o3ha_line} includes a line to indicate the predicted $\OIII/\Ha  = 2.2$ ratio predicted by MAPPING~V models \citep{Sutherland2018, Kewley2019} for an ISM with $Z/Z_{\odot} = 0.2$ and $\log Q/\mathrm{(cm/s)} = 8$. The photometrically inferred line ratios of our EELGs are consistent with this MAPPINGS~V prediction.
% representing a high ionization low metallicity ISM and corresponding to $\frac{[OIII]}{H\alpha} = 2.2$, indicated by the blue line on the graph.
The implied high ionization and moderately low metallicity ISM in EELGs is similar to the inferred ISM conditions of spectroscopically observed galaxies at $z>4$ \citep[e.g.][]{Trump2023, Maiolino2023a, Sanders2023}.
% This agreement with our population demonstrates an ability to derive these ISM conditions directly from photometric \OIII\ / \Ha\ ratios which we will explore in future work. 

\section{Summary and Future Work}

In this work we describe a census of EELGs selected from the 4 reddest NIRCam photometric filters of the CEERS observations.
% We note that the F410M medium-band filter is especially useful for identifying extreme emission, both from emission lines falling in the filter and from emission lines falling in neighboring filters such that F410M samples the continuum.
We use NIRSpec to verify our selection and find excellent spectroscopic agreement with our photometrically predicted extreme emission lines, even in cases where SED fitting struggles to correctly identify photometric redshifts. The EELG population is generally consistent with young, blue stellar populations and SFRs of 1-100 stellar masses per year, reasonable for star-forming and starburst galaxies at these epochs. Images of the EELGs indicate that most have nearby neighbors, although a detailed investigation into the environments of EELGs will be the subject of future work. Distributions of rest-frame EWs, rest-frame color, size, and inferred emission line ratios are all consistent with EELGs being the extreme emission-line tail of bright emission-line galaxies at this epoch. A significant subset of these EELGs have very high EWs not predicted by simulations, and with more compact emission-line sizes that may indicate AGN contribution to the emission lines.
%and are more compact in size at the extreme emission-line than at the continuum, indicating contribution from point source AGN played a role in at least a subset of the population. 
% Our observed census of EELGs has an excellent match to the theoretical prediction from FLARES simulations except for these very extreme cases.
We present examples of EELG SEDs and show that some of these galaxies pose a problem for photometric redshift fitting and can be mistaken for ultra-high redshift galaxies, especially if extreme emission is blended across multiple filters.

Next steps for a better understanding of the EELG population will include a more detailed investigation into their derived properties and environmental relationships. EELG selection would benefit from additional photometric coverage of the F300M filter in CEERS to sample the continuum between \Hb\ + \OIII\ and \Ha\ at $3.5<z<5$ and capture \Hb\ + \OIII\ when it falls in the gap between F277W and F356W, a population that is otherwise missed by the current CEERS filters at $5<z<5.3$. The EELG sample is also an excellent target for follow-up NIRSpec spectroscopy to capture these extreme emission-lines and further identify early broad-line AGN. Understanding the role of EELGs for reionization is another important next step and will be the subject of future work.
% Our next steps will include studies of environmental relationships to EELGs, explorations into improvements for SED based modeling codes, and recovery of \Ha\ at $z>6.5$ with upcoming CEERS MIRI coverage. 

\facility{HST (ACS, WFC3)}
\facility{JWST (NIRCam, NIRSpec)}

\software{\texttt{Astropy} \citep{astropy:2013, astropy:2018, astropy:2022} , \texttt{lmfit} \citep{lmfit}, \texttt{Matplotlib} \citep{Hunter2007}, \texttt{NumPy} \citep{numpy},
 \texttt{Pandas} \citep{pandas1, pandas2},  \texttt{Scipy} \citep{scipy},  \texttt{Seaborn} \citep{seaborn}} %\citep{vanderwalt2011}

%\begin{acknowledgments}

We acknowledge the work of our colleagues in the CEERS collaboration and everyone involved in the JWST mission.
KD acknowledges support from a NSF Graduate Research Fellowship award number 2040433. JRT, RCS, and BEB acknowledge support from NASA grants JWST-ERS-01345, JWST-AR-01721, and NSF grant CAREER-1945546.

This work has made use of the Rainbow Cosmological Surveys Database, which is operated by the Centro de Astrobiología (CAB), CSIC-INTA, partnered with the University of California Observatories at Santa Cruz (UCO/Lick,UCSC).

We thank University of Connecticut graduate students Juliette K. Stecenko, Corvus Koithan, and Sagan Sutherland for their \texttt{Matplotlib} stylistic feedback.

%Add NSF GRFP grant aknowledgement

%\end{acknowledgments}

\bibliographystyle{aasjournal}
\bibliography{ref}{}

\begin{thebibliography}{}
\expandafter\ifx\csname natexlab\endcsname\relax\def\natexlab#1{#1}\fi
\providecommand{\url}[1]{\href{#1}{#1}}
\providecommand{\dodoi}[1]{doi:~\href{http://doi.org/#1}{\nolinkurl{#1}}}
\providecommand{\doeprint}[1]{\href{http://ascl.net/#1}{\nolinkurl{http://ascl.net/#1}}}
\providecommand{\doarXiv}[1]{\href{https://arxiv.org/abs/#1}{\nolinkurl{https://arxiv.org/abs/#1}}}

\bibitem[{{Amor{\'\i}n} {et~al.}(2012){Amor{\'\i}n}, {P{\'e}rez-Montero},
  {V{\'\i}lchez}, \& {Papaderos}}]{Amorin2012}
{Amor{\'\i}n}, R., {P{\'e}rez-Montero}, E., {V{\'\i}lchez}, J.~M., \&
  {Papaderos}, P. 2012, \apj, 749, 185, \dodoi{10.1088/0004-637X/749/2/185}

\bibitem[{{Amor{\'\i}n} {et~al.}(2014){Amor{\'\i}n}, {Sommariva}, {Castellano},
  {Grazian}, {Tasca}, {Fontana}, {Pentericci}, {Cassata}, {Garilli}, {Le Brun},
  {Le F{\`e}vre}, {Maccagni}, {Thomas}, {Vanzella}, {Zamorani}, {Zucca},
  {Bardelli}, {Capak}, {Cassar{\'a}}, {Cimatti}, {Cuby}, {Cucciati}, {de la
  Torre}, {Durkalec}, {Giavalisco}, {Hathi}, {Ilbert}, {Lemaux}, {Moreau},
  {Paltani}, {Ribeiro}, {Salvato}, {Schaerer}, {Scodeggio}, {Talia},
  {Taniguchi}, {Tresse}, {Vergani}, {Wang}, {Charlot}, {Contini}, {Fotopoulou},
  {L{\'o}pez-Sanjuan}, {Mellier}, \& {Scoville}}]{Amorin2014}
{Amor{\'\i}n}, R., {Sommariva}, V., {Castellano}, M., {et~al.} 2014, \aap, 568,
  L8, \dodoi{10.1051/0004-6361/201423816}

\bibitem[{{Amor{\'\i}n} {et~al.}(2015){Amor{\'\i}n}, {P{\'e}rez-Montero},
  {Contini}, {V{\'\i}lchez}, {Bolzonella}, {Tasca}, {Lamareille}, {Zamorani},
  {Maier}, {Carollo}, {Kneib}, {Le F{\`e}vre}, {Lilly}, {Mainieri}, {Renzini},
  {Scodeggio}, {Bardelli}, {Bongiorno}, {Caputi}, {Cucciati}, {de la Torre},
  {de Ravel}, {Franzetti}, {Garilli}, {Iovino}, {Kampczyk}, {Knobel},
  {Kova{\v{c}}}, {Le Borgne}, {Le Brun}, {Mignoli}, {Pell{\`o}}, {Peng},
  {Presotto}, {Ricciardelli}, {Silverman}, {Tanaka}, {Tresse}, {Vergani}, \&
  {Zucca}}]{Amorin2015}
{Amor{\'\i}n}, R., {P{\'e}rez-Montero}, E., {Contini}, T., {et~al.} 2015, \aap,
  578, A105, \dodoi{10.1051/0004-6361/201322786}

\bibitem[{{Amor{\'\i}n} {et~al.}(2010){Amor{\'\i}n}, {P{\'e}rez-Montero}, \&
  {V{\'\i}lchez}}]{Amorin2010}
{Amor{\'\i}n}, R.~O., {P{\'e}rez-Montero}, E., \& {V{\'\i}lchez}, J.~M. 2010,
  \apjl, 715, L128, \dodoi{10.1088/2041-8205/715/2/L128}

\bibitem[{{Arrabal Haro} {et~al.}(2023{\natexlab{a}}){Arrabal Haro},
  {Dickinson}, {Finkelstein}, {Fujimoto}, {Fern{\'a}ndez}, {Kartaltepe},
  {Jung}, {Cole}, {Burgarella}, {Chworowsky}, {Hutchison}, {Morales},
  {Papovich}, {Simons}, {Amor{\'\i}n}, {Backhaus}, {Bagley}, {Bisigello},
  {Calabr{\`o}}, {Castellano}, {Cleri}, {Dav{\'e}}, {Dekel}, {Ferguson},
  {Fontana}, {Gawiser}, {Giavalisco}, {Harish}, {Hathi}, {Hirschmann},
  {Holwerda}, {Huertas-Company}, {Koekemoer}, {Larson}, {Lucas}, {Mobasher},
  {P{\'e}rez-Gonz{\'a}lez}, {Pirzkal}, {Rose}, {Santini}, {Trump}, {de la
  Vega}, {Wang}, {Weiner}, {Wilkins}, {Yang}, {Yung}, \& {Zavala}}]{Haro2023b}
{Arrabal Haro}, P., {Dickinson}, M., {Finkelstein}, S.~L., {et~al.}
  2023{\natexlab{a}}, \apjl, 951, L22, \dodoi{10.3847/2041-8213/acdd54}

\bibitem[{{Arrabal Haro} {et~al.}(2023{\natexlab{b}}){Arrabal Haro},
  {Dickinson}, {Finkelstein}, {Kartaltepe}, {Donnan}, {Burgarella}, {Carnall},
  {Cullen}, {Dunlop}, {Fern{\'a}ndez}, {Fujimoto}, {Jung}, {Krips}, {Larson},
  {Papovich}, {P{\'e}rez-Gonz{\'a}lez}, {Amor{\'\i}n}, {Bagley}, {Buat},
  {Casey}, {Chworowsky}, {Cohen}, {Ferguson}, {Giavalisco}, {Huertas-Company},
  {Hutchison}, {Kocevski}, {Koekemoer}, {Lucas}, {McLeod}, {McLure}, {Pirzkal},
  {Trump}, {Weiner}, {Wilkins}, \& {Zavala}}]{Haro2023a}
---. 2023{\natexlab{b}}, arXiv e-prints, arXiv:2303.15431,
  \dodoi{10.48550/arXiv.2303.15431}

\bibitem[{{Astropy Collaboration} {et~al.}(2013){Astropy Collaboration},
  {Robitaille}, {Tollerud}, {Greenfield}, {Droettboom}, {Bray}, {Aldcroft},
  {Davis}, {Ginsburg}, {Price-Whelan}, {Kerzendorf}, {Conley}, {Crighton},
  {Barbary}, {Muna}, {Ferguson}, {Grollier}, {Parikh}, {Nair}, {Unther},
  {Deil}, {Woillez}, {Conseil}, {Kramer}, {Turner}, {Singer}, {Fox}, {Weaver},
  {Zabalza}, {Edwards}, {Azalee Bostroem}, {Burke}, {Casey}, {Crawford},
  {Dencheva}, {Ely}, {Jenness}, {Labrie}, {Lim}, {Pierfederici}, {Pontzen},
  {Ptak}, {Refsdal}, {Servillat}, \& {Streicher}}]{astropy:2013}
{Astropy Collaboration}, {Robitaille}, T.~P., {Tollerud}, E.~J., {et~al.} 2013,
  \aap, 558, A33, \dodoi{10.1051/0004-6361/201322068}

\bibitem[{{Astropy Collaboration} {et~al.}(2018){Astropy Collaboration},
  {Price-Whelan}, {Sip{\H{o}}cz}, {G{\"u}nther}, {Lim}, {Crawford}, {Conseil},
  {Shupe}, {Craig}, {Dencheva}, {Ginsburg}, {VanderPlas}, {Bradley},
  {P{\'e}rez-Su{\'a}rez}, {de Val-Borro}, {Aldcroft}, {Cruz}, {Robitaille},
  {Tollerud}, {Ardelean}, {Babej}, {Bach}, {Bachetti}, {Bakanov}, {Bamford},
  {Barentsen}, {Barmby}, {Baumbach}, {Berry}, {Biscani}, {Boquien}, {Bostroem},
  {Bouma}, {Brammer}, {Bray}, {Breytenbach}, {Buddelmeijer}, {Burke},
  {Calderone}, {Cano Rodr{\'\i}guez}, {Cara}, {Cardoso}, {Cheedella}, {Copin},
  {Corrales}, {Crichton}, {D'Avella}, {Deil}, {Depagne}, {Dietrich}, {Donath},
  {Droettboom}, {Earl}, {Erben}, {Fabbro}, {Ferreira}, {Finethy}, {Fox},
  {Garrison}, {Gibbons}, {Goldstein}, {Gommers}, {Greco}, {Greenfield},
  {Groener}, {Grollier}, {Hagen}, {Hirst}, {Homeier}, {Horton}, {Hosseinzadeh},
  {Hu}, {Hunkeler}, {Ivezi{\'c}}, {Jain}, {Jenness}, {Kanarek}, {Kendrew},
  {Kern}, {Kerzendorf}, {Khvalko}, {King}, {Kirkby}, {Kulkarni}, {Kumar},
  {Lee}, {Lenz}, {Littlefair}, {Ma}, {Macleod}, {Mastropietro}, {McCully},
  {Montagnac}, {Morris}, {Mueller}, {Mumford}, {Muna}, {Murphy}, {Nelson},
  {Nguyen}, {Ninan}, {N{\"o}the}, {Ogaz}, {Oh}, {Parejko}, {Parley}, {Pascual},
  {Patil}, {Patil}, {Plunkett}, {Prochaska}, {Rastogi}, {Reddy Janga},
  {Sabater}, {Sakurikar}, {Seifert}, {Sherbert}, {Sherwood-Taylor}, {Shih},
  {Sick}, {Silbiger}, {Singanamalla}, {Singer}, {Sladen}, {Sooley},
  {Sornarajah}, {Streicher}, {Teuben}, {Thomas}, {Tremblay}, {Turner},
  {Terr{\'o}n}, {van Kerkwijk}, {de la Vega}, {Watkins}, {Weaver}, {Whitmore},
  {Woillez}, {Zabalza}, \& {Astropy Contributors}}]{astropy:2018}
{Astropy Collaboration}, {Price-Whelan}, A.~M., {Sip{\H{o}}cz}, B.~M., {et~al.}
  2018, \aj, 156, 123, \dodoi{10.3847/1538-3881/aabc4f}

\bibitem[{{Astropy Collaboration} {et~al.}(2022){Astropy Collaboration},
  {Price-Whelan}, {Lim}, {Earl}, {Starkman}, {Bradley}, {Shupe}, {Patil},
  {Corrales}, {Brasseur}, {N{\"o}the}, {Donath}, {Tollerud}, {Morris},
  {Ginsburg}, {Vaher}, {Weaver}, {Tocknell}, {Jamieson}, {van Kerkwijk},
  {Robitaille}, {Merry}, {Bachetti}, {G{\"u}nther}, {Aldcroft},
  {Alvarado-Montes}, {Archibald}, {B{\'o}di}, {Bapat}, {Barentsen},
  {Baz{\'a}n}, {Biswas}, {Boquien}, {Burke}, {Cara}, {Cara}, {Conroy},
  {Conseil}, {Craig}, {Cross}, {Cruz}, {D'Eugenio}, {Dencheva}, {Devillepoix},
  {Dietrich}, {Eigenbrot}, {Erben}, {Ferreira}, {Foreman-Mackey}, {Fox},
  {Freij}, {Garg}, {Geda}, {Glattly}, {Gondhalekar}, {Gordon}, {Grant},
  {Greenfield}, {Groener}, {Guest}, {Gurovich}, {Handberg}, {Hart},
  {Hatfield-Dodds}, {Homeier}, {Hosseinzadeh}, {Jenness}, {Jones}, {Joseph},
  {Kalmbach}, {Karamehmetoglu}, {Ka{\l}uszy{\'n}ski}, {Kelley}, {Kern},
  {Kerzendorf}, {Koch}, {Kulumani}, {Lee}, {Ly}, {Ma}, {MacBride}, {Maljaars},
  {Muna}, {Murphy}, {Norman}, {O'Steen}, {Oman}, {Pacifici}, {Pascual},
  {Pascual-Granado}, {Patil}, {Perren}, {Pickering}, {Rastogi}, {Roulston},
  {Ryan}, {Rykoff}, {Sabater}, {Sakurikar}, {Salgado}, {Sanghi}, {Saunders},
  {Savchenko}, {Schwardt}, {Seifert-Eckert}, {Shih}, {Jain}, {Shukla}, {Sick},
  {Simpson}, {Singanamalla}, {Singer}, {Singhal}, {Sinha}, {Sip{\H{o}}cz},
  {Spitler}, {Stansby}, {Streicher}, {{\v{S}}umak}, {Swinbank}, {Taranu},
  {Tewary}, {Tremblay}, {de Val-Borro}, {Van Kooten}, {Vasovi{\'c}}, {Verma},
  {de Miranda Cardoso}, {Williams}, {Wilson}, {Winkel}, {Wood-Vasey}, {Xue},
  {Yoachim}, {Zhang}, {Zonca}, \& {Astropy Project
  Contributors}}]{astropy:2022}
{Astropy Collaboration}, {Price-Whelan}, A.~M., {Lim}, P.~L., {et~al.} 2022,
  \apj, 935, 167, \dodoi{10.3847/1538-4357/ac7c74}

\bibitem[{{Backhaus} {et~al.}(2023){Backhaus}, {Trump}, {Pirzkal}, {Barro},
  {Finkelstein}, {Arrabal Haro}, {Simons}, {Wessner}, {Cleri}, {Hirschmann},
  {Bagley}, {Nicholls}, {Dickinson}, {Kartaltepe}, {Papovich}, {Kocevski},
  {Koekemoer}, {Bisigello}, {Jaskot}, {Lucas}, {Jung}, {Wilkins}, {Yung},
  {Ferguson}, {Fontana}, {Grazian}, {Grogin}, {Kewley}, {Kirkpatrick}, {Lotz},
  {Pentericci}, {Perez-Gonzalez}, {Ravindranath}, {Somerville}, {Yang},
  {Holwerda}, {Kurczynski}, {Hathi}, {Rose}, \& {Davis}}]{Backhaus2023}
{Backhaus}, B.~E., {Trump}, J.~R., {Pirzkal}, N., {et~al.} 2023, arXiv
  e-prints, arXiv:2307.09503, \dodoi{10.48550/arXiv.2307.09503}

\bibitem[{{Bagley} {et~al.}(2023){Bagley}, {Finkelstein}, {Koekemoer},
  {Ferguson}, {Arrabal Haro}, {Dickinson}, {Kartaltepe}, {Papovich},
  {P{\'e}rez-Gonz{\'a}lez}, {Pirzkal}, {Somerville}, {Willmer}, {Yang}, {Yung},
  {Fontana}, {Grazian}, {Grogin}, {Hirschmann}, {Kewley}, {Kirkpatrick},
  {Kocevski}, {Lotz}, {Medrano}, {Morales}, {Pentericci}, {Ravindranath},
  {Trump}, {Wilkins}, {Calabr{\`o}}, {Cooper}, {Costantin}, {de la Vega},
  {Hilbert}, {Hutchison}, {Larson}, {Lucas}, {McGrath}, {Ryan}, {Wang}, \&
  {Wuyts}}]{Bagley2023}
{Bagley}, M.~B., {Finkelstein}, S.~L., {Koekemoer}, A.~M., {et~al.} 2023,
  \apjl, 946, L12, \dodoi{10.3847/2041-8213/acbb08}

\bibitem[{{Bakx} {et~al.}(2023){Bakx}, {Zavala}, {Mitsuhashi}, {Treu},
  {Fontana}, {Tadaki}, {Casey}, {Castellano}, {Glazebrook}, {Hagimoto},
  {Ikeda}, {Jones}, {Leethochawalit}, {Mason}, {Morishita}, {Nanayakkara},
  {Pentericci}, {Roberts-Borsani}, {Santini}, {Serjeant}, {Tamura}, {Trenti},
  \& {Vanzella}}]{Bakx2023}
{Bakx}, T. J.~L.~C., {Zavala}, J.~A., {Mitsuhashi}, I., {et~al.} 2023, \mnras,
  519, 5076, \dodoi{10.1093/mnras/stac3723}

\bibitem[{{Barro} {et~al.}(2023){Barro}, {Perez-Gonzalez}, {Kocevski},
  {McGrath}, {Trump}, {Simons}, {Somerville}, {Yung}, {Arrabal Haro}, {Bagley},
  {Cleri}, {Costantin}, {Davis}, {Dickinson}, {Finkelstein}, {Giavalisco},
  {Gomez-Guijarro}, {Hathi}, {Hirschmann}, {Akins}, {Holwerda},
  {Huertas-Company}, {Lucas}, {Papovich}, {Seille}, {Tacchella}, {Wilkins}, {de
  la Vega}, {Yang}, \& {Zavala}}]{Barro2023}
{Barro}, G., {Perez-Gonzalez}, P.~G., {Kocevski}, D.~D., {et~al.} 2023, arXiv
  e-prints, arXiv:2305.14418, \dodoi{10.48550/arXiv.2305.14418}

\bibitem[{{Brammer} {et~al.}(2008){Brammer}, {van Dokkum}, \&
  {Coppi}}]{Brammer2008}
{Brammer}, G.~B., {van Dokkum}, P.~G., \& {Coppi}, P. 2008, \apj, 686, 1503,
  \dodoi{10.1086/591786}

\bibitem[{{Brammer} {et~al.}(2010){Brammer}, {van Dokkum}, \&
  {Coppi}}]{Brammer2010}
---. 2010, {EAZY: A Fast, Public Photometric Redshift Code}, Astrophysics
  Source Code Library, record ascl:1010.052.
\newblock \doeprint{1010.052}

\bibitem[{{Brinchmann}(2023)}]{Brinchmann2023}
{Brinchmann}, J. 2023, \mnras, 525, 2087, \dodoi{10.1093/mnras/stad1704}

\bibitem[{{Bunker} {et~al.}(2023){Bunker}, {Saxena}, {Cameron}, {Willott},
  {Curtis-Lake}, {Jakobsen}, {Carniani}, {Smit}, {Maiolino}, {Witstok},
  {Curti}, {D'Eugenio}, {Jones}, {Ferruit}, {Arribas}, {Charlot}, {Chevallard},
  {Giardino}, {de Graaff}, {Looser}, {Luetzgendorf}, {Maseda}, {Rawle}, {Rix},
  {Rodriguez Del Pino}, {Alberts}, {Egami}, {Eisenstein}, {Endsley},
  {Hainline}, {Hausen}, {Johnson}, {Rieke}, {Rieke}, {Robertson}, {Shivaei},
  {Stark}, {Sun}, {Tacchella}, {Tang}, {Williams}, {Willmer}, {Baker}, {Baum},
  {Bhatawdekar}, {Bowler}, {Boyett}, {Chen}, {Circosta}, {Helton}, {Ji}, {Lyu},
  {Nelson}, {Parlanti}, {Perna}, {Sandles}, {Scholtz}, {Suess}, {Topping},
  {Uebler}, {Wallace}, \& {Whitler}}]{Bunker2023}
{Bunker}, A.~J., {Saxena}, A., {Cameron}, A.~J., {et~al.} 2023, arXiv e-prints,
  arXiv:2302.07256, \dodoi{10.48550/arXiv.2302.07256}

\bibitem[{{Bushouse} {et~al.}(2022){Bushouse}, {Eisenhamer}, {Dencheva},
  {Davies}, {Greenfield}, {Morrison}, {Hodge}, {Simon}, {Grumm}, {Droettboom},
  {Slavich}, {Sosey}, {Pauly}, {Miller}, {Jedrzejewski}, {Hack}, {Davis},
  {Crawford}, {Law}, {Gordon}, {Regan}, {Cara}, {MacDonald}, {Bradley},
  {Shanahan}, {Jamieson}, {Teodoro}, \& {Williams}}]{Bushouse2022}
{Bushouse}, H., {Eisenhamer}, J., {Dencheva}, N., {et~al.} 2022, {JWST
  Calibration Pipeline}, 1.8.2, Zenodo,  Zenodo, \dodoi{10.5281/zenodo.7325378}

\bibitem[{{Calabr{\`o}} {et~al.}(2017){Calabr{\`o}}, {Amor{\'\i}n}, {Fontana},
  {P{\'e}rez-Montero}, {Lemaux}, {Ribeiro}, {Bardelli}, {Castellano},
  {Contini}, {De Barros}, {Garilli}, {Grazian}, {Guaita}, {Hathi}, {Koekemoer},
  {Le F{\`e}vre}, {Maccagni}, {Pentericci}, {Schaerer}, {Talia}, {Tasca}, \&
  {Zucca}}]{Calabro2017}
{Calabr{\`o}}, A., {Amor{\'\i}n}, R., {Fontana}, A., {et~al.} 2017, \aap, 601,
  A95, \dodoi{10.1051/0004-6361/201629762}

\bibitem[{{Calzetti} {et~al.}(2000){Calzetti}, {Armus}, {Bohlin}, {Kinney},
  {Koornneef}, \& {Storchi-Bergmann}}]{Calzetti2000}
{Calzetti}, D., {Armus}, L., {Bohlin}, R.~C., {et~al.} 2000, \apj, 533, 682,
  \dodoi{10.1086/308692}

\bibitem[{{Caputi} {et~al.}(2017){Caputi}, {Deshmukh}, {Ashby}, {Cowley},
  {Bisigello}, {Fazio}, {Fynbo}, {Le F{\`e}vre}, {Milvang-Jensen}, \&
  {Ilbert}}]{Caputi2017}
{Caputi}, K.~I., {Deshmukh}, S., {Ashby}, M.~L.~N., {et~al.} 2017, \apj, 849,
  45, \dodoi{10.3847/1538-4357/aa901e}

\bibitem[{{Cardamone} {et~al.}(2009){Cardamone}, {Schawinski}, {Sarzi},
  {Bamford}, {Bennert}, {Urry}, {Lintott}, {Keel}, {Parejko}, {Nichol},
  {Thomas}, {Andreescu}, {Murray}, {Raddick}, {Slosar}, {Szalay}, \&
  {Vandenberg}}]{Cardamone2009}
{Cardamone}, C., {Schawinski}, K., {Sarzi}, M., {et~al.} 2009, \mnras, 399,
  1191, \dodoi{10.1111/j.1365-2966.2009.15383.x}

\bibitem[{{Chary} {et~al.}(2005){Chary}, {Stern}, \& {Eisenhardt}}]{Chary2005}
{Chary}, R.-R., {Stern}, D., \& {Eisenhardt}, P. 2005, \apjl, 635, L5,
  \dodoi{10.1086/499205}

\bibitem[{{Cleri} {et~al.}(2023){Cleri}, {Olivier}, {Hutchison}, {Papovich},
  {Trump}, {Amor{\'\i}n}, {Backhaus}, {Berg}, {Fern{\'a}ndez}, {Finkelstein},
  {Fujimoto}, {Hirschmann}, {Kartaltepe}, {Kocevski}, {Simons}, {Wilkins}, \&
  {Yung}}]{Cleri2023}
{Cleri}, N.~J., {Olivier}, G.~M., {Hutchison}, T.~A., {et~al.} 2023, \apj, 953,
  10, \dodoi{10.3847/1538-4357/acde55}

\bibitem[{{Conroy} \& {Gunn}(2010)}]{Conroy2010}
{Conroy}, C., \& {Gunn}, J.~E. 2010, {FSPS: Flexible Stellar Population
  Synthesis}.
\newblock \doeprint{1010.043}

\bibitem[{{De Barros} {et~al.}(2019){De Barros}, {Oesch}, {Labb{\'e}},
  {Stefanon}, {Gonz{\'a}lez}, {Smit}, {Bouwens}, \&
  {Illingworth}}]{deBarros2019}
{De Barros}, S., {Oesch}, P.~A., {Labb{\'e}}, I., {et~al.} 2019, \mnras, 489,
  2355, \dodoi{10.1093/mnras/stz940}

\bibitem[{{Donnan} {et~al.}(2023){Donnan}, {McLeod}, {Dunlop}, {McLure},
  {Carnall}, {Begley}, {Cullen}, {Hamadouche}, {Bowler}, {Magee}, {McCracken},
  {Milvang-Jensen}, {Moneti}, \& {Targett}}]{Donnan2023}
{Donnan}, C.~T., {McLeod}, D.~J., {Dunlop}, J.~S., {et~al.} 2023, \mnras, 518,
  6011, \dodoi{10.1093/mnras/stac3472}

\bibitem[{{Endsley} {et~al.}(2023){Endsley}, {Stark}, {Whitler}, {Topping},
  {Johnson}, {Robertson}, {Tacchella}, {Alberts}, {Baker}, {Bhatawdekar},
  {Boyett}, {Bunker}, {Cameron}, {Carniani}, {Charlot}, {Chen}, {Chevallard},
  {Curtis-Lake}, {Danhaive}, {Egami}, {Eisenstein}, {Hainline}, {Helton}, {Ji},
  {Looser}, {Maiolino}, {Nelson}, {Pusk{\'a}s}, {Rieke}, {Rieke}, {Rix},
  {Sandles}, {Saxena}, {Simmonds}, {Smit}, {Sun}, {Williams}, {Willmer},
  {Willott}, \& {Witstok}}]{Endsley2023}
{Endsley}, R., {Stark}, D.~P., {Whitler}, L., {et~al.} 2023, arXiv e-prints,
  arXiv:2306.05295, \dodoi{10.48550/arXiv.2306.05295}

\bibitem[{{Fan} {et~al.}(2002){Fan}, {Narayanan}, {Strauss}, {White}, {Becker},
  {Pentericci}, \& {Rix}}]{Fan2002}
{Fan}, X., {Narayanan}, V.~K., {Strauss}, M.~A., {et~al.} 2002, \aj, 123, 1247,
  \dodoi{10.1086/339030}

\bibitem[{{Ferruit} {et~al.}(2022){Ferruit}, {Jakobsen}, {Giardino}, {Rawle},
  {Alves de Oliveira}, {Arribas}, {Beck}, {Birkmann}, {B{\"o}ker}, {Bunker},
  {Charlot}, {de Marchi}, {Franx}, {Henry}, {Karakla}, {Kassin}, {Kumari},
  {L{\'o}pez-Caniego}, {L{\"u}tzgendorf}, {Maiolino}, {Manjavacas}, {Marston},
  {Moseley}, {Muzerolle}, {Pirzkal}, {Rauscher}, {Rix}, {Sabbi}, {Sirianni},
  {te Plate}, {Valenti}, {Willott}, \& {Zeidler}}]{Ferruit2022}
{Ferruit}, P., {Jakobsen}, P., {Giardino}, G., {et~al.} 2022, \aap, 661, A81,
  \dodoi{10.1051/0004-6361/202142673}

\bibitem[{{Finkelstein} {et~al.}(2022){Finkelstein}, {Bagley}, {Arrabal Haro},
  {Dickinson}, {Ferguson}, {Kartaltepe}, {Papovich}, {Burgarella}, {Kocevski},
  {Huertas-Company}, {Iyer}, {Koekemoer}, {Larson}, {P{\'e}rez-Gonz{\'a}lez},
  {Rose}, {Tacchella}, {Wilkins}, {Chworowsky}, {Medrano}, {Morales},
  {Somerville}, {Yung}, {Fontana}, {Giavalisco}, {Grazian}, {Grogin}, {Kewley},
  {Kirkpatrick}, {Kurczynski}, {Lotz}, {Pentericci}, {Pirzkal}, {Ravindranath},
  {Ryan}, {Trump}, {Yang}, {Almaini}, {Amor{\'\i}n}, {Annunziatella},
  {Backhaus}, {Barro}, {Behroozi}, {Bell}, {Bhatawdekar}, {Bisigello}, {Bromm},
  {Buat}, {Buitrago}, {Calabr{\`o}}, {Casey}, {Castellano}, {Ch{\'a}vez Ortiz},
  {Ciesla}, {Cleri}, {Cohen}, {Cole}, {Cooke}, {Cooper}, {Cooray}, {Costantin},
  {Cox}, {Croton}, {Daddi}, {Dav{\'e}}, {de La Vega}, {Dekel}, {Elbaz},
  {Estrada-Carpenter}, {Faber}, {Fern{\'a}ndez}, {Finkelstein}, {Freundlich},
  {Fujimoto}, {Garc{\'\i}a-Argum{\'a}nez}, {Gardner}, {Gawiser},
  {G{\'o}mez-Guijarro}, {Guo}, {Hamblin}, {Hamilton}, {Hathi}, {Holwerda},
  {Hirschmann}, {Hutchison}, {Jaskot}, {Jha}, {Jogee}, {Juneau}, {Jung},
  {Kassin}, {Le Bail}, {Leung}, {Lucas}, {Magnelli}, {Mantha}, {Matharu},
  {McGrath}, {McIntosh}, {Merlin}, {Mobasher}, {Newman}, {Nicholls}, {Pandya},
  {Rafelski}, {Ronayne}, {Santini}, {Seill{\'e}}, {Shah}, {Shen}, {Simons},
  {Snyder}, {Stanway}, {Straughn}, {Teplitz}, {Vanderhoof}, {Vega-Ferrero},
  {Wang}, {Weiner}, {Willmer}, {Wuyts}, {Zavala}, \& {CEERS
  Team}}]{Finkelstein2022}
{Finkelstein}, S.~L., {Bagley}, M.~B., {Arrabal Haro}, P., {et~al.} 2022,
  \apjl, 940, L55, \dodoi{10.3847/2041-8213/ac966e}

\bibitem[{{Finkelstein} {et~al.}(2023){Finkelstein}, {Bagley}, {Ferguson},
  {Wilkins}, {Kartaltepe}, {Papovich}, {Yung}, {Haro}, {Behroozi}, {Dickinson},
  {Kocevski}, {Koekemoer}, {Larson}, {Le Bail}, {Morales},
  {P{\'e}rez-Gonz{\'a}lez}, {Burgarella}, {Dav{\'e}}, {Hirschmann},
  {Somerville}, {Wuyts}, {Bromm}, {Casey}, {Fontana}, {Fujimoto}, {Gardner},
  {Giavalisco}, {Grazian}, {Grogin}, {Hathi}, {Hutchison}, {Jha}, {Jogee},
  {Kewley}, {Kirkpatrick}, {Long}, {Lotz}, {Pentericci}, {Pierel}, {Pirzkal},
  {Ravindranath}, {Ryan}, {Trump}, {Yang}, {Bhatawdekar}, {Bisigello}, {Buat},
  {Calabr{\`o}}, {Castellano}, {Cleri}, {Cooper}, {Croton}, {Daddi}, {Dekel},
  {Elbaz}, {Franco}, {Gawiser}, {Holwerda}, {Huertas-Company}, {Jaskot},
  {Leung}, {Lucas}, {Mobasher}, {Pandya}, {Tacchella}, {Weiner}, \&
  {Zavala}}]{Finkelstein2023}
{Finkelstein}, S.~L., {Bagley}, M.~B., {Ferguson}, H.~C., {et~al.} 2023, \apjl,
  946, L13, \dodoi{10.3847/2041-8213/acade4}

\bibitem[{{Fujimoto} {et~al.}(2023){Fujimoto}, {Arrabal Haro}, {Dickinson},
  {Finkelstein}, {Kartaltepe}, {Larson}, {Burgarella}, {Bagley}, {Behroozi},
  {Chworowsky}, {Hirschmann}, {Trump}, {Wilkins}, {Yung}, {Koekemoer},
  {Papovich}, {Pirzkal}, {Ferguson}, {Fontana}, {Grogin}, {Grazian}, {Kewley},
  {Kocevski}, {Lotz}, {Pentericci}, {Ravindranath}, {Somerville}, {Wilkins},
  {Amor{\'\i}n}, {Backhaus}, {Calabr{\`o}}, {Casey}, {Cooper}, {Fern{\'a}ndez},
  {Franco}, {Giavalisco}, {Hathi}, {Harish}, {Hutchison}, {Iyer}, {Jung},
  {Lucas}, \& {Zavala}}]{Fujimoto2023}
{Fujimoto}, S., {Arrabal Haro}, P., {Dickinson}, M., {et~al.} 2023, \apjl, 949,
  L25, \dodoi{10.3847/2041-8213/acd2d9}

\bibitem[{{Gardner} {et~al.}(2023){Gardner}, {Mather}, {Abbott}, {Abell},
  {Abernathy}, {Abney}, {Abraham}, {Abraham}, {Abul-Huda}, {Acton}, {Adams},
  {Adams}, {Adler}, {Adriaensen}, {Aguilar}, {Ahmed}, {Ahmed}, {Ahmed},
  {Albat}, {Albert}, {Alberts}, {Aldridge}, {Allen}, {Allen}, {Altenburg},
  {Altunc}, {Alvarez}, {{\'A}lvarez-M{\'a}rquez}, {Alves de Oliveira},
  {Ambrose}, {Anandakrishnan}, {Andersen}, {Anderson}, {Anderson}, {Anderson},
  {Anderson}, {Aprea}, {Archer}, {Arenberg}, {Argyriou}, {Arribas}, {Artigau},
  {Arvai}, {Atcheson}, {Atkinson}, {Averbukh}, {Aymergen}, {Bacinski},
  {Baggett}, {Bagnasco}, {Baker}, {Balzano}, {Banks}, {Baran}, {Barker},
  {Barrett}, {Barringer}, {Barto}, {Bast}, {Baudoz}, {Baum}, {Beatty},
  {Beaulieu}, {Bechtold}, {Beck}, {Beddard}, {Beichman}, {Bellagama}, {Bely},
  {Berger}, {Bergeron}, {Bernier}, {Bertch}, {Beskow}, {Betz}, {Biagetti},
  {Birkmann}, {Bjorklund}, {Blackwood}, {Blazek}, {Blossfeld}, {Bluth},
  {Boccaletti}, {Boegner}, {Bohlin}, {Boia}, {B{\"o}ker}, {Bonaventura},
  {Bond}, {Bosley}, {Boucarut}, {Bouchet}, {Bouwman}, {Bower}, {Bowers},
  {Bowers}, {Boyce}, {Boyer}, {Boyer}, {Boyer}, {Boyer}, {Bradley}, {Brady},
  {Brandl}, {Brannen}, {Breda}, {Bremmer}, {Brennan}, {Bresnahan}, {Bright},
  {Broiles}, {Bromenschenkel}, {Brooks}, {Brooks}, {Brown}, {Brown}, {Brown},
  {Bruce}, {Bryson}, {Bujanda}, {Bullock}, {Bunker}, {Bureo}, {Burt}, {Bush},
  {Bushouse}, {Bussman}, {Cabaud}, {Cale}, {Calhoon}, {Calvani}, {Canipe},
  {Caputo}, {Cara}, {Carey}, {Case}, {Cesari}, {Cetorelli}, {Chance},
  {Chandler}, {Chaney}, {Chapman}, {Charlot}, {Chayer}, {Cheezum}, {Chen},
  {Chen}, {Cherinka}, {Chichester}, {Chilton}, {Chittiraibalan}, {Clampin},
  {Clark}, {Clark}, {Clark}, {Claybrooks}, {Cleveland}, {Cohen}, {Cohen},
  {Col{\'o}n}, {Coleman}, {Colina}, {Comber}, {Comeau}, {Comer}, {Conde Reis},
  {Connolly}, {Conroy}, {Contos}, {Contreras}, {Cook}, {Cooper}, {Cooper},
  {Correia}, {Correnti}, {Cossou}, {Costanza}, {Coulais}, {Cox}, {Coyle},
  {Cracraft}, {Crew}, {Curtis}, {Cusveller}, {Da Costa Maciel}, {Dailey},
  {Daugeron}, {Davidson}, {Davies}, {Davis}, {Davis}, {Day}, {de Chambure}, {de
  Jong}, {De Marchi}, {Dean}, {Decker}, {Delisa}, {Dell}, {Dellagatta},
  {Dembinska}, {Demosthenes}, {Dencheva}, {Deneu}, {DePriest}, {Deschenes},
  {Dethienne}, {Detre}, {Diaz}, {Dicken}, {DiFelice}, {Dillman}, {Disharoon},
  {Dixon}, {Doggett}, {Dominguez}, {Donaldson}, {Doria-Warner}, {Santos},
  {Doty}, {Douglas}, {Doyon}, {Dressler}, {Driggers}, {Driggers}, {Dunn},
  {DuPrie}, {Dupuis}, {Durning}, {Dutta}, {Earl}, {Eccleston}, {Ecobichon},
  {Egami}, {Ehrenwinkler}, {Eisenhamer}, {Eisenhower}, {Eisenstein}, {El
  Hamel}, {Elie}, {Elliott}, {Elliott}, {Engesser}, {Espinoza}, {Etienne},
  {Etxaluze}, {Evans}, {Fabreguettes}, {Falcolini}, {Falini}, {Fatig},
  {Feeney}, {Feinberg}, {Fels}, {Ferdous}, {Ferguson}, {Ferrarese}, {Ferreira},
  {Ferruit}, {Ferry}, {Filippazzo}, {Firre}, {Fix}, {Flagey}, {Flanagan},
  {Fleming}, {Florian}, {Flynn}, {Foiadelli}, {Fontaine}, {Fontanella},
  {Forshay}, {Fortner}, {Fox}, {Framarini}, {Francisco}, {Franck}, {Franx},
  {Franz}, {Friedman}, {Friend}, {Frost}, {Fu}, {Fullerton}, {Gaillard},
  {Galkin}, {Gallagher}, {Galyer}, {Garc{\'\i}a Mar{\'\i}n}, {Gardner},
  {Garland}, {Garrett}, {Gasman}, {G{\'a}sp{\'a}r}, {Gastaud}, {Gaudreau},
  {Gauthier}, {Geers}, {Geithner}, {Gennaro}, {Gerber}, {Gereau}, {Giampaoli},
  {Giardino}, {Gibbons}, {Gilbert}, {Gilman}, {Girard}, {Giuliano}, {Gkountis},
  {Glasse}, {Glassmire}, {Glauser}, {Glazer}, {Goldberg}, {Golimowski},
  {Gonzaga}, {Gordon}, {Gordon}, {Goudfrooij}, {Gough}, {Graham}, {Grau},
  {Green}, {Greene}, {Greene}, {Greenfield}, {Greenhouse}, {Greve}, {Greville},
  {Grimaldi}, {Groe}, {Groebner}, {Grumm}, {Grundy}, {G{\"u}del}, {Guillard},
  {Guldalian}, {Gunn}, {Gurule}, {Gutman}, {Guy}, {Guyot}, {Hack}, {Haderlein},
  {Hagan}, {Hagedorn}, {Hainline}, {Haley}, {Hami}, {Hamilton}, {Hammann},
  {Hammel}, {Hanley}, {Hansen}, {Hardy}, {Harnisch}, {Harr}, {Harris}, {Hart},
  {Hartig}, {Hasan}, {Hashim}, {Hashimoto}, {Haskins}, {Hawkins}, {Hayden},
  {Hayden}, {Healy}, {Hecht}, {Heeg}, {Hejal}, {Helm}, {Hengemihle}, {Henning},
  {Henry}, {Henry}, {Henshaw}, {Hernandez}, {Herrington}, {Heske}, {Hesman},
  {Hickey}, {Hilbert}, {Hines}, {Hinz}, {Hirsch}, {Hitcho}, {Hodapp}, {Hodge},
  {Hoffman}, {Holfeltz}, {Holler}, {Hoppa}, {Horner}, {Howard}, {Howard},
  {Huber}, {Hunkeler}, {Hunter}, {Hunter}, {Hurd}, {Hurst}, {Hutchings},
  {Hylan}, {Ignat}, {Illingworth}, {Irish}, {Isaacs}, {Jackson}, {Jaffe},
  {Jahic}, {Jahromi}, {Jakobsen}, {James}, {James}, {James}, {Jamieson},
  {Jandra}, {Jayawardhana}, {Jedrzejewski}, {Jeffers}, {Jensen}, {Joanne},
  {Johns}, {Johnson}, {Johnson}, {Johnson}, {Johnson}, {Johnson}, {Johnson},
  {Johnstone}, {Jollet}, {Jones}, {Jones}, {Jones}, {Jones}, {Jones}, {Jordan},
  {Jordan}, {Jue}, {Jurkowski}, {Justis}, {Justtanont}, {Kaleida}, {Kalirai},
  {Kalmanson}, {Kaltenegger}, {Kammerer}, {Kan}, {Kanarek}, {Kao}, {Karakla},
  {Karl}, {Kassin}, {Kauffman}, {Kavanagh}, {Kelley}, {Kelly}, {Kendrew},
  {Kennedy}, {Kenny}, {Keski-Kuha}, {Keyes}, {Khan}, {Kidwell}, {Kimble},
  {King}, {King}, {Kinzel}, {Kirk}, {Kirkpatrick}, {Klaassen}, {Klingemann},
  {Klintworth}, {Knapp}, {Knight}, {Knollenberg}, {Knutsen}, {Koehler},
  {Koekemoer}, {Kofler}, {Kontson}, {Kovacs}, {Kozhurina-Platais}, {Krause},
  {Kriss}, {Krist}, {Kristoffersen}, {Krogel}, {Krueger}, {Kulp}, {Kumari},
  {Kwan}, {Kyprianou}, {Labador}, {Labiano}, {Lafreni{\`e}re}, {Lagage},
  {Laidler}, {Laine}, {Laird}, {Lajoie}, {Lallo}, {Lam}, {LaMassa}, {Lambros},
  {Lampenfield}, {Lander}, {Langston}, {Larson}, {Larson}, {LaVerghetta},
  {Law}, {Lawrence}, {Lee}, {Lee}, {Lee}, {Leisenring}, {Leveille}, {Levenson},
  {Levi}, {Levine}, {Lewis}, {Lewis}, {Lewis}, {Libralato}, {Lidon},
  {Liebrecht}, {Lightsey}, {Lilly}, {Lim}, {Lim}, {Ling}, {Link}, {Link},
  {Lipinski}, {Liu}, {Lo}, {Lobmeyer}, {Logue}, {Long}, {Long}, {Long}, {Long},
  {L{\'o}pez-Caniego}, {Lotz}, {Love-Pruitt}, {Lubskiy}, {Luers}, {Luetgens},
  {Luevano}, {Lui}, {Lund}, {Lundquist}, {Lunine}, {L{\"u}tzgendorf}, {Lynch},
  {MacDonald}, {MacDonald}, {Macias}, {Macklis}, {Maghami}, {Maharaja},
  {Maiolino}, {Makrygiannis}, {Malla}, {Malumuth}, {Manjavacas}, {Marini},
  {Marrione}, {Marston}, {Martel}, {Martin}, {Martin}, {Martinez}, {Maschmann},
  {Masci}, {Masetti}, {Maszkiewicz}, {Matthews}, {Matuskey}, {McBrayer},
  {McCarthy}, {McCaughrean}, {McClare}, {McClare}, {McCloskey}, {McClurg},
  {McCoy}, {McElwain}, {McGregor}, {McGuffey}, {McKay}, {McKenzie}, {McLean},
  {McMaster}, {McNeil}, {De Meester}, {Mehalick}, {Meixner}, {Mel{\'e}ndez},
  {Menzel}, {Menzel}, {Merz}, {Mesterharm}, {Meyer}, {Meyett}, {Meza},
  {Midwinter}, {Milam}, {Miller}, {Miller}, {Miskey}, {Misselt}, {Mitchell},
  {Mohan}, {Montoya}, {Moran}, {Morishita}, {Moro-Mart{\'\i}n}, {Morrison},
  {Morrison}, {Morse}, {Moschos}, {Moseley}, {Mosier}, {Mosner}, {Mountain},
  {Muckenthaler}, {Mueller}, {Mueller}, {Muhiem}, {M{\"u}hlmann}, {Mullally},
  {Mullen}, {Munger}, {Murphy}, {Murray}, {Muzerolle}, {Mycroft}, {Myers},
  {Myers}, {Myers}, {Myers}, {Myrick}, {Nagle}, {Nayak}, {Naylor}, {Neff},
  {Nelan}, {Nella}, {Nguyen}, {Nguyen}, {Nickson}, {Nidhiry}, {Niedner},
  {Nieto-Santisteban}, {Nikolov}, {Nishisaka}, {Noriega-Crespo}, {Nota},
  {O'Mara}, {Oboryshko}, {O'Brien}, {Ochs}, {Offenberg}, {Ogle}, {Ohl},
  {Olmsted}, {Osborne}, {O'Shaughnessy}, {{\"O}stlin}, {O'Sullivan}, {Otor},
  {Ottens}, {Ouellette}, {Outlaw}, {Owens}, {Pacifici}, {Page}, {Paranilam},
  {Park}, {Parrish}, {Paschal}, {Patapis}, {Patel}, {Patrick}, {Pattishall},
  {Paul}, {Paul}, {Pauly}, {Pavlovsky}, {Pe{\~n}a-Guerrero}, {Pedder}, {Peek},
  {Pelham}, {Penanen}, {Perriello}, {Perrin}, {Perrine}, {Perrygo}, {Peslier},
  {Petach}, {Peterson}, {Pfarr}, {Pierson}, {Pietraszkiewicz}, {Pilchen},
  {Pipher}, {Pirzkal}, {Pitman}, {Player}, {Plesha}, {Plitzke}, {Pohner},
  {Poletis}, {Pollizzi}, {Polster}, {Pontius}, {Pontoppidan}, {Porges},
  {Potter}, {Prescott}, {Proffitt}, {Pueyo}, {Quispe Neira}, {Radich}, {Rager},
  {Rameau}, {Ramey}, {Ramos Alarcon}, {Rampini}, {Rapp}, {Rashford},
  {Rauscher}, {Ravindranath}, {Rawle}, {Rawlings}, {Ray}, {Regan}, {Rehm},
  {Rehm}, {Reid}, {Reis}, {Renk}, {Reoch}, {Ressler}, {Rest}, {Reynolds},
  {Richon}, {Richon}, {Ridgaway}, {Riedel}, {Rieke}, {Rieke}, {Rifelli},
  {Rigby}, {Riggs}, {Ringel}, {Ritchie}, {Rix}, {Robberto}, {Robinson},
  {Robinson}, {Robinson}, {Rock}, {Rodriguez}, {Rodr{\'\i}guez del Pino},
  {Roellig}, {Rohrbach}, {Roman}, {Romelfanger}, {Romo}, {Rosales}, {Rose},
  {Roteliuk}, {Roth}, {Rothwell}, {Rouzaud}, {Rowe}, {Rowlands}, {Roy},
  {Royer}, {Rui}, {Rumler}, {Rumpl}, {Russ}, {Ryan}, {Ryan}, {Saad}, {Sabata},
  {Sabatino}, {Sabbi}, {Sabelhaus}, {Sabia}, {Sahu}, {Saif}, {Salvignol},
  {Samara-Ratna}, {Samuelson}, {Sanders}, {Sappington}, {Sargent}, {Sauer},
  {Savadkin}, {Sawicki}, {Schappell}, {Scheffer}, {Scheithauer}, {Scherer},
  {Schiff}, {Schlawin}, {Schmeitzky}, {Schmitz}, {Schmude}, {Schneider},
  {Schreiber}, {Schroeven-Deceuninck}, {Schultz}, {Schwab}, {Schwartz},
  {Scoccimarro}, {Scott}, {Scott}, {Seaton}, {Seely}, {Seery}, {Seidleck},
  {Sembach}, {Shanahan}, {Shaughnessy}, {Shaw}, {Shay}, {Sheehan}, {Sheth},
  {Shih}, {Shivaei}, {Siegel}, {Sienkiewicz}, {Simmons}, {Simon}, {Sirianni},
  {Sivaramakrishnan}, {Slade}, {Sloan}, {Slocum}, {Slowinski}, {Smith},
  {Smith}, {Smith}, {Smith}, {Smith}, {Smith}, {Smolik}, {Soderblom}, {Sohn},
  {Sokol}, {Sonneborn}, {Sontag}, {Sooy}, {Soummer}, {Southwood}, {Spain},
  {Sparmo}, {Speer}, {Spencer}, {Sprofera}, {Stallcup}, {Stanley},
  {Stansberry}, {Stark}, {Starr}, {Stassi}, {Steck}, {Steeley}, {Stephens},
  {Stephenson}, {Stewart}, {Stiavelli}, {}, {Strada}, {Straughn}, {Streetman},
  {Strickland}, {Strobele}, {Stuhlinger}, {Stys}, {Such}, {Sukhatme},
  {Sullivan}, {Sullivan}, {Sumner}, {Sun}, {Sunnquist}, {Swade}, {Swam},
  {Swenton}, {Swoish}, {Tam Litten}, {Tamas}, {Tao}, {Taylor}, {Taylor}, {te
  Plate}, {Van Tea}, {Teague}, {Telfer}, {Temim}, {Texter}, {Thatte},
  {Thompson}, {Thompson}, {Thomson}, {Thronson}, {Tierney}, {Tikkanen},
  {Tinnin}, {Tippet}, {Todd}, {Tran}, {Trauger}, {Trejo}, {Vinh Truong},
  {Tsukamoto}, {Tufail}, {Tumlinson}, {Tustain}, {Tyra}, {Ubeda}, {Underwood},
  {Uzzo}, {Vaclavik}, {Valenduc}, {Valenti}, {Van Campen}, {van de Wetering},
  {Van Der Marel}, {van Haarlem}, {Vandenbussche}, {van Dishoeck},
  {Vanterpool}, {Vernoy}, {Vila Costas}, {Volk}, {Voorzaat}, {Voyton}, {Vydra},
  {Waddy}, {Waelkens}, {Wahlgren}, {Walker}, {Wander}, {Warfield}, {Warner},
  {Wasiak}, {Wasiak}, {Wehner}, {Weiler}, {Weilert}, {Weiss}, {Wells}, {Welty},
  {Wheate}, {Wheeler}, {White}, {Whitehouse}, {Whiteleather}, {Whitman},
  {Williams}, {Willmer}, {Willott}, {Willoughby}, {Wilson}, {Wilson}, {Wilson},
  {Windhorst}, {Wislowski}, {Wolfe}, {Wolfe}, {Wolff}, {Wondel}, {Woo},
  {Woods}, {Worden}, {Workman}, {Wright}, {Wu}, {Wu}, {Wun}, {Wymer},
  {Yadetie}, {Yan}, {Yang}, {Yates}, {Yeager}, {Yerger}, {Young}, {Young},
  {Yu}, {Yu}, {Zak}, {Zeidler}, {Zepp}, {Zhou}, {Zincke}, {Zonak}, \&
  {Zondag}}]{Gardner2023}
{Gardner}, J.~P., {Mather}, J.~C., {Abbott}, R., {et~al.} 2023, \pasp, 135,
  068001, \dodoi{10.1088/1538-3873/acd1b5}

\bibitem[{{Grogin} {et~al.}(2011){Grogin}, {Kocevski}, {Faber}, {Ferguson},
  {Koekemoer}, {Riess}, {Acquaviva}, {Alexander}, {Almaini}, {Ashby}, {Barden},
  {Bell}, {Bournaud}, {Brown}, {Caputi}, {Casertano}, {Cassata}, {Castellano},
  {Challis}, {Chary}, {Cheung}, {Cirasuolo}, {Conselice}, {Roshan Cooray},
  {Croton}, {Daddi}, {Dahlen}, {Dav{\'e}}, {de Mello}, {Dekel}, {Dickinson},
  {Dolch}, {Donley}, {Dunlop}, {Dutton}, {Elbaz}, {Fazio}, {Filippenko},
  {Finkelstein}, {Fontana}, {Gardner}, {Garnavich}, {Gawiser}, {Giavalisco},
  {Grazian}, {Guo}, {Hathi}, {H{\"a}ussler}, {Hopkins}, {Huang}, {Huang},
  {Jha}, {Kartaltepe}, {Kirshner}, {Koo}, {Lai}, {Lee}, {Li}, {Lotz}, {Lucas},
  {Madau}, {McCarthy}, {McGrath}, {McIntosh}, {McLure}, {Mobasher},
  {Moustakas}, {Mozena}, {Nandra}, {Newman}, {Niemi}, {Noeske}, {Papovich},
  {Pentericci}, {Pope}, {Primack}, {Rajan}, {Ravindranath}, {Reddy}, {Renzini},
  {Rix}, {Robaina}, {Rodney}, {Rosario}, {Rosati}, {Salimbeni}, {Scarlata},
  {Siana}, {Simard}, {Smidt}, {Somerville}, {Spinrad}, {Straughn}, {Strolger},
  {Telford}, {Teplitz}, {Trump}, {van der Wel}, {Villforth}, {Wechsler},
  {Weiner}, {Wiklind}, {Wild}, {Wilson}, {Wuyts}, {Yan}, \& {Yun}}]{Grogin2011}
{Grogin}, N.~A., {Kocevski}, D.~D., {Faber}, S.~M., {et~al.} 2011, \apjs, 197,
  35.
\newblock \doarXiv{1105.3753}

\bibitem[{{Harikane} {et~al.}(2022){Harikane}, {Inoue}, {Mawatari},
  {Hashimoto}, {Yamanaka}, {Fudamoto}, {Matsuo}, {Tamura}, {Dayal}, {Yung},
  {Hutter}, {Pacucci}, {Sugahara}, \& {Koekemoer}}]{Harikane2022}
{Harikane}, Y., {Inoue}, A.~K., {Mawatari}, K., {et~al.} 2022, \apj, 929, 1,
  \dodoi{10.3847/1538-4357/ac53a9}

\bibitem[{{Harikane} {et~al.}(2023){Harikane}, {Ouchi}, {Oguri}, {Ono},
  {Nakajima}, {Isobe}, {Umeda}, {Mawatari}, \& {Zhang}}]{Harikane2023}
{Harikane}, Y., {Ouchi}, M., {Oguri}, M., {et~al.} 2023, \apjs, 265, 5,
  \dodoi{10.3847/1538-4365/acaaa9}

\bibitem[{Harris {et~al.}(2020)Harris, Millman, van~der Walt, Gommers,
  Virtanen, Cournapeau, Wieser, Taylor, Berg, Smith, Kern, Picus, Hoyer, van
  Kerkwijk, Brett, Haldane, del R{\'{i}}o, Wiebe, Peterson,
  G{\'{e}}rard-Marchant, Sheppard, Reddy, Weckesser, Abbasi, Gohlke, \&
  Oliphant}]{numpy}
Harris, C.~R., Millman, K.~J., van~der Walt, S.~J., {et~al.} 2020, Nature, 585,
  357, \dodoi{10.1038/s41586-020-2649-2}

\bibitem[{{Heintz} {et~al.}(2023){Heintz}, {Brammer}, {Gim{\'e}nez-Arteaga},
  {Strait}, {del P. Lagos}, {Vijayan}, {Matthee}, {Watson}, {Mason}, {Hutter},
  {Toft}, {Fynbo}, \& {Oesch}}]{Heintz2023}
{Heintz}, K.~E., {Brammer}, G.~B., {Gim{\'e}nez-Arteaga}, C., {et~al.} 2023,
  Nature Astronomy, \dodoi{10.1038/s41550-023-02078-7}

\bibitem[{{Hsiao} {et~al.}(2023){Hsiao}, {Abdurro'uf}, {Coe}, {Larson}, {Jung},
  {Mingozzi}, {Dayal}, {Kumari}, {Kokorev}, {Vikaeus}, {Brammer}, {Furtak},
  {Adamo}, {Andrade-Santos}, {Antwi-Danso}, {Bradac}, {Bradley}, {Broadhurst},
  {Carnall}, {Conselice}, {Diego}, {Donahue}, {Eldridge}, {Fujimoto}, {Henry},
  {Hernandez}, {Hutchison}, {James}, {Norman}, {Park}, {Pirzkal}, {Postman},
  {Ricotti}, {Rigby}, {Vanzella}, {Welch}, {Wilkins}, {Windhorst}, {Xu},
  {Zackrisson}, \& {Zitrin}}]{Hsiao2023}
{Hsiao}, T. Y.-Y., {Abdurro'uf}, {Coe}, D., {et~al.} 2023, arXiv e-prints,
  arXiv:2305.03042, \dodoi{10.48550/arXiv.2305.03042}

\bibitem[{Hunter(2007)}]{Hunter2007}
Hunter, J.~D. 2007, Computing in Science \& Engineering, 9, 90,
  \dodoi{10.1109/MCSE.2007.55}

\bibitem[{{Izotov} {et~al.}(2011){Izotov}, {Guseva}, \& {Thuan}}]{Izotov2011}
{Izotov}, Y.~I., {Guseva}, N.~G., \& {Thuan}, T.~X. 2011, \apj, 728, 161,
  \dodoi{10.1088/0004-637X/728/2/161}

\bibitem[{{Jakobsen} {et~al.}(2022){Jakobsen}, {Ferruit}, {Alves de Oliveira},
  {Arribas}, {Bagnasco}, {Barho}, {Beck}, {Birkmann}, {B{\"o}ker}, {Bunker},
  {Charlot}, {de Jong}, {de Marchi}, {Ehrenwinkler}, {Falcolini}, {Fels},
  {Franx}, {Franz}, {Funke}, {Giardino}, {Gnata}, {Holota}, {Honnen}, {Jensen},
  {Jentsch}, {Johnson}, {Jollet}, {Karl}, {Kling}, {K{\"o}hler}, {Kolm},
  {Kumari}, {Lander}, {Lemke}, {L{\'o}pez-Caniego}, {L{\"u}tzgendorf},
  {Maiolino}, {Manjavacas}, {Marston}, {Maschmann}, {Maurer}, {Messerschmidt},
  {Moseley}, {Mosner}, {Mott}, {Muzerolle}, {Pirzkal}, {Pittet}, {Plitzke},
  {Posselt}, {Rapp}, {Rauscher}, {Rawle}, {Rix}, {R{\"o}del}, {Rumler},
  {Sabbi}, {Salvignol}, {Schmid}, {Sirianni}, {Smith}, {Strada}, {te Plate},
  {Valenti}, {Wettemann}, {Wiehe}, {Wiesmayer}, {Willott}, {Wright}, {Zeidler},
  \& {Zincke}}]{Jakobsen2022}
{Jakobsen}, P., {Ferruit}, P., {Alves de Oliveira}, C., {et~al.} 2022, \aap,
  661, A80, \dodoi{10.1051/0004-6361/202142663}

\bibitem[{{Kennicutt} \& {Evans}(2012)}]{Kennicutt2012}
{Kennicutt}, R.~C., \& {Evans}, N.~J. 2012, \araa, 50, 531,
  \dodoi{10.1146/annurev-astro-081811-125610}

\bibitem[{{Kewley} {et~al.}(2019){Kewley}, {Nicholls}, \&
  {Sutherland}}]{Kewley2019}
{Kewley}, L.~J., {Nicholls}, D.~C., \& {Sutherland}, R.~S. 2019, \araa, 57,
  511, \dodoi{10.1146/annurev-astro-081817-051832}

\bibitem[{{Kniazev} {et~al.}(2004){Kniazev}, {Pustilnik}, {Grebel}, {Lee}, \&
  {Pramskij}}]{Kniazev2004}
{Kniazev}, A.~Y., {Pustilnik}, S.~A., {Grebel}, E.~K., {Lee}, H., \&
  {Pramskij}, A.~G. 2004, \apjs, 153, 429, \dodoi{10.1086/421519}

\bibitem[{{Kocevski} {et~al.}(2023){Kocevski}, {Onoue}, {Inayoshi}, {Trump},
  {Arrabal Haro}, {Grazian}, {Dickinson}, {Finkelstein}, {Kartaltepe},
  {Hirschmann}, {Fujimoto}, {Juneau}, {Amorin}, {Bagley}, {Barro}, {Bell},
  {Bisigello}, {Calabro}, {Cleri}, {Cooper}, {Ding}, {Grogin}, {Ho}, {Inoue},
  {Jiang}, {Jones}, {Koekemoer}, {Li}, {Li}, {McGrath}, {Molina}, {Papovich},
  {Perez-Gonzalez}, {Pirzkal}, {Wilkins}, {Yang}, \& {Yung}}]{Kocevski2023}
{Kocevski}, D.~D., {Onoue}, M., {Inayoshi}, K., {et~al.} 2023, arXiv e-prints,
  arXiv:2302.00012, \dodoi{10.48550/arXiv.2302.00012}

\bibitem[{{Koekemoer} {et~al.}(2011){Koekemoer}, {Faber}, {Ferguson}, {Grogin},
  {Kocevski}, {Koo}, {Lai}, {Lotz}, {Lucas}, {McGrath}, {Ogaz}, {Rajan},
  {Riess}, {Rodney}, {Strolger}, {Casertano}, {Castellano}, {Dahlen},
  {Dickinson}, {Dolch}, {Fontana}, {Giavalisco}, {Grazian}, {Guo}, {Hathi},
  {Huang}, {van der Wel}, {Yan}, {Acquaviva}, {Alexander}, {Almaini}, {Ashby},
  {Barden}, {Bell}, {Bournaud}, {Brown}, {Caputi}, {Cassata}, {Challis},
  {Chary}, {Cheung}, {Cirasuolo}, {Conselice}, {Roshan Cooray}, {Croton},
  {Daddi}, {Dav{\'e}}, {de Mello}, {de Ravel}, {Dekel}, {Donley}, {Dunlop},
  {Dutton}, {Elbaz}, {Fazio}, {Filippenko}, {Finkelstein}, {Frazer}, {Gardner},
  {Garnavich}, {Gawiser}, {Gruetzbauch}, {Hartley}, {H{\"a}ussler},
  {Herrington}, {Hopkins}, {Huang}, {Jha}, {Johnson}, {Kartaltepe},
  {Khostovan}, {Kirshner}, {Lani}, {Lee}, {Li}, {Madau}, {McCarthy},
  {McIntosh}, {McLure}, {McPartland}, {Mobasher}, {Moreira}, {Mortlock},
  {Moustakas}, {Mozena}, {Nandra}, {Newman}, {Nielsen}, {Niemi}, {Noeske},
  {Papovich}, {Pentericci}, {Pope}, {Primack}, {Ravindranath}, {Reddy},
  {Renzini}, {Rix}, {Robaina}, {Rosario}, {Rosati}, {Salimbeni}, {Scarlata},
  {Siana}, {Simard}, {Smidt}, {Snyder}, {Somerville}, {Spinrad}, {Straughn},
  {Telford}, {Teplitz}, {Trump}, {Vargas}, {Villforth}, {Wagner}, {Wandro},
  {Wechsler}, {Weiner}, {Wiklind}, {Wild}, {Wilson}, {Wuyts}, \&
  {Yun}}]{Koekemoer2011}
{Koekemoer}, A.~M., {Faber}, S.~M., {Ferguson}, H.~C., {et~al.} 2011, \apjs,
  197, 36.
\newblock \doarXiv{1105.3754}

\bibitem[{{Labbe} {et~al.}(2023){Labbe}, {Greene}, {Bezanson}, {Fujimoto},
  {Furtak}, {Goulding}, {Matthee}, {Naidu}, {Oesch}, {Atek}, {Brammer},
  {Chemerynska}, {Coe}, {Cutler}, {Dayal}, {Feldmann}, {Franx}, {Glazebrook},
  {Leja}, {Marchesini}, {Maseda}, {Nanayakkara}, {Nelson}, {Pan}, {Papovich},
  {Price}, {Suess}, {Wang}, {Whitaker}, {Williams}, \& {Zitrin}}]{Labbe2023}
{Labbe}, I., {Greene}, J.~E., {Bezanson}, R., {et~al.} 2023, arXiv e-prints,
  arXiv:2306.07320, \dodoi{10.48550/arXiv.2306.07320}

\bibitem[{{Larson} {et~al.}(2022){Larson}, {Finkelstein}, {Hutchison},
  {Papovich}, {Bagley}, {Dickinson}, {Rojas-Ruiz}, {Ferguson}, {Jung},
  {Giavalisco}, {Grazian}, {Pentericci}, \& {Tacchella}}]{Larson22}
{Larson}, R.~L., {Finkelstein}, S.~L., {Hutchison}, T.~A., {et~al.} 2022, \apj,
  930, 104, \dodoi{10.3847/1538-4357/ac5dbd}

\bibitem[{{Larson} {et~al.}(2023){Larson}, {Finkelstein}, {Kocevski},
  {Hutchison}, {Trump}, {Haro}, {Bromm}, {Cleri}, {Dickinson}, {Fujimoto},
  {Kartaltepe}, {Koekemoer}, {Papovich}, {Pirzkal}, {Tacchella}, {Zavala},
  {Bagley}, {Behroozi}, {Champagne}, {Cole}, {Jung}, {Morales}, {Yang},
  {Zhang}, {Zitrin}, {Amor{\'\i}n}, {Burgarella}, {Casey}, {Ch{\'a}vez Ortiz},
  {Cox}, {Chworowsky}, {Fontana}, {Gawiser}, {Grazian}, {Grogin}, {Harish},
  {Hathi}, {Hirschmann}, {Holwerda}, {Juneau}, {Leung}, {Lucas}, {McGrath},
  {P{\'e}rez-Gonz{\'a}lez}, {Rigby}, {Seill{\'e}}, {Simons}, {de La Vega},
  {Weiner}, {Wilkins}, {Yung}, \& {Ceers Team}}]{Larson2023}
{Larson}, R.~L., {Finkelstein}, S.~L., {Kocevski}, D.~D., {et~al.} 2023, \apjl,
  953, L29, \dodoi{10.3847/2041-8213/ace619}

\bibitem[{{Lovell} {et~al.}(2021){Lovell}, {Vijayan}, {Thomas}, {Wilkins},
  {Barnes}, {Irodotou}, \& {Roper}}]{FLARES-I}
{Lovell}, C.~C., {Vijayan}, A.~P., {Thomas}, P.~A., {et~al.} 2021, \mnras, 500,
  2127, \dodoi{10.1093/mnras/staa3360}

\bibitem[{{Madau} \& {Dickinson}(2014)}]{Madau2014}
{Madau}, P., \& {Dickinson}, M. 2014, \araa, 52, 415,
  \dodoi{10.1146/annurev-astro-081811-125615}

\bibitem[{{Madau} \& {Haardt}(2015)}]{Madau2015}
{Madau}, P., \& {Haardt}, F. 2015, \apjl, 813, L8,
  \dodoi{10.1088/2041-8205/813/1/L8}

\bibitem[{{Maiolino} {et~al.}(2023{\natexlab{a}}){Maiolino}, {Scholtz},
  {Curtis-Lake}, {Carniani}, {Baker}, {de Graaff}, {Tacchella}, {{\"U}bler},
  {D'Eugenio}, {Witstok}, {Curti}, {Arribas}, {Bunker}, {Charlot},
  {Chevallard}, {Eisenstein}, {Egami}, {Ji}, {Jones}, {Lyu}, {Rawle},
  {Robertson}, {Rujopakarn}, {Perna}, {Sun}, {Venturi}, {Williams}, \&
  {Willott}}]{Maiolino2023a}
{Maiolino}, R., {Scholtz}, J., {Curtis-Lake}, E., {et~al.} 2023{\natexlab{a}},
  arXiv e-prints, arXiv:2308.01230, \dodoi{10.48550/arXiv.2308.01230}

\bibitem[{{Maiolino} {et~al.}(2023{\natexlab{b}}){Maiolino}, {Scholtz},
  {Witstok}, {Carniani}, {D'Eugenio}, {de Graaff}, {Uebler}, {Tacchella},
  {Curtis-Lake}, {Arribas}, {Bunker}, {Charlot}, {Chevallard}, {Curti},
  {Looser}, {Maseda}, {Rawle}, {Rodriguez Del Pino}, {Willott}, {Egami},
  {Eisenstein}, {Hainline}, {Robertson}, {Williams}, {Willmer}, {Baker},
  {Boyett}, {DeCoursey}, {Fabian}, {Helton}, {Ji}, {Jones}, {Kumari},
  {Laporte}, {Nelson}, {Perna}, {Sandles}, {Shivaei}, \& {Sun}}]{Maiolino2023b}
{Maiolino}, R., {Scholtz}, J., {Witstok}, J., {et~al.} 2023{\natexlab{b}},
  arXiv e-prints, arXiv:2305.12492, \dodoi{10.48550/arXiv.2305.12492}

\bibitem[{{Maseda} {et~al.}(2014){Maseda}, {van der Wel}, {Rix}, {da Cunha},
  {Pacifici}, {Momcheva}, {Brammer}, {Meidt}, {Franx}, {van Dokkum},
  {Fumagalli}, {Bell}, {Ferguson}, {F{\"o}rster-Schreiber}, {Koekemoer}, {Koo},
  {Lundgren}, {Marchesini}, {Nelson}, {Patel}, {Skelton}, {Straughn}, {Trump},
  \& {Whitaker}}]{Maseda2014}
{Maseda}, M.~V., {van der Wel}, A., {Rix}, H.-W., {et~al.} 2014, \apj, 791, 17,
  \dodoi{10.1088/0004-637X/791/1/17}

\bibitem[{{Matthee} {et~al.}(2022){Matthee}, {Mackenzie}, {Simcoe}, {Kashino},
  {Lilly}, {Bordoloi}, \& {Eilers}}]{Matthee2022}
{Matthee}, J., {Mackenzie}, R., {Simcoe}, R.~A., {et~al.} 2022, arXiv e-prints,
  arXiv:2211.08255, \dodoi{10.48550/arXiv.2211.08255}

\bibitem[{{Newville} {et~al.}(2016){Newville}, {Stensitzki}, {Allen}, {Rawlik},
  {Ingargiola}, \& {Nelson}}]{lmfit}
{Newville}, M., {Stensitzki}, T., {Allen}, D.~B., {et~al.} 2016, {Lmfit:
  Non-Linear Least-Square Minimization and Curve-Fitting for Python},
  Astrophysics Source Code Library, record ascl:1606.014.
\newblock \doeprint{1606.014}

\bibitem[{{Oke} \& {Gunn}(1983)}]{Oke1983}
{Oke}, J.~B., \& {Gunn}, J.~E. 1983, \apj, 266, 713, \dodoi{10.1086/160817}

\bibitem[{{Osterbrock}(1989)}]{Osterbrock1989}
{Osterbrock}, D.~E. 1989, {Astrophysics of gaseous nebulae and active galactic
  nuclei}

\bibitem[{pandas~development team(2020)}]{pandas1}
pandas~development team, T. 2020, pandas-dev/pandas: Pandas, 1.3.1,  Zenodo,
  \dodoi{10.5281/zenodo.3509134}

\bibitem[{{Peng} {et~al.}(2010){Peng}, {Ho}, {Impey}, \& {Rix}}]{Peng2010}
{Peng}, C.~Y., {Ho}, L.~C., {Impey}, C.~D., \& {Rix}, H.-W. 2010, \aj, 139,
  2097, \dodoi{10.1088/0004-6256/139/6/2097}

\bibitem[{{P{\'e}rez-Gonz{\'a}lez} {et~al.}(2023){P{\'e}rez-Gonz{\'a}lez},
  {Barro}, {Annunziatella}, {Costantin}, {Garc{\'\i}a-Argum{\'a}nez},
  {McGrath}, {M{\'e}rida}, {Zavala}, {Haro}, {Bagley}, {Backhaus}, {Behroozi},
  {Bell}, {Bisigello}, {Buat}, {Calabr{\`o}}, {Casey}, {Cleri}, {Coogan},
  {Cooper}, {Cooray}, {Dekel}, {Dickinson}, {Elbaz}, {Ferguson}, {Finkelstein},
  {Fontana}, {Franco}, {Gardner}, {Giavalisco}, {G{\'o}mez-Guijarro},
  {Grazian}, {Grogin}, {Guo}, {Huertas-Company}, {Jogee}, {Kartaltepe},
  {Kewley}, {Kirkpatrick}, {Kocevski}, {Koekemoer}, {Long}, {Lotz}, {Lucas},
  {Papovich}, {Pirzkal}, {Ravindranath}, {Somerville}, {Tacchella}, {Trump},
  {Wang}, {Wilkins}, {Wuyts}, {Yang}, \& {Yung}}]{Perez-Gonzalez2023}
{P{\'e}rez-Gonz{\'a}lez}, P.~G., {Barro}, G., {Annunziatella}, M., {et~al.}
  2023, \apjl, 946, L16, \dodoi{10.3847/2041-8213/acb3a5}

\bibitem[{{Planck Collaboration} {et~al.}(2020){Planck Collaboration},
  {Aghanim}, {Akrami}, {Ashdown}, {Aumont}, {Baccigalupi}, {Ballardini},
  {Banday}, {Barreiro}, {Bartolo}, {Basak}, {Battye}, {Benabed}, {Bernard},
  {Bersanelli}, {Bielewicz}, {Bock}, {Bond}, {Borrill}, {Bouchet}, {Boulanger},
  {Bucher}, {Burigana}, {Butler}, {Calabrese}, {Cardoso}, {Carron},
  {Challinor}, {Chiang}, {Chluba}, {Colombo}, {Combet}, {Contreras}, {Crill},
  {Cuttaia}, {de Bernardis}, {de Zotti}, {Delabrouille}, {Delouis}, {Di
  Valentino}, {Diego}, {Dor{\'e}}, {Douspis}, {Ducout}, {Dupac}, {Dusini},
  {Efstathiou}, {Elsner}, {En{\ss}lin}, {Eriksen}, {Fantaye}, {Farhang},
  {Fergusson}, {Fernandez-Cobos}, {Finelli}, {Forastieri}, {Frailis},
  {Fraisse}, {Franceschi}, {Frolov}, {Galeotta}, {Galli}, {Ganga},
  {G{\'e}nova-Santos}, {Gerbino}, {Ghosh}, {Gonz{\'a}lez-Nuevo}, {G{\'o}rski},
  {Gratton}, {Gruppuso}, {Gudmundsson}, {Hamann}, {Handley}, {Hansen},
  {Herranz}, {Hildebrandt}, {Hivon}, {Huang}, {Jaffe}, {Jones}, {Karakci},
  {Keih{\"a}nen}, {Keskitalo}, {Kiiveri}, {Kim}, {Kisner}, {Knox},
  {Krachmalnicoff}, {Kunz}, {Kurki-Suonio}, {Lagache}, {Lamarre}, {Lasenby},
  {Lattanzi}, {Lawrence}, {Le Jeune}, {Lemos}, {Lesgourgues}, {Levrier},
  {Lewis}, {Liguori}, {Lilje}, {Lilley}, {Lindholm}, {L{\'o}pez-Caniego},
  {Lubin}, {Ma}, {Mac{\'\i}as-P{\'e}rez}, {Maggio}, {Maino}, {Mandolesi},
  {Mangilli}, {Marcos-Caballero}, {Maris}, {Martin}, {Martinelli},
  {Mart{\'\i}nez-Gonz{\'a}lez}, {Matarrese}, {Mauri}, {McEwen}, {Meinhold},
  {Melchiorri}, {Mennella}, {Migliaccio}, {Millea}, {Mitra},
  {Miville-Desch{\^e}nes}, {Molinari}, {Montier}, {Morgante}, {Moss}, {Natoli},
  {N{\o}rgaard-Nielsen}, {Pagano}, {Paoletti}, {Partridge}, {Patanchon},
  {Peiris}, {Perrotta}, {Pettorino}, {Piacentini}, {Polastri}, {Polenta},
  {Puget}, {Rachen}, {Reinecke}, {Remazeilles}, {Renzi}, {Rocha}, {Rosset},
  {Roudier}, {Rubi{\~n}o-Mart{\'\i}n}, {Ruiz-Granados}, {Salvati}, {Sandri},
  {Savelainen}, {Scott}, {Shellard}, {Sirignano}, {Sirri}, {Spencer},
  {Sunyaev}, {Suur-Uski}, {Tauber}, {Tavagnacco}, {Tenti}, {Toffolatti},
  {Tomasi}, {Trombetti}, {Valenziano}, {Valiviita}, {Van Tent}, {Vibert},
  {Vielva}, {Villa}, {Vittorio}, {Wandelt}, {Wehus}, {White}, {White},
  {Zacchei}, \& {Zonca}}]{Planck2020}
{Planck Collaboration}, {Aghanim}, N., {Akrami}, Y., {et~al.} 2020, \aap, 641,
  A6, \dodoi{10.1051/0004-6361/201833910}

\bibitem[{{Qin} {et~al.}(2021){Qin}, {Mesinger}, {Bosman}, \& {Viel}}]{Qin2021}
{Qin}, Y., {Mesinger}, A., {Bosman}, S. E.~I., \& {Viel}, M. 2021, \mnras, 506,
  2390, \dodoi{10.1093/mnras/stab1833}

\bibitem[{{Robertson} {et~al.}(2010){Robertson}, {Ellis}, {Dunlop}, {McLure},
  \& {Stark}}]{Robertson2010}
{Robertson}, B.~E., {Ellis}, R.~S., {Dunlop}, J.~S., {McLure}, R.~J., \&
  {Stark}, D.~P. 2010, \nat, 468, 49, \dodoi{10.1038/nature09527}

\bibitem[{{Sanders} {et~al.}(2023){Sanders}, {Shapley}, {Topping}, {Reddy}, \&
  {Brammer}}]{Sanders2023}
{Sanders}, R.~L., {Shapley}, A.~E., {Topping}, M.~W., {Reddy}, N.~A., \&
  {Brammer}, G.~B. 2023, arXiv e-prints, arXiv:2301.06696,
  \dodoi{10.48550/arXiv.2301.06696}

\bibitem[{{Shibuya} {et~al.}(2015){Shibuya}, {Ouchi}, \&
  {Harikane}}]{Shibuya2015}
{Shibuya}, T., {Ouchi}, M., \& {Harikane}, Y. 2015, \apjs, 219, 15,
  \dodoi{10.1088/0067-0049/219/2/15}

\bibitem[{{Smit} {et~al.}(2014){Smit}, {Bouwens}, {Labb{\'e}}, {Zheng},
  {Bradley}, {Donahue}, {Lemze}, {Moustakas}, {Umetsu}, {Zitrin}, {Coe},
  {Postman}, {Gonzalez}, {Bartelmann}, {Ben{\'\i}tez}, {Broadhurst}, {Ford},
  {Grillo}, {Infante}, {Jimenez-Teja}, {Jouvel}, {Kelson}, {Lahav}, {Maoz},
  {Medezinski}, {Melchior}, {Meneghetti}, {Merten}, {Molino}, {Moustakas},
  {Nonino}, {Rosati}, \& {Seitz}}]{Smit2014}
{Smit}, R., {Bouwens}, R.~J., {Labb{\'e}}, I., {et~al.} 2014, \apj, 784, 58,
  \dodoi{10.1088/0004-637X/784/1/58}

\bibitem[{{Spergel} {et~al.}(2007){Spergel}, {Bean}, {Dor{\'e}}, {Nolta},
  {Bennett}, {Dunkley}, {Hinshaw}, {Jarosik}, {Komatsu}, {Page}, {Peiris},
  {Verde}, {Halpern}, {Hill}, {Kogut}, {Limon}, {Meyer}, {Odegard}, {Tucker},
  {Weiland}, {Wollack}, \& {Wright}}]{Spergel2007}
{Spergel}, D.~N., {Bean}, R., {Dor{\'e}}, O., {et~al.} 2007, \apjs, 170, 377,
  \dodoi{10.1086/513700}

\bibitem[{{Storey} \& {Zeippen}(2000)}]{Storey2000}
{Storey}, P.~J., \& {Zeippen}, C.~J. 2000, \mnras, 312, 813,
  \dodoi{10.1046/j.1365-8711.2000.03184.x}

\bibitem[{{Sutherland} {et~al.}(2018){Sutherland}, {Dopita}, {Binette}, \&
  {Groves}}]{Sutherland2018}
{Sutherland}, R., {Dopita}, M., {Binette}, L., \& {Groves}, B. 2018, {MAPPINGS
  V: Astrophysical plasma modeling code}, Astrophysics Source Code Library,
  record ascl:1807.005.
\newblock \doeprint{1807.005}

\bibitem[{{Tang} {et~al.}(2023){Tang}, {Stark}, {Chen}, {Mason}, {Topping},
  {Endsley}, {Senchyna}, {Plat}, {Lu}, {Whitler}, {Robertson}, \&
  {Charlot}}]{Tang2023}
{Tang}, M., {Stark}, D.~P., {Chen}, Z., {et~al.} 2023, arXiv e-prints,
  arXiv:2301.07072, \dodoi{10.48550/arXiv.2301.07072}

\bibitem[{{Terlevich} {et~al.}(1991){Terlevich}, {Melnick}, {Masegosa},
  {Moles}, \& {Copetti}}]{Terlevich1991}
{Terlevich}, R., {Melnick}, J., {Masegosa}, J., {Moles}, M., \& {Copetti},
  M.~V.~F. 1991, \aaps, 91, 285

\bibitem[{{Trump} {et~al.}(2023){Trump}, {Arrabal Haro}, {Simons}, {Backhaus},
  {Amor{\'\i}n}, {Dickinson}, {Fern{\'a}ndez}, {Papovich}, {Nicholls},
  {Kewley}, {Brunker}, {Salzer}, {Wilkins}, {Almaini}, {Bagley}, {Berg},
  {Bhatawdekar}, {Bisigello}, {Buat}, {Burgarella}, {Calabr{\`o}}, {Casey},
  {Ciesla}, {Cleri}, {Cole}, {Cooper}, {Cooray}, {Costantin}, {Croton},
  {Ferguson}, {Finkelstein}, {Fujimoto}, {Gardner}, {Gawiser}, {Giavalisco},
  {Grazian}, {Grogin}, {Hathi}, {Hirschmann}, {Holwerda}, {Huertas-Company},
  {Hutchison}, {Jogee}, {Juneau}, {Jung}, {Kartaltepe}, {Kirkpatrick},
  {Kocevski}, {Koekemoer}, {Lotz}, {Lucas}, {Magnelli}, {Matharu},
  {P{\'e}rez-Gonz{\'a}lez}, {Pirzkal}, {Rafelski}, {Rose}, {Seill{\'e}},
  {Somerville}, {Straughn}, {Tacchella}, {Vanderhoof}, {Weiner}, {Wuyts},
  {Yung}, \& {Zavala}}]{Trump2023}
{Trump}, J.~R., {Arrabal Haro}, P., {Simons}, R.~C., {et~al.} 2023, \apj, 945,
  35, \dodoi{10.3847/1538-4357/acba8a}

\bibitem[{{van der Wel} {et~al.}(2011){van der Wel}, {Straughn}, {Rix},
  {Finkelstein}, {Koekemoer}, {Weiner}, {Wuyts}, {Bell}, {Faber}, {Trump},
  {Koo}, {Ferguson}, {Scarlata}, {Hathi}, {Dunlop}, {Newman}, {Dickinson},
  {Jahnke}, {Salmon}, {de Mello}, {Kocevski}, {Lai}, {Grogin}, {Rodney}, {Guo},
  {McGrath}, {Lee}, {Barro}, {Huang}, {Riess}, {Ashby}, \&
  {Willner}}]{vanderWel2011}
{van der Wel}, A., {Straughn}, A.~N., {Rix}, H.~W., {et~al.} 2011, \apj, 742,
  111, \dodoi{10.1088/0004-637X/742/2/111}

\bibitem[{{Vijayan} {et~al.}(2021){Vijayan}, {Lovell}, {Wilkins}, {Thomas},
  {Barnes}, {Irodotou}, {Kuusisto}, \& {Roper}}]{FLARES-II}
{Vijayan}, A.~P., {Lovell}, C.~C., {Wilkins}, S.~M., {et~al.} 2021, \mnras,
  501, 3289, \dodoi{10.1093/mnras/staa3715}

\bibitem[{Virtanen {et~al.}(2020)Virtanen, Gommers, Oliphant, Haberland, Reddy,
  Cournapeau, Burovski, Peterson, Weckesser, Bright, {van der Walt}, Brett,
  Wilson, Millman, Mayorov, Nelson, Jones, Kern, Larson, Carey, Polat, Feng,
  Moore, {VanderPlas}, Laxalde, Perktold, Cimrman, Henriksen, Quintero, Harris,
  Archibald, Ribeiro, Pedregosa, {van Mulbregt}, \& {SciPy 1.0
  Contributors}}]{scipy}
Virtanen, P., Gommers, R., Oliphant, T.~E., {et~al.} 2020, Nature Methods, 17,
  261, \dodoi{10.1038/s41592-019-0686-2}

\bibitem[{{Wang} {et~al.}(2023){Wang}, {Fujimoto}, {Labb{\'e}}, {Furtak},
  {Miller}, {Setton}, {Zitrin}, {Atek}, {Bezanson}, {Brammer}, {Leja}, {Oesch},
  {Price}, {Chemerynska}, {Cutler}, {Dayal}, {van Dokkum}, {Goulding},
  {Greene}, {Fudamoto}, {Khullar}, {Kokorev}, {Marchesini}, {Pan}, {Weaver},
  {Whitaker}, \& {Williams}}]{Wang2023}
{Wang}, B., {Fujimoto}, S., {Labb{\'e}}, I., {et~al.} 2023, \apjl, 957, L34,
  \dodoi{10.3847/2041-8213/acfe07}

\bibitem[{Waskom(2021)}]{seaborn}
Waskom, M.~L. 2021, Journal of Open Source Software, 6, 3021,
  \dodoi{10.21105/joss.03021}

\bibitem[{{W}es {M}c{K}inney(2010)}]{pandas2}
{W}es {M}c{K}inney. 2010, in {P}roceedings of the 9th {P}ython in {S}cience
  {C}onference, ed. {S}t\'efan van~der {W}alt \& {J}arrod {M}illman, 56 -- 61,
  \dodoi{10.25080/Majora-92bf1922-00a}

\bibitem[{{Wilkins} {et~al.}(2023){Wilkins}, {Lovell}, {Vijayan}, {Irodotou},
  {Adams}, {Roper}, {Caruana}, {Matthee}, {Seeyave}, {Conselice},
  {P{\'e}rez-Gonz{\'a}lez}, {Turner}, {Donnellan}, {Verma}, \&
  {Trussler}}]{Wilkins2023}
{Wilkins}, S.~M., {Lovell}, C.~C., {Vijayan}, A.~P., {et~al.} 2023, \mnras,
  522, 4014, \dodoi{10.1093/mnras/stad1126}

\bibitem[{{Yan} {et~al.}(2023){Yan}, {Ma}, {Ling}, {Cheng}, \&
  {Huang}}]{Yan2023}
{Yan}, H., {Ma}, Z., {Ling}, C., {Cheng}, C., \& {Huang}, J.-S. 2023, \apjl,
  942, L9, \dodoi{10.3847/2041-8213/aca80c}

\bibitem[{{Zackrisson} {et~al.}(2008){Zackrisson}, {Bergvall}, \&
  {Leitet}}]{Zackrisson2008}
{Zackrisson}, E., {Bergvall}, N., \& {Leitet}, E. 2008, \apjl, 676, L9,
  \dodoi{10.1086/587030}

\bibitem[{{Zavala} {et~al.}(2023){Zavala}, {Buat}, {Casey}, {Finkelstein},
  {Burgarella}, {Bagley}, {Ciesla}, {Daddi}, {Dickinson}, {Ferguson}, {Franco},
  {Jim{\'e}nez-Andrade}, {Kartaltepe}, {Koekemoer}, {Le Bail}, {Murphy},
  {Papovich}, {Tacchella}, {Wilkins}, {Aretxaga}, {Behroozi}, {Champagne},
  {Fontana}, {Giavalisco}, {Grazian}, {Grogin}, {Kewley}, {Kocevski},
  {Kirkpatrick}, {Lotz}, {Pentericci}, {P{\'e}rez-Gonz{\'a}lez}, {Pirzkal},
  {Ravindranath}, {Somerville}, {Trump}, {Yang}, {Yung}, {Almaini},
  {Amor{\'\i}n}, {Annunziatella}, {Haro}, {Backhaus}, {Barro}, {Bell},
  {Bhatawdekar}, {Bisigello}, {Buitrago}, {Calabr{\`o}}, {Castellano},
  {Ch{\'a}vez Ortiz}, {Chworowsky}, {Cleri}, {Cohen}, {Cole}, {Cooke},
  {Cooper}, {Cooray}, {Costantin}, {Cox}, {Croton}, {Dav{\'e}}, {de La Vega},
  {Dekel}, {Elbaz}, {Estrada-Carpenter}, {Fern{\'a}ndez}, {Finkelstein},
  {Freundlich}, {Fujimoto}, {Garc{\'\i}a-Argum{\'a}nez}, {Gardner}, {Gawiser},
  {G{\'o}mez-Guijarro}, {Guo}, {Hamilton}, {Hathi}, {Holwerda}, {Hirschmann},
  {Huertas-Company}, {Hutchison}, {Iyer}, {Jaskot}, {Jha}, {Jogee}, {Juneau},
  {Jung}, {Kassin}, {Kurczynski}, {Larson}, {Leung}, {Long}, {Lucas},
  {Magnelli}, {Mantha}, {Matharu}, {McGrath}, {McIntosh}, {Medrano}, {Merlin},
  {Mobasher}, {Morales}, {Newman}, {Nicholls}, {Pandya}, {Rafelski}, {Ronayne},
  {Rose}, {Ryan}, {Santini}, {Seill{\'e}}, {Shah}, {Shen}, {Simons}, {Snyder},
  {Stanway}, {Straughn}, {Teplitz}, {Vanderhoof}, {Vega-Ferrero}, {Wang},
  {Weiner}, {Willmer}, {Wuyts}, \& {CEERS Team}}]{Zavala2023}
{Zavala}, J.~A., {Buat}, V., {Casey}, C.~M., {et~al.} 2023, \apjl, 943, L9,
  \dodoi{10.3847/2041-8213/acacfe}

\end{thebibliography}
\end{document}